%% file: paper_main.tex
\documentclass[12pt]{article}
\input{Preamble/Preamble_main}

\input{Preamble/Preamble_BibLaTeX}
\usepackage[flushleft]{threeparttable}
\usepackage{subcaption}
\usepackage{pdfpages}
\usepackage{longtable}
\usepackage{verbatim}
\usepackage{placeins}
\definecolor{lightblue}{RGB}{100,149,237}
\bibliography{Preamble/library_luca}
\bibliography{Preamble/library_brian}

\renewcommand\footnotemark{}
\begin{document}
\newgeometry{top=1.1in, bottom=0.6in, left=0.7in, right=0.7in} 
\title{ 
\vspace{-1cm}
{\fontsize{22}{22}\selectfont Voice AI in Firms: \\ \vspace{-0.2cm}
\fontsize{18}{18}\selectfont A Natural Field Experiment on Automated Job Interviews}}

\thanks{\newline \scriptsize
        \textbf{Contact:} Brian Jabarian: Carnegie Mellon University, CAAI, CESifo, J-PAL, IGL, JILAEE, \mbox{jabarian@cmu.edu}. Luca Henkel: Erasmus University Rotterdam, ERIM, CESifo, JILAEE, IZA@LISER, Tinbergen Institute; \mbox{henkel@ese.eur.nl}.  \newline
        \textbf{Acknowledgements:} We thank Peter Andre, Iav Bojinov, Felix Chopra, Joshua Deutschmann, Drew Fudenberg, Avi Goldfarb, Mitchell Hoffman, John Horton, Alex Imas, Matthew Jackson, Jon Kleinberg, Hyunjin Kim, Rem Koning, Andrew Koh, David Laibson, Devin Pope, John A. List, Paolo Piacquadio, Raphael Raux, P\"ellumb Reshidi, Frederik Schwerter, Frank Schilbach, Benjamin Schoefer, Sofia Shchukina, Richard Thaler, Catherine Tucker, George Wu, and Luigi Zingales for valuable feedback and suggestions. %
        Bernard Shen, Rishane Dassanayake, Ziyue Feng, Marco Di Giacomo, Shubhaankar Gupta, and Andrew James provided outstanding research assistance.
        \newline
        \textbf{Partnership \& support:} At PSG Global Solutions, we are deeply grateful to Vivek Padmanabhan, David Koch, David Bethea, and Anand Krishnan for their innovative leadership and unwavering scientific partnership. We also thank Glaiza Acevedo, Pritam Gawai, Rob McIntosh, Anand Narole, Jon-Ray Riveras, Ram Sabaru, James Shen, and Shraddha Bhujbal for their indispensable logistical support. At Google Cloud, we are grateful for the infrastructure and engineering support provided by Karim Atek, Brad Burns, and John Cecala. At Chicago Booth, we thank John Buenger and Michael Wu, Rich Cortez and Yvonne McConico, and Cheryl Danton for their guidance on legal, compliance, ethical, and data governance matters. Finally, we are grateful to Emily Bembeneck and Amy Boonstra for their administrative support.
        \newline
        \textbf{Funding:} Brian Jabarian gratefully acknowledges funding support from the UChicago Booth Center for Applied Artificial Intelligence, the Becker-Friedman Institute Program in Behavioral Economics Research, Google Cloud Research Program, Effective Venture US and the Swiss National Science Foundation Grant \#10001768. This project is supported by Schmidt Sciences.
       \newline
        \textbf{Research transparency:}  The study was approved by the University of Chicago IRB (\#IRB24-1894; \#IRB25-1002) and pre-registered on the AEA RCT Registry (\href{https://www.socialscienceregistry.org/trials/15385}{\#15385}). See also Appendix \ref{app_sec:research_transparency}. This study was conducted under the Data-Use Agreement \#160600008471, established by Brian Jabarian through the Booth School of Business at the University of Chicago, with PSG Global Solutions, a subsidiary of Teleperformance.%
         \newline
         \textbf{Disclosure:} The partner firm supplied the data but had no role in analyses, manuscript preparation, or the decision to publish. Both authors and research assistants declare no financial or personal conflicts of interest during the conduct of this study. The pre-registered data collection was completed on June 7, 2025. Nearly two months later, on August 1, 2025, Brian Jabarian accepted an unpaid role as Chief Economist at PSG Global Solutions (a Teleperformance subsidiary), renewing the existing research partnership over the next five years, providing extended access for future data and field experiments within the same pre-agreed publication rights as in the original DUA \#160600008471.
         }

\author{
\begin{tabular}{c cc c}
  Brian Jabarian &&& Luca Henkel \\
\end{tabular}
}

\date{September 10, 2026}

\normalsize
\maketitle

\vspace{-1cm}
\begin{abstract}

\noindent We study AI agents as information-collection technologies: automated systems that elicit decision-relevant signals from humans through live interactions. We test how such AI automation impacts information collection and organizational outcomes using a natural field experiment with 70,000 applicants applying for real jobs. Applicants were randomly assigned to be interviewed by either human recruiters or AI voice agents. Afterward, human recruiters evaluate the interviews and make hiring decisions. Applicants interviewed by AI agents are 12\% more likely to receive job offers, and these gains translate into higher job starts and worker retention, with no decline in the productivity of hired workers. Analyzing interview transcripts reveals that AI voice agents achieve \textit{controlled variance}: their interviews are more structured and consistent while remaining responsive to individual applicants, which is associated with more hiring-relevant information collected. Our results suggest that a key advantage of AI automation lies in environments where information collection is delegated across many human workers and repeated such that variance in task execution becomes noise in decision-relevant signals, which AI compresses through adaptive standardization.

\end{abstract}


\newpage
\restoregeometry

\input{Sections_paper/1_introduction}

\input{Sections_paper/2_background}

\input{Sections_paper/3_experiment}
\input{Sections_paper/4_results}
\input{Sections_paper/5_mechanism}
\input{Sections_paper/6_extensions}
\input{Sections_paper/7_conclusion}

\phantomsection 

\begin{refcontext}[sorting=nyt] 
    \printbibliography[heading=bibintoc]
\end{refcontext}

\clearpage

\appendix
\counterwithin{figure}{section}
\counterwithin{table}{section}

{\LARGE\section*{Appendix}}
\input{Sections_appendix/appendix_additional_figures}

\input{Sections_appendix/appendix_additional_tables}
\input{Sections_appendix/appendix_recruitmentprocess}

\input{Sections_appendix/appendix_interview_transcript_variables}

\input{Sections_appendix/appendix_recruiter_experience}
\input{Sections_appendix/appendix_applicant_survey_instructions}

\input{Sections_appendix/appendix_recruiter_survey_instructions}
\input{Sections_appendix/appendix_prompts}

\input{Sections_appendix/appendix_transparency}
\end{document}

%% file: Preamble/Preamble_main.tex
\usepackage{booktabs}
\usepackage{comment}
\usepackage{rotating}
\usepackage{epsfig}
\usepackage{graphicx}
\usepackage{lscape}
\usepackage{setspace}
\usepackage{amsthm, amssymb,amsmath}
\usepackage{tikz}
\usepackage{enumitem}
\usepackage{soul}
\usepackage[normalem]{ulem}
\usetikzlibrary{arrows.meta,positioning,calc}
\usetikzlibrary{decorations.markings}
\usetikzlibrary{positioning}
\usetikzlibrary{calc}
\usepackage{graphicx}
\usepackage{colortbl}
\usepackage{dcolumn}
\usepackage{appendix}

\definecolor{UBonnBlue}{RGB}{0,6,179}
\definecolor{orange}{RGB}{224, 105, 11}
\definecolor{briqblue}{RGB}{0, 85, 160}
\definecolor{Booth}{RGB}{136, 0, 0}

\definecolor{tempcolor1}{rgb}{0.13,0.44,0.71} 
\definecolor{tempcolor2}{rgb}{0.77,0.86,0.93} 
\definecolor{tempcolor3}{rgb}{0.96,0.98,1} 
\definecolor{tempcolor4}{rgb}{0.42,0.68,0.84} 
\definecolor{tempcolor5}{rgb}{0.26,0.57,0.77} 
\definecolor{tempcolor6}{rgb}{0.62,0.79,0.88} 
\definecolor{tempcolor7}{rgb}{0.87,0.92,0.96} 
\definecolor{tempcolor8}{rgb}{0.03,0.19,0.42} 
\definecolor{black}{rgb}{0.0, 0.0, 0.0} 

\usepackage[
	colorlinks=true,
	linkcolor=UBonnBlue,
	citecolor=UBonnBlue,
	filecolor=black,
	urlcolor=UBonnBlue,
	bookmarks=true,
	bookmarksnumbered=true,
	bookmarksopenlevel=2,
	pdfstartview=Fit,
	pdfpagelayout=SinglePage,
	plainpages=false,
	pdfpagelabels=true
]{hyperref}
\usepackage{makecell}
\usepackage{pict2e}
\usepackage{listings}
\usepackage{titling}
\usepackage[T1]{fontenc}
\usepackage[utf8]{inputenc}
\usepackage[ngerman, american]{babel}
\usepackage{tcolorbox}
\tcbuselibrary{breakable}

\usepackage{array}
\usepackage{multirow}
\usepackage{csquotes}

\usepackage{chngcntr} %

\usepackage{silence}
\usepackage[font=footnotesize]{caption}

\usepackage{listings}
\usepackage{xcolor}

\lstdefinelanguage{JavaScript}{
  keywords={await, async, return, var, if, else, break, for, foreach, null, int, string, Task, Tuple, List, PropertyInfo, GetValue, SetValue},
  keywordstyle=\color{blue}\bfseries,
  ndkeywords={class, export, boolean, throw, implements, import, this},
  ndkeywordstyle=\color{blue}\bfseries,
  identifierstyle=\color{black},
  sensitive=false,
  comment=[l]{//},
  morecomment=[s]{/*}{*/},
  commentstyle=\color{gray}\ttfamily,
  stringstyle=\color{red}\ttfamily,
  morestring=[b]',
  morestring=[b]"
}

\makeatletter

\usepackage{bbm}

\makeatother
\usepackage[bottom, multiple]{footmisc}

\makeatother
\makeatletter
\renewcommand\footnoterule{%
  \vspace{0.8em}%
 \kern-3\p@\hrule\@width.4\columnwidth%
   \kern2.6\p@}
\makeatother
\usepackage[margin=1in]{geometry}

\def\sym#1{\ifmmode^{#1}\else\(^{#1}\)\fi}

\usepackage[semibold, lf, tabular, scale=0.91]{FiraSans} \usepackage[lf, scale=0.91]{FiraMono}	%
\usepackage[protrusion=true, expansion=false, kerning=true]{microtype}
\usepackage{mathpazo}

\usepackage{ctable}
\usepackage{eurosym}

\usepackage[font={sf, small}]{floatrow}	%
\renewcommand{\floatfootskip}{10mm}

\newcommand{\figurenotes}[2][Notes:]{%
	\floatfoot{%
		\setlength{\baselineskip}{11pt}%
		\vspace{-\floatfootskip}%
		\\[-\baselineskip]
		\textit{#1} #2%
	}%
}

\makeatletter
\renewcommand{\paragraph}{%
  \@startsection{paragraph}{4}%
  {\z@}{1.75ex \@plus 1ex \@minus .2ex}{-1em}%
  {\normalfont\normalsize\bfseries}%
}
\makeatother

\newtheorem*{lemma*}{Lemma}

%% file: Preamble/Preamble_BibLaTeX.tex
\usepackage[backend=biber, natbib=true, bibencoding=inputenc, bibstyle=authoryear, citestyle=authoryear-comp, mincitenames=1, maxcitenames=2, minbibnames=99, maxbibnames=99, uniquename=false, uniquelist=false, url=true, backrefstyle=three, doi=false, isbn=false, dashed=false, sorting=ynt, sortcites=true, mergedate=true, dateabbrev=false, eprint=false, backref=true, %
]{biblatex}

\DeclareBibliographyAlias{newspaper}{article}

\renewcommand{\cite}{\citet}

\defbibheading{subbibliography}[\refname]{%
	\section*{#1}%
	\sectionmark{#1}%
	\addcontentsline{toc}{section}{#1}%
}

\DefineBibliographyStrings{english}{%
	andothers = {et~al\adddot}
}

\usepackage{xpatch}
\xpretobibmacro{author}{\mkbibbold\bgroup}{}{}
\xapptobibmacro{author}{\egroup}{}{}
\renewbibmacro{in:}{}

\DeclareFieldFormat{pages}{\mkcomprange{#1}}

\AtEveryBibitem{

}

\renewcommand*{\bibnamedash}{---{\kern -2.25pt}---{\kern -2.25pt}---\addcomma\space}

\DefineBibliographyStrings{english}{byeditor = {\addcomma\space ed\adddot}}
\renewbibmacro*{byeditor}{%
	\ifnameundef{editor}
	{}
	{
		\printnames[byeditor]{editor}%
		\setunit{\addspace}%
		\usebibmacro{bytypestrg}{editor}{editor}%
		\newunit}%
	\usebibmacro{byeditorx}}
\renewbibmacro*{byeditor+others}{%
	\ifnameundef{editor}
	{}
	{
		\printnames[byeditor]{editor}%
		\clearname{editor}%
		\setunit{\addspace}%
		\usebibmacro{byeditor+othersstrg}%
		\newunit}%
	\usebibmacro{byeditorx}%
	\usebibmacro{bytranslator+others}}

\renewcommand*{\intitlepunct}{\space}
\renewbibmacro{in:}{%
	\ifentrytype{article}%
		{}%
		{\ifentrytype{newspaper}%
			{}%
			{\printtext{\bibstring{in}\intitlepunct}}%
		}%
}

\DeclareFieldFormat[article]{journaltitle}{%
	\mkbibemph{#1}%
	\iffieldundef{volume}{\addperiod}{}%
}

\DeclareFieldFormat[newspaper]{journaltitle}{%
	\mkbibemph{#1} \mkbibparens{\thefield{edition}}\addcomma\addspace%
	\mkbibmonth{\thefield{month}}\addnbspace\thefield{day}\addcomma\addspace\thefield{year}%
	\iffieldundef{volume}{\addcolon}{\addcomma}%
}

\DeclareFieldFormat{month}{\mkbibmonth{#1}}

\DeclareFieldFormat[thesis, unpublished, report, misc, newspaper]{title}{\mkbibquote{#1}}

\DeclareFieldFormat[thesis, unpublished, report, misc]{type}{\MakeCapital{#1}}

\DeclareFieldFormat[incollection]{series}{#1\nopunct}

\DeclareFieldFormat[book, inbook, incollection, inproceedings]{volume}{vol.\addnbspace#1\addcomma}
\DeclareFieldFormat[article]{volume}{%
	\iffieldundef{volume}{}{%
		\addspace#1%
		\iffieldundef{number}{\addcolon}{\nopunct}%
	}
}
\DeclareFieldFormat[article]{number}{\addnbthinspace\mkbibparens{#1}\addcolon}
\DeclareFieldFormat[article]{date}{#1}

\DeclareSourcemap{
    \maps[datatype=bibtex]{
        \map{
            \pertype{article}
			\step[notfield=number, final]
			\step[fieldsource=month, fieldtarget=number]
            \step[fieldset=month, null]
            \step[fieldset=month_numeric, null]
        }
    }
}

\DefineBibliographyStrings{english}{%
	backrefpage  = {},	%
	backrefpages = {}	%
}

\renewbibmacro*{pageref}{%
	\addperiod
	\iflistundef{pageref}
	{}
	{\printtext[brackets]{%
			\ifnumgreater{\value{pageref}}{1}
			{\bibstring{backrefpages}}
			{\bibstring{backrefpage}}%
			\printlist[pageref][-\value{listtotal}]{pageref}%
		}%
	}%
}

\DeclareSourcemap{
	\maps[datatype=bibtex]{
		\map[overwrite]{
		 	\step[fieldsource=annote]
		 	\step[fieldset=note, origfieldval, append]
			\step[fieldset=annote, null]
		}
		\map{
			\step[fieldsource=note, final]
			\step[fieldset=addendum, origfieldval, final]
			\step[fieldset=note, null]
		}
	}
}

\AtEveryBibitem{%
	\ifentrytype{newspaper}%
		{\clearfield{addendum}}%
		{\clearfield{month}}%
}

\DeclareSourcemap{
	\maps[datatype=bibtex]{
		\map{%
			\pertype{techreport}
			\step[fieldset=type, fieldvalue={Working paper}]
		}
	}
}

\DeclareSourcemap{ 
	\maps[datatype=bibtex]{
		\map{
			\step[fieldsource=journal, match={\regexp{^The\s}}, replace={}]
		}
	}      
}
\renewbibmacro*{url+urldate}{}

\AtEveryBibitem{%
  \clearfield{month}%
  \clearfield{day}%
}

%% file: Sections_paper/1_introduction.tex
\section{Introduction}
\label{main_sec:introduction}

Artificial intelligence (AI) can enter economic activity by augmenting or automating tasks, and by providing or collecting information. Most existing empirical work studies augmented information provision: AI assists humans by providing advice, feedback, or recommendations in tasks such as writing, coding, customer support, or entrepreneurial activity \parencite{noyExperimentalEvidenceProductivity2023, wilesMoreWorseImpact2024, Brynjolfsson2025, otis2025uneven, Chen2024creative}. Much less is known about AI systems that automate the live collection of information from humans. Existing studies show that AI can serve as a methodical tool for tasks such as conducting qualitative interviews \parencite{geieckeConversationsScaleRobust2024,chopraConductingQualitativeInterviews2024} or annotating tasks \parencite{shleifer2026gpt}. However, evidence on the real-world impact of AI automating information collection from humans remains scarce. 

This paper provides evidence from a large-scale natural field experiment in which AI voice agents automate job interviews at scale. Our setting is a high-volume hiring market in which applicants for customer service jobs are randomly assigned to have their job interview either with an AI voice agent or with a human recruiter. We find that interviews conducted by an AI voice agent rather than by human recruiters lead to higher job offer rates, job starts, and worker retention, with no decline in the productivity of hires. We trace these gains to \textit{controlled variance} as mechanism: AI-led interviews are more structured and consistent, which reduces interviewer-driven dispersion. At the same time, they remain adaptive to individual applicants, allowing the AI voice agent to collect more hiring-relevant information. These results suggest that when information collection is delegated across many human workers and repeated at scale, execution variance enters decision-relevant signals. AI agents compress this variance through standardization while preserving adaptiveness, thereby improving information collection and thus organizational outcomes. 

Job interviews at scale exemplify how variance in information collection can undermine hiring outcomes. Interviews are designed to collect information about applicants’ skills, motivation, and fit \parencite{PeoplePracticesProductivity2025}, yet substantial evidence shows that human-led interviews are prone to both between-recruiter and within-recruiter variation, especially when discretion is high.\footnote{For instance, recruiters differ in their degree of bias towards specific applicants \parencite{Shukla2025,Mocanu2024,kauselOverconfidenceP2016} and their judgment is influenced by previous applicants \parencite{radbruchInterviewSequences2025}. More generally, meta-analyses have shown that human-led interviews exhibit low validity in detecting talent, especially when unstructured \parencite[e.g.,][]{mcdanielValidityEmploymentInterviews1994,judgeEmploymentInterviewReview2000,levashinaStructuredEmploymentInterview2014}.\label{footnote_interview_biases}} Such variation can distort hiring outcomes by introducing noise unrelated to applicant quality. Firms typically try to reduce this noise by standardizing interviews through guidelines and structured processes. Yet standardization can also weaken a key advantage of interviews: the ability to ask follow-up questions that clarify, probe, and expand on applicant responses. When hiring requires many human recruiters, firms therefore face a trade-off between standardization and adaptiveness. Employing AI agents may help firms better navigate this trade-off, as these agents can follow guidelines while remaining responsive to each applicant in a more consistent and controlled manner. Whether this capability transfers to real-world settings, and whether it reduces undesirable variation in practice remains an open empirical question.

We conducted our natural field experiment in partnership with a recruitment firm\footnote{PSG Global Solutions --  a recruitment process outsourcing firm and a subsidiary of Teleperformance, a global provider of business process outsourcing services.} involving 70,884 applications for entry-level customer service jobs. Applicants who pass an initial screening are randomly assigned to be interviewed by a human recruiter (\textit{Human interviewer} condition) or by an AI voice agent (\textit{AI interviewer}). In a third condition, they choose between the two (\textit{Choice of interviewer}). Crucially, all hiring decisions are made by human recruiters. The design therefore separates the information collection stage -- automated by AI in one treatment -- from the evaluation stage, which remains under human control. We collected comprehensive data at the interview, applicant, and recruiter levels, including interview transcripts and linked applicant-worker data, to examine how AI-led interviews change the information collected in interviews and how these changes translate into hiring outcomes.

Recruiters are instructed to follow interview guidelines that specify the topics to be covered, their recommended sequence, and sample questions. These guidelines nevertheless grant recruiters substantial flexibility: they may tailor topic coverage, wording, and follow-up questions to individual applicants. The AI voice agent uses the same guidelines. After the interview, applicants take a standardized test assessing language and analytical skills. Recruiters then assess interview and test performance and determine whether the applicant meets the firm's standard.\footnote{Recruiters evaluate both AI-led and human-led interviews. When human-led, they assess the interviews conducted by themselves.} The firm's performance metrics for hiring quality are the likelihood that a hire starts the job and remains employed for at least a month. 

We find that applicants in the \textit{Human Interviewer} condition receive a job offer in 8.70\% of cases, compared with 9.73\% in the \textit{AI Interviewer} condition -- a 12\% higher likelihood of receiving a job offer. Among all applicants randomized into either treatment, applicants in the \textit{AI Interviewer} condition also have an 18\% higher likelihood of starting their job ($p<0.001$) and an 18\% higher likelihood of having an employment spell lasting at least one month ($p<0.001$). These positive effects persist at longer horizons: applicants are 19\% more likely to be employed after two months, 18\% after three months, and 19\% after four months. Conditional on accepting an offer, applicants in the \textit{AI Interviewer} condition have a 7\% higher likelihood of starting their job ($p=0.003$) and a 6\% higher likelihood of still being employed after one month ($p=0.025$), showing that the increases in retention are not solely driven by offer-stage selection.

These retention gains do not appear to come at the cost of worker productivity. First, we find no significant differences in the reasons for separation between hired workers. The fraction of workers who leave involuntarily versus voluntarily does not differ between workers hired in the \textit{AI Interviewer} and the \textit{Human Interviewer} condition. Second, for a subset of hires, we observe performance measures. We find neither statistically significant nor economically meaningful differences in the average time a worker spends handling customers, the satisfaction scores of the customers they handle, or the quality assurance scores assigned by their employer. As these three measures jointly determine productivity in our work context, there appear to be no differences in productivity between hiring modes.

To understand why offer rates are higher under AI, we analyze interview transcripts using natural language processing.\footnote{Interview performance matters in our setting: interview scores strongly predict offer decisions, even after controlling for test scores. In addition, 96\% of recruiters state in our recruiter survey that they consider interview performance at least as important as test scores in their offer decisions.} We focus on how AI and humans navigate the trade-off between standardization and adaptiveness and its consequences for information collection. Along the standardization dimension, the AI agent implements a more structured interview process: compared to human recruiters, it adheres more closely to the guideline topic order, covers a more consistent number of topics across interviews, and uses more standardized wording. The AI agent is more consistent than the average human recruiter on each variable, and more consistent than the top quartile of recruiters on following guidelines. Along the adaptiveness dimension, both AI- and human-led interviews contain a significant share of follow-up questions that respond to applicant input. Human recruiters ask more follow-up questions, but these questions are more concentrated on applicant responses that already provide sufficient information. This pattern suggests that the AI navigates the standardization-adaptiveness trade-off more effectively. Indeed, we find that AI-led interviews yield more hiring-relevant information about applicants. Our results further suggest that this information is beneficial: applicants in AI-led interviews display stronger communication signals that, in human-led interviews, predict higher offer rates, such as sustained conversational exchange, and fewer signals associated with lower offers, such as backchannel cues.

Together, these patterns suggest that a key advantage of AI-led information collection lies in its ability to achieve what we call \textit{controlled variance}: the AI voice agent adapts its flow to each applicant within a standardized framework, reducing interviewer-driven dispersion while preserving within-interview adaptiveness. This feature allows the AI voice agent to navigate the standardization-adaptiveness trade-off so that more and higher-quality information is collected. 

We then examine applicants' behavioral responses to AI voice agents and find no evidence of backlash. Applicants accept job offers with similar likelihoods in the \textit{AI Interviewer} condition relative to the \textit{Human Interviewer} condition. The industry's key applicant satisfaction metric -- the likelihood of recommending the firm to a friend (Net Promoter Score) -- is almost identical across treatments. In a detailed candidate experience survey, applicants also rate interview quality variables such as perceived stress, comfort, follow-up fluency, and feedback quality similarly between treatments. One notable exception is perceived naturalness: AI-led interviews are rated as significantly less natural.%

When given the choice, most applicants prefer an AI interviewer over a human recruiter: in the \textit{Choice of interviewer} condition, 78\% choose the AI voice agent. Survey evidence shows generally positive attitudes toward AI in this sample, and these attitudes predict interviewer choice. However, along the quality dimension, we find evidence of negative sorting into AI: applicants who choose the AI voice agent have significantly lower language and analytical scores than those who choose a human recruiter. 

Finally, we analyze how recruiters evaluate applicants. A total of 131 recruiters evaluated applicants, with a core group of 43 handling most of them. Recruiters assign significantly higher interview scores to AI-interviewed applicants than to those they interview themselves. Aggregating to the recruiter level shows that these differences are widespread: they are not driven by a few outliers, but by a majority of recruiters who provide higher scores and extend more offers to applicants interviewed by AI. Yet recruiters also weigh the quality signals they receive from interviews and standardized tests differently across conditions. When evaluating AI-interviewed applicants, human recruiters place less weight on interview scores and more on language scores than when evaluating applicants interviewed by themselves. 

\paragraph{Related literature.} Our paper contributes to three strands of research. First, we contribute to the literature on the economic impact of generative AI tools. A growing body of evidence shows that these tools can enhance productivity in diverse tasks such as writing, customer support, coding or software development \parencite{noyExperimentalEvidenceProductivity2023, choiAIAssistanceLegal2024, wilesMoreWorseImpact2024, Brynjolfsson2025, pengImpactAIDeveloper2023, doshi2023generative, kumar2025math, otis2025uneven, Chen2024creative, dellacquaNavigatingJaggedTechnological2026,kanazawaAISkillProductivity2026}. In these studies, AI tools primarily augment human work through assistance and information provision. In contrast, we examine a setting in which an AI agent replaces humans in an expert task: conducting job interviews. Here, the AI agent autonomously collects and interprets information generated during a natural-language conversation. Thus, we provide causal evidence from a natural setting on the impact of automating a production stage using AI agents, both in terms of economic consequences and human behavioral responses. Our results highlight that a key advantage of AI-based systems may lie in their ability to implement \textit{controlled variance}, which informs the debate on which tasks or environments AI automation may have the largest effects.\footnote{Recruiters typically perform two expert tasks: conducting job interviews and evaluating applicants. Our evidence thus offers empirical input to the debate on whether AI shifts human labor to higher expertise \parencite{brynjolfsson2014second, acemoglu2019automation, athey2020allocation, gruber2020managing, alam2024automation, autor2024applying, autor_thompson_2025_expertise}, in this case by redirecting recruiter expertise toward evaluation and potentially raising standards in low-entry job assessments.}

Second, our hiring setting contributes to the literature on the use of AI and algorithms in labor markets. A large body of work examines technologies that help recruiters select applicants before job interviews or evaluate them afterward \parencite[see, e.g.,][for a review]{PeoplePracticesProductivity2025}. Algorithmic recommendations increase match quality and fill rates \parencite{hortonEffectsAlgorithmicLabor2017}, AI screening algorithms increase hiring success \parencite{awuahAutomation2025}, and optimized algorithms influence the quality of applicants selected for interviews \parencite{liHiringExploration2026} and hires \parencite{dargniesBehavioralMeasuresImprove2025}. AI evaluation tools change how recruiters assess applicants \parencite{averyDoesArtificialIntelligence2024}, how underrepresented groups are assessed \parencite{averyBraveNewWorld}, and the composition of hired workers \parencite{cowgillBiasProductivityHumans2020}. In these papers, however, the interview process remains under human control. In our setting, AI agents conduct the interview itself without any human supervision. \textcite{aka2025better} conduct a field experiment in which applicants are screened either by a resume-scoring algorithm or by completing first an AI-led interview, before a human recruiter conducts a second human-led interview. They find that AI-screened applicants are more likely to pass the subsequent interviews. In contrast, our comparison is not between an algorithm and AI, but between human and AI interviewers, which allows us to study our novel controlled-variance mechanism. Moreover, we have linked hiring-employment data, allowing us to analyze recruiter and applicant behavior as well as quantify the impact of randomized AI employment on organizational and economic outcomes, including job starts, retention, and productivity. Accordingly, we relate to a broader literature that studies how recruiters use quality signals and make decisions based on them, see e.g., \textcite{hoffmanDiscretionHiring2018,chapmanApplicantAttractionOrganizations2005}.\footnote{Relatedly, recent work studies screening technology choice as an endogenous signal of applicant type \parencite{JabarianReshidiChoice}.}

Third, we contribute to behavioral research on how humans, and in particular workers and managers in labor settings, perceive, trust, and interact with AI systems. On the information collection side, recent work on AI in persuasion finds that AI agents are less effective than humans in debt collection calls \parencite{choiHowGoodAI2025}, while being capable of changing beliefs in conspiracy theories \parencite{costelloDurablyReducingConspiracy2024} and motivating behavioral change in people \parencite{chopraEvaluatingBehavioralInterventions2026}. On the information provision side, \textcite{agarwalCombiningHumanExpertise2024,agarwalDesigningHumanAICollaboration2025} show that humans under-respond to AI predictions in the contexts of radiology and fact-checking. \textcite{angelovaAlgorithmicRecommendationsHuman2025} and \textcite{stevensonAlgorithmicRiskAssessment2024} show that judges frequently deviate from or override AI recommendations.\footnote{Moreover, \textcite{dargniesAversionHiringAlgorithms2026,fumagalliOKComputerWorker2022} study people's perceptions and preferences towards algorithms in hiring.} We add field evidence on real-world economic outcomes and rich behavioral data on human-AI interactions.

%% file: Sections_paper/2_background.tex
\section{Background and experimental design}
\label{main_sec:background}

We study the impact of introducing an AI voice agent that collects information from humans through spoken conversations. Specifically, the agent conducts hiring interviews, thus interacting with the applicants in a conversation. The goal is to collect information from this interaction that recruiters can use later to make hiring decisions. In the following, we describe the AI system deployed, the economic environment in which the hiring takes place, as well as the design of the field experiment. 
  
\subsection{AI voice agents}

\paragraph{Technical architecture.} AI voice agents are a specific class of recently developed ``generative AI'' tools that generate new data after being trained on existing data using machine learning. The purpose of AI voice agents is to communicate with humans through natural language conversations. To enable conversations, the agent generates human-like speech and responds to human speech based on three interacting technological systems. 

First, to generate the content of AI-based speech, the agent generates text using a large language model (LLM). LLMs are trained on large amounts of text and are tasked with predicting the next word in a sequence based on prior context. This enables them to generate contextually relevant content in natural language. Second, to produce the speech itself, the agent is equipped with a text-to-speech system, which converts the text generated by the LLM into audible speech using a multi-step process. It first models pronunciation by translating text into phonetic representations. Then, it determines the appropriate prosodic features, such as stress and intonation, and regulates the pace for natural delivery. Finally, a neural voice encoder (vocoder) generates natural-sounding speech. Third, to respond to speech inputs from the human counterpart, the voice agent converts the spoken input into text using an automatic speech recognition system. This process involves acoustic modeling of the input waveform to identify phonetic units, followed by lexical and language modeling to infer the most likely word sequence. In this step, again, a large language model is used to infer content from speech with sufficient accuracy, which is particularly important in cases of noisy or poorly articulated speech. The first system then uses this input to generate a response. 

\paragraph{Challenges to implementation.} Using AI voice agents for spoken conversation poses several challenges to AI systems, which are particularly pronounced in interviewing. Human language is multifaceted, layered, and complex. Any lapses, misunderstandings, or errors on the AI side will reduce the experience of the applicants. They also make it more difficult or even impossible for recruiters to evaluate the interview, leading to information loss. Moreover, it is important that the AI remains on topic during the entire conversation. This requires minimizing instances of ``hallucinations,'' where LLM-based tools unpredictably generate coherent but factually incorrect or nonsensical output. Similarly, guardrails need to be in place to prevent the AI from going off-topic. Moreover, the AI needs to be secure against attempts by applicants to game it, e.g., if the applicant parrots buzzwords or reads from a script. 

While most of these challenges are shared with text-based AI tools, additional challenges arise for spoken conversations. Any functioning tool must deal with ambient noise, variations in speech rate, and differences in accents and intonation. In addition, the time to respond -- latency -- becomes even more important as multiple systems need to work together, potentially increasing latency. At the same time, substantial delays in questions and responses break the flow of a conversation. This, in turn, may decrease comfort and increase stress for the human counterpart. 

\subsection{Economic environment}

\paragraph{Data partner.} We partner with a firm that has employed an AI voice agent in its hiring process. The firm is the recruitment process outsourcing (RPO) firm PSG Global Solutions (hereafter referred to as ``the firm''), integrated into the \$11 billion global business process outsourcing (BPO) firm Teleperformance. The firm specializes in high-volume recruitment for Fortune 500 clients across the healthcare, IT, and industrial sectors, with recruiting centers worldwide. 

\paragraph{Job descriptions.} Our setting is the Philippines, where the firm recruits customer service representatives for large US-based and European clients. The jobs for which our firm recruits pay between Php 16,000 and Php 25,000 per month ($\approx\$280$ to $\$435$).\footnote{As reference, at the time of the experiment, the minimum wage in the Philippines ranged from \$125 to \$260 per month, depending on the region.} Required skills include English fluency, communication skills, flexibility to work in changing shifts, strong analytical and logical thinking, and problem-solving skills. See Appendix Section \ref{app_sec:jobdescriptions} for an example of a detailed job description.

\paragraph{Industry background.}  The customer service industry in the Philippines is a major sector, estimated to employ more than 1.5 million workers \parencite{Hernandez2024}. The Philippines has become the world's leading provider of call center support, with a large proportion working to assist customers in the US. Factors driving this growth include, on the labor supply side, (i) a large share of young and comparatively well-educated individuals, particularly in terms of English language proficiency, and (ii) comparatively high wages, as call center jobs offer better pay given their skill requirements compared to alternative employment options within the Philippines. On the demand side, wages are relatively low compared to high-income countries such as the US, and the Philippine accent is close to a typical US-American accent. These factors make it attractive for US-based firms to outsource customer support to call centers in the Philippines. At the same time, the call center industry is characterized by high turnover. Estimates from the US suggest that up to 60\% of call center workers leave each year \parencite{Buesing2020}, and similar numbers have been suggested for the Philippines \parencite{Sallaz2019}. Possible reasons workers voluntarily leave include the repetitive and stressful nature of the work, combined with tight monitoring. The high attrition rates lead firms to devote substantial resources to continuous recruitment and training of new workers \parencite{Berg2018}.

Together, these factors mean recruiting firms face large volumes of applicants. This has resulted in a competitive market in which multiple recruitment firms compete to identify and recruit suitable applicants. Within the large pool of applications, there is a substantial fraction of applicants who are unfit for the job and need to be screened out by the recruiters. There also exists a smaller but significant fraction of applicants who are highly qualified, but who often use the job as a temporary bridge to their next career step (e.g., going to university), thus leaving the job after only a short time. These similarly need to be identified and screened out by the recruiters.

\subsection{Recruitment process}
\label{sub_sec:background_recruitment_process}

The firm has established low entry barriers to interview as many applicants as possible. Applicants learn about job openings through job advertisements posted on various platforms such as Indeed, Facebook, the firm's website, its digital campaigns, or through referrals. 

\paragraph{Recruitment modes.} After learning about a job opening, applicants have two ways to apply, which determine the subsequent recruitment mode. In the \textit{Remote} mode, applicants submit an expression of interest online, which contains contact details and some basic information. In the \textit{Walk-in} mode, applicants come in person to the nearest firm recruitment site. Once the firm receives an application, a profile is created in the recruitment software and the formal recruitment process begins. Appendix Figure \ref{fig:hiring-funnel} provides an overview of the process, which we will discuss in detail next.

\paragraph{Screening process.} The firm's approach is to conduct the primary screening during the interview itself. Hence, little to no screening is performed prior to the interview. In the \textit{Remote} mode, the primary variable for selection is the engagement score of an applicant. The engagement score is generated algorithmically based on the level of detail provided by the applicants in their expression of interest. Applicants whose scores are below a certain threshold are screened out.\footnote{See Appendix Section \ref{app_sec:engagement_score_algorithm}
 for more details.} The rest are invited to an interview. For applicants in the \textit{Walk-in} mode, no engagement score is calculated; instead, all eligible applicants are invited to an interview after expressing their interest in person. Interview invitations are sent by telephone, text messages, and email. See Appendix Section \ref{app_sec:invitation_texts} for the content of the invitations. Applicants are assigned randomly to recruiters using a round-robin scheduling algorithm \parencite{SilberschatzEtAl2018}.

\paragraph{Job interviews.} Interviews are conducted in two modes. In the \textit{Walk-in} mode, interviews take place in person, while in the \textit{Remote} mode, the interview takes place remotely by telephone. A full-length interview takes between 10 and 20 minutes. Recruiters follow structured interview guidelines designed to ensure a standardized interview process. There is a maximum of 14 topics that can be covered in each interview (see Appendix Table \ref{tab:storyboard_full} for details), and questions are a mix of verification and open-ended questions. Interviews start with questions about the suitability of applicants for the position, such as their current location, willingness to commute, and flexibility with respect to the work schedule. Recruiters then ask applicants about their career goals and motivations, before shifting to questions about previous work experience and their education level. Toward the end of the interview, recruiters provide additional details about the position. They also provide applicants with the opportunity to ask questions about the position and the recruitment process. The guidelines allow recruiters significant flexibility to adapt their approach. For instance, recruiters are asked to tailor their questions and follow-up questions to the applicant's background to assess aspects like gaps in employment or transitions between jobs. We provide two full-length examples of interviews in Appendix Section \ref{app_sec:interview_full_examples}.

\paragraph{Standardized language and analytical tests.} If an interview is successfully completed, applicants are invited to a standardized test. The test takes about 30 minutes and contains a language and a quantitative skill component, each featuring adaptive questions. The language component assesses applicants' writing and reading capabilities in English through classic language testing tasks. %
Scores are based on the CEFR framework (six categories from A1-beginner to C2-proficient). The quantitative skill component consists of three individual parts: attention to detail, verbal reasoning, and numerical ability. %
Performance is aggregated across the three individual parts into a score from 0 to 100. Completing the test is mandatory to advance to the hiring decision stage.

\paragraph{Review and hiring decisions.} Recruiters make hiring decisions based on applicants' interview and test performance. To do so, they review the (i) interview transcript, (ii) audio recording, and (iii) standardized test scores. Interview performance is assessed according to four main categories. First, recruiters consider the applicant's level of experience in customer service roles. Second, they assess how proficient the applicant is in communicating in English. Third, they evaluate the risk of attrition, that is, the likelihood that the applicant will not work in the prospective job for an extended period. Fourth, they determine whether the applicant’s salary expectations align with the offered wage range. Recruiters rate interviews on a three-point scale and provide a short justification in an open-ended text format. For details on the scoring, see Appendix Table \ref{tab:score_grid}. They then determine whether an applicant is suitable to be hired by considering applicants' interview scores and test performance. Although monthly hiring targets exist for each application site, the high volume and continuous flow of applicants as well as sustained demand from client firms ensure that individual hiring decisions have essentially no external effects on other applicants. If a recruiter judges an applicant to be suitable for the job, they assess whether their location and qualifications match a job opening supplied by the client account. If a match is found, an email with the job offer is sent to the applicant. If not, the applicant is kept in the system and may be contacted at a later stage. 

\paragraph{Onboarding.} If an applicant accepts an offer, they are forwarded to the respective client company. Depending on the client and the job profile, the applicants undergo additional validation and medical checks. Once passed, the applicant begins the onboarding process with a job training period. A key challenge at this step is that usually a substantial fraction of applicants accept the offer but do not show up for training, are absent during training, or do not pass the additional validation checks employed by the client. 

\paragraph{Employment.} After completing training, applicants begin regular employment with the client. The key employment outcome for the firm is job match quality, i.e., the compatibility between the skills and preferences of the worker and the requirements and conditions of the position. %
Because match quality is generally not observable, we use worker retention, i.e., whether a worker remains employed after a given period, as a proxy \parencite[as is standard in the literature, see e.g.,][]{Jovanovic1979,hoffmanDiscretionHiring2018,Autor2008}. The hiring firm's main objective is to maximize the retention rate among applicants, as the client firms compensate them for hires who remain employed for at least one month. We argue that the retention rate is a suitable measure of match quality in our context for the following reasons. The business-process outsourcing market we study features minimal frictions on both worker and firm sides. Because performance is quantifiable and continuously tracked, it is easy for firms to identify poor matches quickly. At the same time, workers can be replaced at short notice. These factors enable firms to easily fire workers for lack of job performance.\footnote{In our setting, more than 30\% of all involuntary separations are due to failed performance reviews according to the company's records.} Workers, on the other hand, can voluntarily separate from the firm quickly and without incurring negative consequences. In fact, about half of all separations are initiated by the worker in our setting. Due to these factors, we argue that short-term retention is a sufficient statistic for match quality.

%% file: Sections_paper/3_experiment.tex
\subsection{Experimental design}

Our pre-registered experiment\footnote{See Appendix Section \ref{app_sec:research_transparency} for details on the mapping between pre-registration and paper.} aims to test the causal impact of automating job interviews with AI voice agents. Accordingly, our treatment variation concerns the interview stage, where we vary who conducts the interview. Once an applicant qualifies for an interview, they are randomized into one of three experimental conditions: \textit{Human Interviewer}, \textit{AI Interviewer}, and \textit{Choice of Interviewer}. 

\paragraph{\textit{Human Interviewer}.} In the \textit{Human Interviewer} condition, applicants are interviewed by a human recruiter. In the \textit{Remote} mode, human recruiters interview applicants remotely via phone; in the \textit{Walk-in} mode, they conduct the interview in person at the nearest application center of the firm. 

\paragraph{\textit{AI Interviewer}.} In the \textit{AI Interviewer} condition, applicants are instead interviewed by an AI voice agent. The agent is prompted to follow the same structured interview guidelines that are given to the human interviewers. The AI voice agent conducts the interview in both interview modes -- \textit{Remote} and \textit{Walk-in} -- via phone. While in \textit{Remote} the phone call takes place remotely, in \textit{Walk-in} the phone call takes place at the application center in which the applicant applied. That is, in both modes, the external environment in which the interview takes place is the same across the \textit{Human} and \textit{AI Interviewer} conditions. When the call starts, the AI voice agent immediately discloses its artificial identity to avoid any deception, according to firm compliance, and explicitly states that a human recruiter will review the interview, evaluate it, and make the hiring decision, not the AI itself. 

\paragraph{\textit{Choice of Interviewer}.} Lastly, in the \textit{Choice of Interviewer} condition, applicants can choose whether a human or an AI voice agent will interview them. They are offered the choice upon receiving the interview invitation (in both \textit{Remote} and \textit{Walk-in}). If they do not choose within a certain time frame, the AI voice agent calls them. The AI agent briefly explains the process, provides them with insights into how the AI agent works, and then asks them to make the choice.\footnote{The firm implemented this feature because while most applicants have experience with human recruiters, they have none with AI voice agents. This design provides applicants with initial exposure to the AI agent, potentially alleviating some apprehension about being interviewed by an AI.}

\paragraph{Assessment across treatments.} Importantly, interviews are evaluated and subsequent hiring decisions are made by humans, irrespective of treatment condition. That is, regardless of whether the AI voice agent or a human recruiter conducted the interview, a human recruiter reviews the audio, transcript, and test scores of an applicant and then makes a hiring decision. Each applicant is assigned a unique recruiter. In particular, when a human recruiter conducts the interview, the same recruiter also later evaluates the applicant. When an AI voice agent conducts the interview, the same round-robin scheduling algorithm that assigns recruiters to applicants who face a human interviewer also assigns recruiters to evaluate applicants who face the AI agent. This process means that recruiters know whether an applicant was interviewed by \textit{AI} or not. However, in the \textit{Remote} mode, recruiters do not observe whether the applicant was randomly assigned or self-selected for an interview with the AI voice agent or a human. In contrast, in the \textit{Walk-in} mode, since recruiters themselves ask the applicants assigned to the \textit{Choice of Interviewer} condition to make a choice, they know if the candidates self-selected for an interview with the AI voice agent or with a human recruiter. Importantly, all recruiters evaluate applications of all three treatment conditions. They are further instructed to apply the same assessment criteria for interviews and hiring, regardless of treatment or mode of interview. 

By varying only who conducts the interview, the experimental design isolates the direct effect of automating the interview stage. This feature allows us to analyze the performance of AI in collecting information from humans in natural conversations, as well as the human response to AI before, during, and after the interview.

\subsection{Experimental sample}

Our full sample consists of 70,884 applications that the firm received from March 7 to June 7, 2025. In total, 17,621 applications are in the \textit{Walk-in} mode and 53,263 in the \textit{Remote} mode. Applications were received for 48 different job postings and 41 different client accounts, which operate in technology, insurance, telecommunications, retail, finance, healthcare, and transportation (see Appendix Table \ref{tab_industries_clients} for the full breakdown). All applications for the job postings were part of the experiment. Applications were processed by 26 different application sites in 19 cities. The sites are distributed across several regions in the Philippines.\footnote{See Appendix Figure \ref{fig:map_of_applications_by_center} for information on the regions and the distribution of applications per region.} Most of the applications were processed by sites located in Metro Manila and the Central Visayas region.

\paragraph{Randomized sample.} Of the 70,884 applications, 67,056 were found to be eligible and were therefore randomized to one of the three treatment conditions described above. As pre-registered, we will use this sample for our analysis. In total, 40,103 applications ($59.81\%$) were randomized into the \textit{AI Interviewer} condition (10,421 in \textit{Walk-in}, 29,682 in \textit{Remote}), 13,557 applications ($20.22\%$) were randomized into the \textit{Human Interviewer} condition (3,478 in \textit{Walk-in}, 10,079 in \textit{Remote}), and 13,396 ($19.98\%$) into the \textit{Choice of Interviewer} condition (3,469 in \textit{Walk-in}, 9,927 in \textit{Remote}). In Appendix Table \ref{tab_balance_check_main_treatment}, we provide evidence that randomization was successful, as pre-treatment variables are balanced between the three treatments. In total, 6,319 applications received job offers, and 4,294 successfully started their job and passed the training so that we can match their applications with employee data. 

\paragraph{Applicant characteristics.} Of all applications, 64,556 were submitted by unique individuals, which means that 5,341 applicants submitted more than one application during the experiment. In total, 58\% of applicants learn about the job through digital job postings, 19\% of applications are submitted through referrals, and the rest through other sources such as word of mouth. In total, 60\% of applicants were female. Most applicants were between 20 and 30 years old and had some prior experience in customer service jobs.

\paragraph{Applicant survey.} To collect more detailed data on applicants' beliefs and experiences, the firm invited applicants to participate in a customer experience survey. Invitations were sent by email. The survey contained five blocks of questions; see Appendix \ref{app_sec:survey} for the instructions. 
Applicants were randomly assigned a short or long version of the survey. The long version took about 10 minutes and contained a total of 23 questions within the five blocks. Applicants were compensated with \$2 for answering all questions in the survey, a payment that implies an hourly wage six times higher than the minimum wage. The short version took approximately 2 minutes to complete, was offered without compensation, and contained only a subset of questions (11 in total). The survey was sent to 19,200 applicants, of whom 2,764 completed it. This implies a completion rate of 14\%.

\paragraph{Recruiters.} Our sample of applications was assessed by a total of 131 recruiters. On average, each recruiter had 512 applicants assigned to them throughout the experiment (Median = 121, SD = 1,153). However, this average masks substantial heterogeneity in the number of applications assigned to recruiters. A core team of 43 recruiters was assigned 90\% of all applications.

\paragraph{Recruiter survey.} To collect additional data on the beliefs and opinions of the recruiters, the firm conducted a firm-wide survey after the experiment but before the recruiters could learn about the experiment’s results. The survey lasted about five minutes and contained two blocks. The first asked recruiters to predict the impact of AI, the second asked about their interview and evaluation experiences; only recruiters who had previously evaluated AI-led interviews received this block. See Appendix \ref{app_sec:survey_recruiter} for the instructions. The survey was completed by 173 recruiters. Of these, 133 evaluated AI-led interviews and 98 evaluated applicants who participated in the experiment. These 98 recruiters evaluated 82\% of the applicants in our experiment.

%% file: Sections_paper/4_results.tex
\section{Main results on employment outcomes}
\label{main_sec:main_results}

In this section, we compare key recruitment outcomes between the \textit{AI Interviewer} and \textit{Human Interviewer} conditions. We first examine whether applicants receive a job offer, start the job, and remain employed one to four months after starting. We then assess whether these effects are accompanied by changes in separation reasons or on-the-job productivity among hired applicants. 

\paragraph{Recruiters' forecasts.} We begin by benchmarking our findings against recruiters' expectations. In the recruiter survey, we asked the firm's recruiters to forecast the impact of introducing AI-led interviews. Overall, recruiters expected applicants interviewed by the AI voice agent to perform worse. For offer rates, 36\% of recruiters expected applicants to receive lower rates under AI, 49\% expected equal rates, and 15\% expected higher rates. For retention, 48\% expected workers hired through AI to have lower retention, 39\% expected equal retention, and 13\% expected higher retention. We find similar fractions when asking recruiters to predict workers' productivity. Moreover, 61\% expected AI-led interviews to be of lower quality.\footnote{Forecasts may be driven by genuine predictions about the AI's technical capabilities but also influenced by factors such as incumbent bias or motivated reasoning, i.e., recruiters downweight the AI's capabilities because it threatens their expertise.} 

\subsection{Results on offer rates and match quality}
\label{subsec:main_recruitment_results}

\begin{figure}[!t]
    \caption{Treatment effect on key recruiting outcomes in the unconditional sample}
    \label{fig:main_sec_main_outcomes_full_figure_unconditional}
    \centering
    \includegraphics[width=0.9\textwidth]{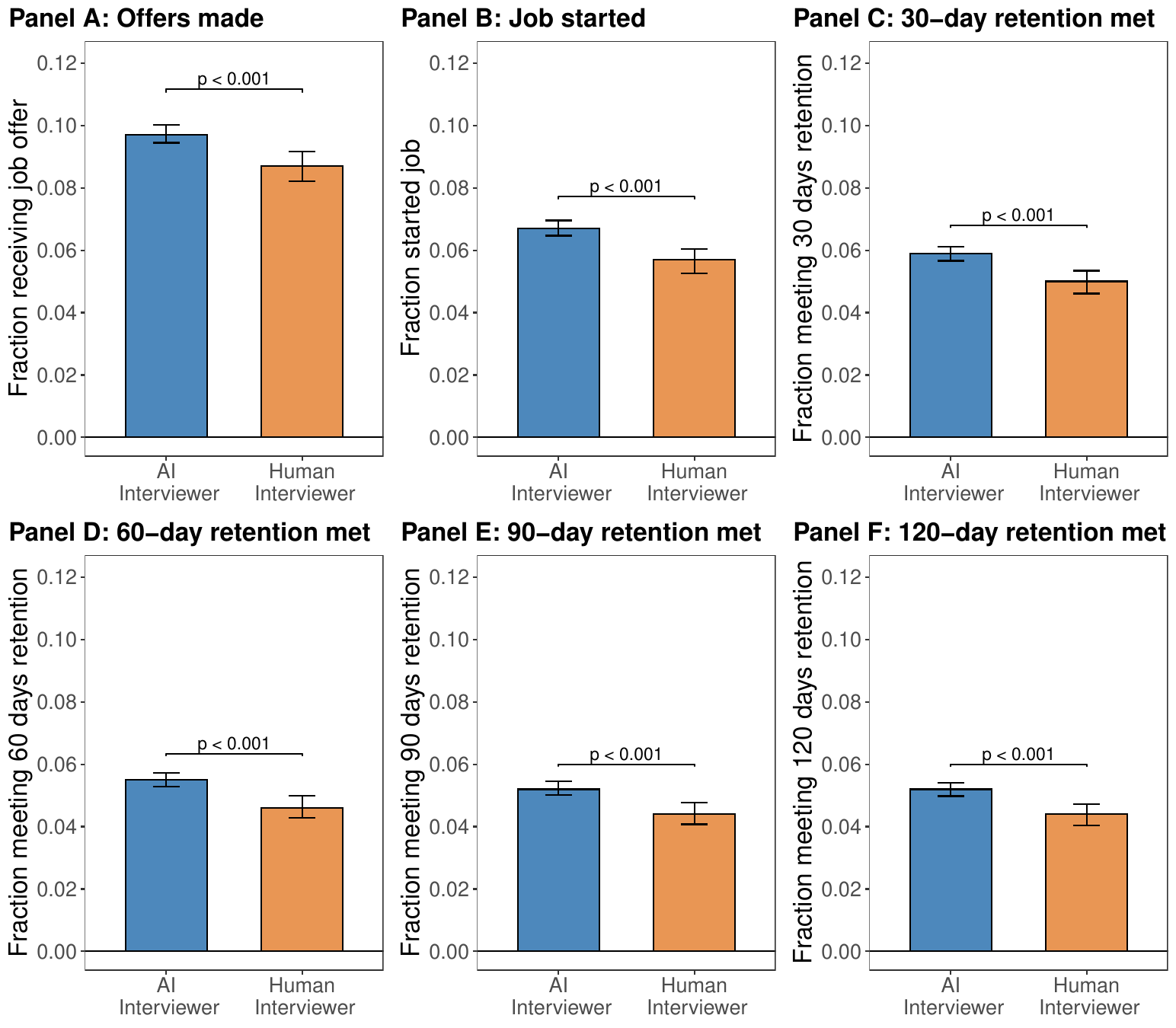}
    \figurenotes[Notes:]{The figure displays the recruiting outcomes of applicants. Each panel displays the fraction of applicants who realize the specific outcome. Fractions are displayed separately for the \textit{Human Interviewer} condition, in which applicants are interviewed by a human, and for the \textit{AI Interviewer} condition, in which applicants are interviewed by an AI voice agent. Bars indicate 95\% confidence intervals; p-values are calculated from a two-sample proportion test.
    }
\end{figure}

For the analysis, we consider two samples. The first is the unconditional sample, which compares all applicants randomized into either the \textit{Human Interviewer} or \textit{AI Interviewer} condition. The second is the conditional sample, which includes only applicants who accepted a job offer.

\paragraph{Sample: all applicants.} Figure \ref{fig:main_sec_main_outcomes_full_figure_unconditional} displays the treatment effects in the unconditional sample. The likelihood of receiving a job offer is 8.70\% in the \textit{Human Interviewer} condition (1,179 out of 13,557 applicants) and 9.73\% in the \textit{AI Interviewer} condition (3,904 out of 40,103 applicants). Thus, applicants interviewed by the AI voice agent have a 1.03 percentage point, or 12\%, higher likelihood of receiving a job offer ($p<0.001$, two-sample proportion test). The same pattern appears for job starts: 5.65\% of applicants in \textit{Human Interviewer} (766) and 6.71\% in \textit{AI Interviewer} (2,691) start their job ($p<0.001$, two-sample proportion test). One month after starting, 4.98\% of applicants in \textit{Human Interviewer} (675) and 5.89\% in \textit{AI Interviewer} (2,363) are still employed. Applicants interviewed by the AI voice agent therefore have an 18\% higher unconditional retention rate, corresponding to 0.91 percentage points ($p<0.001$, two-sample proportion test). The positive effect persists over longer horizons (Panels D--F): the likelihood that applicants are still employed is 19\% higher after two months, 18\% higher after three months, and 19\% higher after four months ($p<0.001$, $p<0.001$, and $p<0.001$, respectively, two-sample proportion tests).

\paragraph{Sample: only applicants who accepted an offer.} We next repeat the analysis among applicants who accepted a job offer (see Appendix Figure \ref{fig:main_sec_main_outcomes_full_figure_conditional} for a graphical presentation). In total, 8.14\% of applicants in \textit{Human Interviewer} accept a job offer (1,104), while this fraction is 8.99\% in the \textit{AI Interviewer} treatment (3,604). Among applicants who accepted an offer, 68.84\% in \textit{Human Interviewer} and 73.36\% in \textit{AI Interviewer} successfully start their job. After one month, 60.60\% and 64.32\%, respectively, are still employed. Thus, applicants in \textit{AI Interviewer} who accepted their job offer have significantly higher fractions of job starts and one-month employment ($p=0.003$ and $p=0.025$, respectively, two-sample proportion test). The same pattern holds at longer horizons: after two months, 56.34\% in \textit{Human Interviewer} and 59.93\% in \textit{AI Interviewer} are still employed ($p=0.034$); after three months, these numbers are 53.71\% and 56.96\% ($p=0.057$); and after four months, they are 53.26\% and 56.60\% ($p=0.050$). Accordingly, retention rates are consistently higher in the \textit{AI Interviewer} condition, and the percentage improvement is stable over time (one month: 6.14\%, two months: 6.37\%, three months: 6.05\%, four months: 6.27\%).

\paragraph{Robustness.} These results are robust to the inclusion of additional pre-treatment controls and fixed effects. Specifically, we regress each key recruitment outcome on treatment status. We control for an applicant's gender, source of application (i.e., referral, online job posting, etc.), pre-treatment engagement score, and whether the application is from an applicant who submitted more than one application to any of the firm’s job postings during the period from six months before the experiment began until its conclusion. We also include fixed effects for the week in which the application was received, the recruiter assigned to the application, the city of the application site, and the specific job posting targeted by the application. Across all outcomes -- whether an application led to an offer, a successful job start, and an employment spell of at least one, two, three, or four months -- treatment effects are very similar with and without added controls and fixed effects. For details, see Appendix Table \ref{tab_main_sec_job_offers_treatment_main_outcomes_uncond} (sample of all applicants) and \ref{tab_main_sec_job_offers_treatment_main_outcomes_cond} (sample of applicants who accepted an offer).

\subsection{Results on separation reasons and productivity}

\paragraph{Separation reasons.} In addition to retention, we observe, for employees who left their job after starting, whether separation was voluntary (i.e., initiated by the employee) or involuntary (i.e., initiated by the employer). The primary reasons for voluntary leaves are taking another job, returning to school or university, or family obligations or childcare responsibilities. The primary reasons for involuntary leaves are excessive absenteeism or failed performance checks. Among employees who left their jobs, 58.25\% of those hired from the \textit{Human Interviewer} condition separate voluntarily, compared with 58.81\% of those hired from the \textit{AI Interviewer} condition. The corresponding involuntary separation rates are 41.75\% and 41.19\%, respectively. Hence, there is no difference in separation reasons between conditions ($p=0.95$, two-sample proportion test).

\paragraph{Job performance.} For a subset of employees, we also observe job performance. We focus on three variables that are commonly considered to jointly determine a customer service worker's productivity \parencite[see e.g., ][]{Brynjolfsson2025}. The first is \textit{average handle time}, the average number of minutes a worker spends handling a customer request from start to finish. The second is the \textit{quality assurance score} (between 0 and 100), the employer's assessment of a worker's quality in handling customer requests based on a review of selected interactions. The third is the \textit{customer satisfaction score} (between 0 and 100),\footnote{Depending on the client, different scales are used. We perform a linear transformation to create a common 0-100 scale.} which measures customer feedback elicited after a request. Productive workers manage to achieve low average handle times while maintaining high customer satisfaction and quality assurance scores.

We compare these productivity measures between workers hired in the \textit{AI Interviewer} condition and those hired in the \textit{Human Interviewer} condition. We estimate OLS regressions controlling for client, month and worker tenure fixed effects. Table \ref{tab_main_results_productivity} shows the results. Across all three performance variables, we find no significant differences between workers hired in the \textit{AI Interviewer} condition and those hired in the \textit{Human Interviewer} condition.\footnote{We do not observe all three variables for every worker. Our results are similar if we focus only on workers for whom we observe all three variables.} Workers in the \textit{AI Interviewer} condition take slightly longer to handle requests, while customers are also slightly more likely to be satisfied. The differences therefore do not show a consistent directional tendency across variables and are economically very small. This result suggests that the positive effects on job offers, starts, and retention are not accompanied by decreased employee productivity.  

\begin{table}[!t]
\caption{Job performance of applicants hired}
\label{tab_main_results_productivity}
\resizebox*{1\textwidth}{!}{
\begin{threeparttable}[c]
    \begin{tabular}{@{\extracolsep{5pt}}lccc ccc} 
        \\[-1.8ex]\hline 
        \hline \\[-1.8ex] 
         & \multicolumn{6}{c}{\textit{Dependent variable:}} \\ 
         & \multicolumn{2}{c}{Average handle time (minutes)}  & \multicolumn{2}{c}{Quality assurance (0-100)} & \multicolumn{2}{c}{Customer satisfaction (0-100)}\\ 
            \cline{2-3} \cline{4-5}  \cline{6-7} \\[-1.8ex] 
                                   & (1) & (2) & (3)  & (4) & (5) & (6)  \\ 
           \hline \\[-1.8ex] 
           \textit{AI Interviewer} & 1.519   & 1.332   & -0.293  & -0.318  & 1.618   & 1.444\\   
                                   & (1.027) & (1.005) & (0.535) & (0.516) & (0.990) & (0.971)\\  
           \midrule
           Dependent variable mean & 23.992  & 23.992  & 86.265  & 86.265  & 66.946  & 66.946\\  
           Observations            & 7,560   & 7,560   & 4,512   & 4,512   & 3,877   & 3,877\\  
           R$^2$                   & 0.295   & 0.303   & 0.361   & 0.380   & 0.450   & 0.470\\ 
           Client FE               & Yes     & Yes     & Yes     & Yes     & Yes     & Yes    \\  
           Month \& tenure FE      & --      & Yes     & --      & Yes     & --      & Yes\\  
        \\[-1.8ex] 
        \hline 
        \hline \\[-1.8ex] 
    \end{tabular} 
            \begin{tablenotes}
                    \item \footnotesize \textit{Notes:} 
                    The table shows OLS estimates of differences in worker job performance by treatment condition. \textit{AI interviewer} is an indicator equal to one if the interview leading to the hiring of the worker was AI-led, and zero if it was human-led. The dependent variable in columns (1)-(2) denotes the time (in minutes) a worker spent handling a customer service request, averaged across a month among all requests handled by the worker. The dependent variable in (3)-(4) is a monthly quality assurance score (from 0 to 100) by the employer about the worker, where higher scores indicate better performance by the worker. The dependent variable in (5)-(6) is satisfaction reported by customers for requests handled by a worker (from 0 to 100), averaged across a month. Observations for this regression are at the worker-month level, and all standard errors are clustered at the worker level. Significance levels: $^{*}\,p<0.1$,$^{**}\,p<0.05$, $^{***}\,p<0.01$.
            \end{tablenotes}
\end{threeparttable}
       }
\end{table}

%% file: Sections_paper/5_mechanism.tex
\section{Mechanism: controlled variance in information collection}
\label{sec:interview_transcript_analysis}

Why are applicants in the \textit{AI Interviewer} condition more likely to receive offers and have higher retention rates? We investigate a key potential mechanism: AI voice agents conduct interviews more consistently than human recruiters, resulting in a more standardized interview process. If this standardization leads to interviews that more precisely reveal information about applicants, such as their abilities, motivation, and fit, it improves recruiters' hiring decisions and thus the quality of hired applicants. 

\subsection{The standardization-adaptiveness trade-off}
\label{sec:interview_transcript_analysis_mechanism_description}

To understand the mechanism, we first describe a central challenge that employers face in designing optimal interviews. Employers use interviews to assess applicants' suitability for employment. A large literature on interviews shows that recruiters' interview behavior and how interviews are factored into hiring decisions can be influenced by idiosyncratic and bias-inducing factors (see Footnote \ref{footnote_interview_biases}). Differences in recruiters' experience, training, and habits may then induce differences in how interviews are conducted and applicants are evaluated, which can negatively affect hiring quality. In particular, recruiter characteristics can shape interviews more strongly the more discretion recruiters have over the interviews' content and structure. Moreover, these effects may be amplified in high-volume hiring, which requires a large team of recruiters to assess the large number of applicants.
 
A natural solution to recruiter-driven variance is to standardize the interview process, reducing interview discretion. When choosing the optimal degree of standardization, firms face a trade-off. On the one hand, full discretion leads to unstructured interviews, generating recruiter-induced variance.  On the other hand, consider the boundary case of an entirely standardized interview process. In this case, recruiters are required to ask all applicants the same questions without follow-ups by either the applicant or the recruiter. This eliminates recruiter variance, but also the interview as a conversation. However, adapting to individual applicants, particularly by asking follow-up questions, can be valuable because it allows recruiters to tailor their inquiries into employment histories, clarify ambiguous answers, and better understand applicants’ motivations for the job. Likewise, it allows applicants to better understand the job and company they are applying to. Hence, the optimal interview design needs to balance standardization with adaptiveness: preserving a common structure while allowing responses and follow-up questions tailored to each applicant.

AI voice agents offer a way to combine standardization with a sufficient degree of adaptiveness. A single agentic system, guided by a common protocol, can interview all applicants. This feature thus has the potential to drastically reduce between-recruiter variance. At the same time, as AI voice agents can engage in live spoken conversation, they can ask probing and follow-up questions based on applicant input. Hence, AI agents can perform what we call \textit{controlled variance}: a structured interview process that reduces interviewer-driven variation while allowing response-contingent follow-ups. 

We investigate this potential mechanism by analyzing interview transcripts. Specifically, we proceed in three steps. First, we analyze the controlled dimension of the mechanism: Are AI-led interviews more standardized than human-led? Second, we turn to the adaptiveness dimension: Is the AI agent still able to respond to applicant input even if it is more standardized? Third, we analyze how applicants respond to recruiter behavior that varies in controlled ways: Do potential differences in structure and adaptiveness change the job-relevant signals applicants reveal during the interview?

\paragraph{Interview transcript data.}  Our dataset consists of raw verbatim transcripts from 34,109 interviews conducted in the \textit{Human Interviewer} and \textit{AI Interviewer} conditions. These interviews are a subset of all interviews conducted. Using LLMs and standard prompt engineering techniques, we first anonymized and labeled the transcripts. We then construct three distinct sets of variables. First, a general classification of interview types. Second, a set of variables capturing recruiter behavior in interviews. Third, variables that capture applicants' language and behavior. We describe the transcript variables briefly in the main text and provide a detailed description of all transcript variables in Appendix Section \ref{app_sec:transcript_variables}. 

\paragraph{Relevance of interview content.} Before investigating the mechanism, we first assess the relevance of job interviews for offer decisions -- a necessary condition for differences in interview content to matter. As evident from recruiters' survey responses, job interviews are a crucial step in the hiring process: when asked about the importance of interview performance relative to test scores in determining offers, 33\% of recruiters say interview performance is more important, 63\% say they are equally important, and only 4\% consider test scores more important. 

To empirically test the relevance of interviews for hiring, we examine the baseline predictive power of interview assessments for offer decisions in the \textit{Human Interviewer} condition. Our outcome variable is whether an application led to a job offer. For interview assessments, we use the score (1, 2, 3) with which recruiters rate applicants' interview performance, and the open-ended text in which they describe applicants' performance. For the latter, we use natural language processing to classify sentiment (negative = -1, neutral = 0, positive = 1).

We find that both variables strongly predict offer decisions. Each increase in interview score is associated with a 25 percentage-point increase in offer likelihood. Conditional on test score performance and our other controls and fixed effects, the increase remains highly significant at 15 percentage points. The sentiment of the recruiter text assessment is similarly highly predictive of offer decisions. For details, see Appendix Table \ref{tab_interview_vars_predicting_outcomes}.

\subsection{Recruiter behavior in interviews: standardization}
\label{subsec:transcript_analysis_recrutiers_standardization}

\begin{figure}[!t]
    \caption{Recruiter behavior with respect to standardization}
    \label{fig:recruiter_distribution_topic_coverage}
    \centering
    \includegraphics[width=1\textwidth]{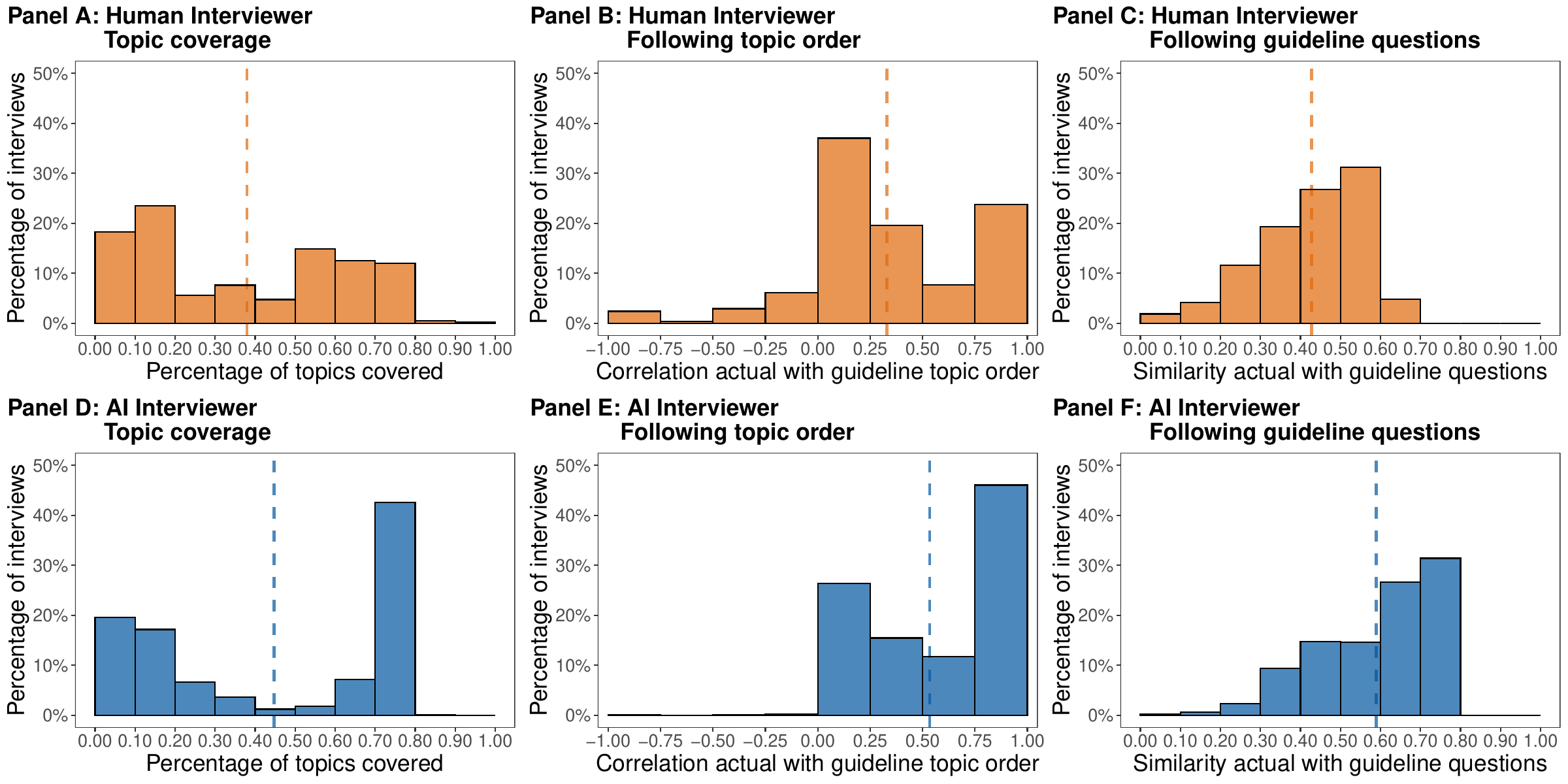}
    \figurenotes[Notes:]{The figure displays the distribution of several recruiter variables inferred from transcript data. ''Topic coverage'' denotes the percentage share of the 14 guideline topics that are covered during an interview. ''Following topic order'' is defined as the Kendall's $\tau$ correlation between the actual and guideline topic sequences. ''Following guideline questions'' denotes the similarity between the questions asked and the guideline questions. Distributions are displayed separately for the \textit{Human Interviewer} condition, in which applicants are interviewed by a human, and for the \textit{AI Interviewer} condition, in which applicants are interviewed by an AI voice agent. The dashed lines display mean values.
  }
\end{figure}

\paragraph{Measuring standardization.} As described in Section \ref{sub_sec:background_recruitment_process}, recruiters are tasked with following structured interview guidelines to ensure a standardized interview process. Specifically, they have a list of 14 topics, a recommended topic order, and representative questions and conversation starters for each topic. To measure the extent to which human recruiters and the AI voice agent implemented structured, standardized interviews, we construct the following three variables for each interview: (1) the \textit{percentage of topics covered}, i.e., how many out of all topics the recruiter covered, (2) the \textit{correlation of the actual with guideline topic order} (using Kendall's $\tau$), measuring how closely recruiters followed the guideline topic order, and (3) the \textit{similarity of the actual with guideline questions}, measuring how closely recruiters phrased questions following the guidelines. For details on the variable construction, see Appendix Section \ref{app_sec:transcript_recruiter_feature_variables}.

\paragraph{Results.} Figure \ref{fig:recruiter_distribution_topic_coverage} provides the distributions of the variables separately for both treatments. Starting with Panels A and D, which display the \textit{percentage of topics covered}, we find that an average AI-led interview covers 45\% of all possible topics, while the average human-led interview covers 38\%, a significant difference ($p<0.001$, two-sample t-test). Importantly, the distribution of topic coverage is more distinctly bimodal in the \textit{AI Interviewer} condition, with a concentration around either high coverage or very low coverage. Moreover, the variance of topics covered is significantly lower relative to the \textit{Human Interviewer} condition ($p<0.001$, Levene test).

We next turn to the \textit{correlation of the actual with guideline topic order}, displayed in Panels B and E. While the correlation between actual and guideline order is $\tau= 0.53$ in \textit{AI Interviewer}, it is $\tau= 0.33$ in \textit{Human Interviewer} ($p<0.001$, two-sample t-test). Hence, as expected in an interactive conversation with probing and follow-ups, both AI and human recruiters do not mechanically implement the topic order in every instance and instead frequently deviate from it. However, AI implements it more consistently across interviews, leading to significantly lower variance in the correlation between the actual and guideline order of topics ($p<0.001$, Levene test).

Third, in Panels C and F, we compare the \textit{similarity of the actual with guideline questions} asked by the recruiters across conditions. The average similarity per interview is $0.59$ in \textit{AI Interviewer} and $0.43$ in \textit{Human Interviewer}, again a significant difference ($p<0.001$, two-sample t-test). As with topic order, neither recruiter mechanically reads from a script but tailors questions and follow-ups to individual applicants and their responses, resulting in correlations of less than one in each interview. Yet, the AI voice agent appears to adhere more closely to the guidelines. 

\paragraph{Robustness.} So far, we have compared the transcript content with the firm’s interview guidelines and found that the AI voice agent follows them more closely. The benchmark in that analysis is thus the firm standard. As a complementary test, we examine, within each condition, how similar recruiters' questions are to the \textit{average question} for each topic -- that is, the typical language used for that topic within the same condition. Specifically, we compute the distance between each recruiter's question embedding and the centroid embedding of all recruiters for that topic; higher values indicate greater deviation from the average. Across nearly all topics, we find that the distance is significantly lower for the \textit{AI Interviewer} than for the \textit{Human Interviewer} (see Appendix Table \ref{tab_transcript_recruiter_analysis_topic_question_similarity} for details).

\paragraph{Heterogeneity in recruiter behavior.} The previous results were at the transcript level, comparing the average interview performance of AI and human recruiters. A complementary analysis is to compare their respective performance across interviews. That is, for each human recruiter, we calculate the average value of our four recruiter behavior variables across all interviews they conducted during the experiment. We find substantial heterogeneity in behavior. Some recruiters achieve substantially higher topic coverage and similarity with the guideline topic order and questions. For instance, some recruiters cover on average 60\% of topics in their interviews, while others cover less than 40\%. Importantly, when comparing the distribution of recruiter averages with the average of the AI voice agent, we find that the latter achieves a higher similarity to the guideline topic order and questions than 83\% and 100\% of human recruiters, respectively. For the percentage of topic coverage, the AI voice agent achieves higher scores than 61\% of human recruiters.\footnote{For comparison, the 25\% of human recruiters scoring highest within the respective variable score on average a topic coverage of 61\%, a similarity with the guideline topic order and questions of 0.61 and 0.49, respectively.} For the full distribution, see Appendix Figure \ref{fig:plot_recruiter_features_average_standardization}.

\subsection{Recruiter behavior in interviews: adaptiveness}
\label{subsec:transcript_analysis_recrutiers_adaptiveness}

\begin{figure}[!t]
     \caption{Recruiter behavior with respect to adaptiveness}
    \label{fig:transcript_analysis_recrutiers_adaptiveness}
    \centering
    \includegraphics[width=1\textwidth]{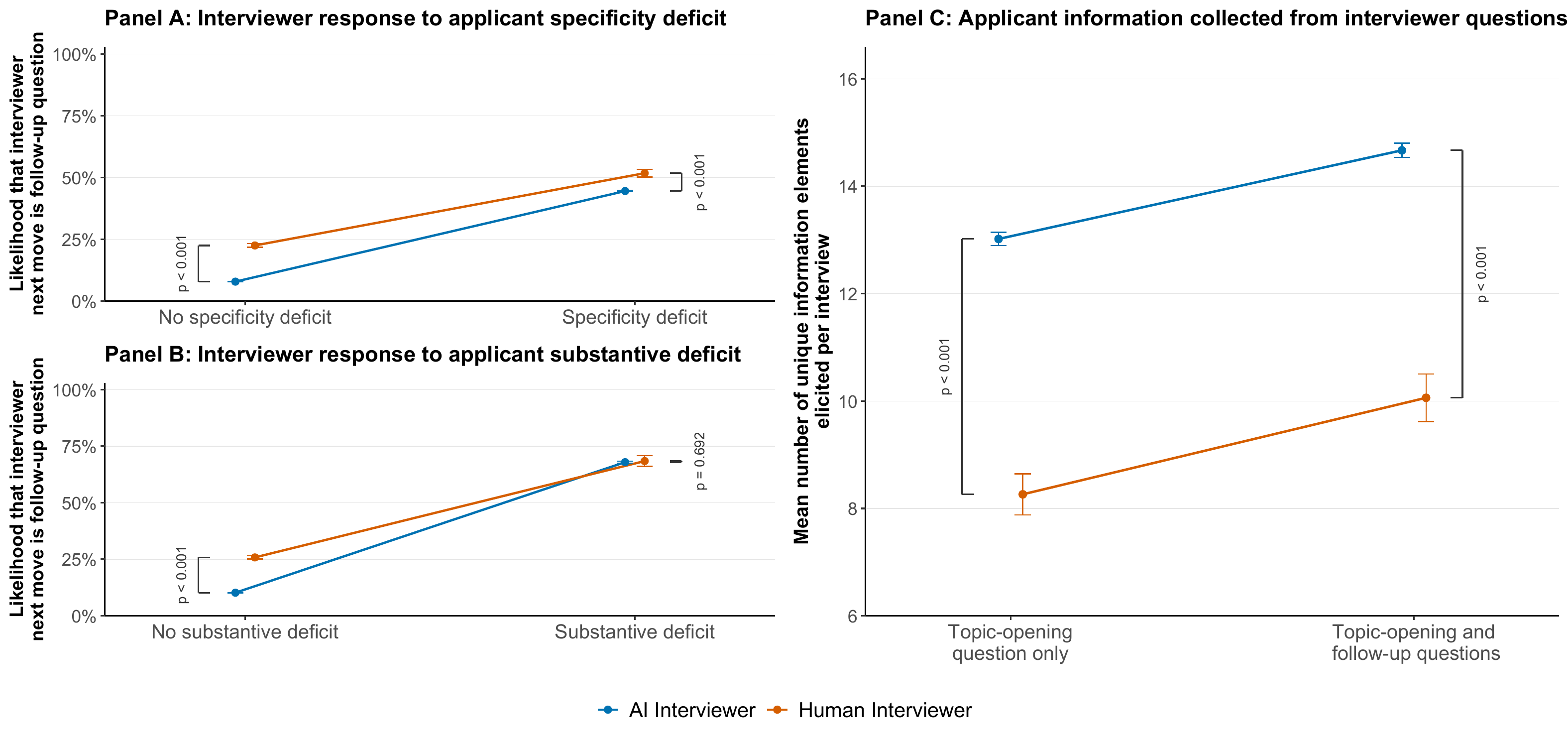}
    \figurenotes[Notes:]{The figure compares interviewer behavior in the \textit{AI Interviewer} and \textit{Human Interviewer} conditions. In Panels A and B, the unit of observation is an initial applicant response to a topic-opening question. The outcome is an indicator equal to one if the interviewer responded with a follow-up question rather than opening a new topic. Panel A splits applicant responses based on whether they exhibit a specificity deficit, defined as insufficient detail. Panel B splits responses based on whether they exhibit a substantive deficit, defined as insufficient engagement with the question's topic. In Panel C, the unit of observation is an interview. The panel reports the mean number of unique information elements elicited per interview, first considering only applicant responses to topic-opening questions, then including information elicited through adaptive follow-up questions. Error bars show 95\% confidence intervals. In Panels A and B, standard errors are clustered at the interview level. 
    }
\end{figure}

Next, we turn to the second part of the standardization-adaptiveness trade-off. Having shown that the AI voice agent conducts interviews in a more standardized way, we analyze the extent to which this standardization comes at the cost of adaptiveness to applicant input. 

\paragraph{Dataset.} To study adaptiveness, we analyze the conversational dynamics of the interviews at the level of individual conversational moves. We define a conversational move as a spoken contribution by either the interviewer or the applicant, bounded by a change in speaker. A move may consist of a question, an answer, a statement, or a combination of these. Using Gemini 2.5 Flash, we identify moves from the transcript data and record their sequential position within the interview. This produces a dataset containing 747,696 conversational moves. 

We use this move-level dataset to capture three objects. First, we measure the interviewer's conversational behavior, focusing on the extent to which questions are independent of, or adapted to, the applicant. Second, we measure the quality of applicant responses relative to the interviewers' questions, which in turn allows us to test whether interviewers' follow-up behavior depends on incomplete or vague responses. Third, we measure the hiring-relevant information collected through the conversation, which allows us to assess whether differences in interviewer and applicant behavior translate into more information elicited. We again use Gemini 2.5 Flash to construct the variables that capture these three objects, as explained next. 

\paragraph{Measuring adaptiveness.} We measure the interviewer's conversational behavior through the \textit{interviewer question type} variable. An interviewer question can be a \textit{Topic-opening} question, an \textit{Adaptive follow-up} question, or an off-topic remark. A \textit{Topic-opening} question starts the conversation about one of the 14 interview topics (see Appendix Table \ref{tab:storyboard_full}) or a subtopic within them. For instance, one of the guideline questions starting the \textit{Motivation \& attrition risk} topic is ``What attracted you to this opportunity and how does it align with your longer-term goals?''. After hearing the applicant's response, the interviewer can either pose the next \textit{Topic-opening} question or ask an \textit{Adaptive follow-up}, that is, a question adapted to the applicant's specific response to the opening question. For instance, the interviewer may ask the applicant to clarify an aspect of the stated motivation or to elaborate on a mentioned goal. Lastly, an interviewer's move is classified as an off-topic remark if it is not related to any interview topic. 

To measure the quality of applicants’ responses, we classify each response along two dimensions. The first variable, \textit{specificity deficit}, indicates whether a response is vague, unclear, or lacks relevant details. For example, when asked about previous work experience, a response is classified as having a specificity deficit if it does not provide concrete information such as dates, responsibilities, or details about the position. The second variable, \textit{substantive deficit}, indicates whether the applicant fails to engage with the substance of the question. This may occur because the applicant asks for clarification without providing an answer, avoids the question, or deflects to a different topic. Each variable is coded as an indicator: a response either exhibits a \textit{specificity deficit} or \textit{substantive deficit}, or it does not.\footnote{Our results are robust to using continuous measures of both response characteristics.}

Lastly, to assess the hiring-relevant information collected through the interview, we construct a dictionary of relevant information elements using the company's guidelines. Each element represents a distinct piece of information within an interview topic that is relevant to assessing the applicant. For example, within the \textit{Motivation \& attrition risk} topic, one information element is the applicant's main motivation for applying for the role, while another is their short- and long-term career goals. For each interviewer-question and applicant-response pair, we classify which information element the question targets and whether the response provides sufficient information to assess the applicant’s skills or fit for that element. Based on these classifications, we construct the variable \textit{number of unique information elements elicited}, which measures the number of distinct information elements for which the interview collected sufficient information.

Before moving to analyzing the distribution of our variables and how they differ across treatments, we first assess their validity. Specifically, we need to verify that our variables measuring the quality of applicants' responses are negatively valued by recruiters, and that more information collected is a positive property, i.e., positively valued by recruiters. As shown in Appendix Table \ref{tab_information_elements_predicting_outcomes}, the two deficit variables indeed negatively predict job offer and retention rates, whereas the number of information elements positively predicts these outcomes. Accordingly, these variables appear to be hiring-relevant. 

\paragraph{Results.} We start by comparing interviewer moves across treatments. Among all interviewer question types from the AI voice agent, 76.53\% are \textit{Topic-opening} questions, and 22.95\% are \textit{Adaptive follow-up} questions. Among human interviewers, these shares are 62.35\% and 34.03\%, respectively.\footnote{The remaining 0.52\% of conversational moves by the AI voice agent and 3.62\% of moves by human interviewers are off-topic.} We find similar shares if we instead look at average interviewer question type shares per interview. Accordingly, both human interviewers and AI voice agents adapt a substantial share of their moves in response to applicants' responses. At the same time, human interviewers engage more in \textit{Adaptive follow-up} behavior than the AI voice agent. 

We next ask whether adaptive follow-ups are targeted toward lower-quality responses. In Panels A and B of Figure \ref{fig:transcript_analysis_recrutiers_adaptiveness}, we compare, across treatments, the likelihood that an interviewer poses an \textit{Adaptive follow-up} question depending on whether the applicant's response showed a \textit{specificity deficit} or a \textit{substantive deficit}. Reassuringly, the likelihood of an \textit{Adaptive follow-up} is substantially higher when the applicant's answer is insufficient under either variable. Importantly, however, human interviewers have a significantly higher \textit{Adaptive follow-up} rate for answers classified as sufficient, i.e., without either a \textit{specificity deficit} or a \textit{substantive deficit} ($p<0.001$). For insufficient answers, we find no difference between AI and human interviewers for \textit{substantive deficit} ($p=0.69$), and only a small, albeit statistically significant, difference for \textit{specificity deficit} ($p<0.001$). Accordingly, we find that the AI voice agent engages less in follow-up behavior, but this is mainly driven by human interviewers following up on applicant responses classified as sufficient. 

Finally, Panel C of Figure \ref{fig:transcript_analysis_recrutiers_adaptiveness} connects interviewer behavior to the information collected through the interview. In \textit{Human Interviewer}, \textit{Topic-opening} questions elicit an average of 8.26 unique information elements per interview; including \textit{Adaptive follow-up} questions increases this number to 10.06. In \textit{AI Interviewer}, the corresponding numbers are 13.02 and 14.67. We thus find that follow-up behavior improves information collection. At the same time, the AI voice agent elicits significantly more unique information elements, both at the initial stage and when including the adaptive stage ($p<0.001$). See Appendix Figure \ref{fig:transcript_analysis_recrutiers_adaptiveness_distribution_information} for the distributions of the unique information elements elicited per interview. 

\paragraph{Heterogeneity in recruiter behavior.} We again complement our transcript-level analysis with a recruiter-level analysis. Specifically, for each recruiter, we calculate the average rate at which they follow up on applicant responses across all interviews they conducted. Mirroring our recruiter-level results for the standardization variables, we find substantial heterogeneity in recruiter follow-up behavior for all four response types, i.e., whether responses exhibit a substantive deficit and whether they exhibit a specificity deficit. See Appendix Figure \ref{fig:plot_recruiter_features_average_adaptiveness} for details. Comparing the AI voice agent with human recruiters, we find that only 5\% (8\%) of human recruiters follow up less often than the AI agent after responses with no specificity (substantive) deficit. By contrast, after responses with a specificity (substantive) deficit, the AI agent follows up more often than 18\% (51\%) of human recruiters. 

\medskip

Taken together, these results suggest that AI voice agents solve the standardization-adaptiveness trade-off differently from human recruiters. AI conducts interviews that are more consistent than human recruiters, both in adhering to firm guidelines and in terms of average behavior on each topic. In contrast, human recruiters are more likely to ask follow-up questions, thereby having a higher degree of adaptiveness. Yet, the AI does not implement standardization by following a fixed script. Instead, it remains responsive to applicant responses. Since the AI collects more hiring-relevant information and human recruiters focus more on sufficient applicant responses, our results indicate that the AI resolves the trade-off more optimally than human recruiters. 

\subsection{Applicant behavior in interviews}
\label{subsec:transcript_analysis_applicants}

In our previous analysis, we have investigated treatment differences in the \textit{quantity} of information collected as part of the standardization-adaptiveness trade-off. Next, we analyze the \textit{quality} of information collected. To assess quality, we construct variables that capture key communication signals in applicants' responses. We focus on communication signals because applicants are hired for customer service jobs, and thus one key purpose of the interview is to assess their conversational abilities.  

\paragraph{Applicants' communication signals.} For each interview, we construct the following eight variables using standard natural language processing methods: (1) vocabulary richness, (2) syntactic complexity, (3) frequency of discourse markers (sequential, causal, and clarifying words), (4) frequency of filler words, and (5) frequency of backchannel cues (short cues indicating attention or agreement). We also record (6) the number of exchanges between the interviewer and the applicant and (7) the number of questions posed by the applicant. Finally, we construct (8) an index of linguistic style matching between the interviewer and the applicant. Appendix \ref{app_sec:transcript_applicant_feature_variables} provides details on the construction of each variable. We selected and preregistered these eight variables to capture a broad spectrum of components relevant to conversational abilities. 

\begin{figure}[!t]
    \caption{Transcript communication signals analysis}
    \label{fig:transcript_features_main_results}
    \centering
    \includegraphics[width=1\textwidth]{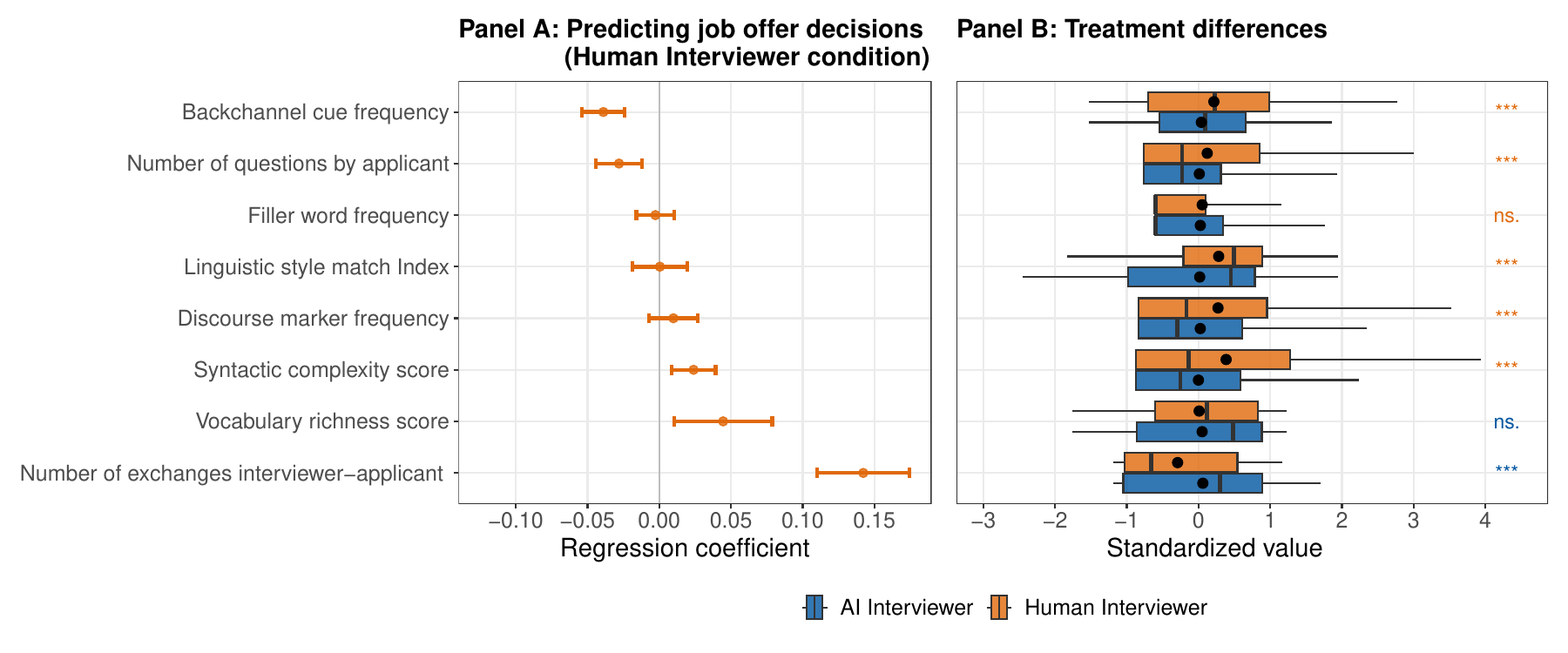}
    \figurenotes[Notes:]{Panel A displays the coefficients of an OLS regression of job offer decisions on the displayed variables measuring the communication signals of applicants (all standardized) in the \textit{Human Interviewer} condition. Error bars indicate the 95\% confidence interval, obtained using standard errors clustered at the applicant level. For details on the variables, see Appendix Table \ref{app_sec:transcript_applicant_feature_variables}. Panel B displays the distribution of the variables split by treatment using box plots. Each box represents the interquartile range (25th-75th percentiles), with the horizontal line indicating the median. Whiskers extend to 1.5 times the interquartile range, and dots represent means. Stars indicate statistically significant differences in means between the two treatments ($^{*}\,p<0.1$,$^{**}\,p<0.05$, $^{***}\,p<0.01$), based on two-sided t-tests with Bonferroni correction.
    }
\end{figure}

\paragraph{Treatment differences in communication signals.} We develop a two-step approach to investigate how applicants' language differs between AI-led and human-led interviews. In the first step, we identify which communication signals displayed by applicants are positively and which are negatively valued by human recruiters. A feature is defined as positively (negatively) valued if it positively (negatively) predicts job-offer decisions. Accordingly, we estimate a joint regression of job-offer decisions on all linguistic variables within the \textit{Human Interviewer} condition. Results are reported in Panel A of Figure  \ref{fig:transcript_features_main_results}. We find that the number of exchanges between the interviewer and the applicant, as well as the richness of applicants' vocabulary and the syntactic complexity, are significantly positive predictors. In contrast, the frequency with which applicants use backchannel cues and the number of questions they pose are negative predictors.

In the second step, we compare how the communication signals differ between treatments. As shown in Panel B of Figure \ref{fig:transcript_features_main_results}, the average scores differ significantly for six of the eight features, and we reject the equality of distributions in every instance ($p<0.01$, Kolmogorov–Smirnov test, Bonferroni corrected). Importantly, features that positively predict job offers are higher in the \textit{AI Interviewer} condition, while those that negatively predict job offers -- or have no significant predictive power -- are higher in the \textit{Human Interviewer} condition. These patterns suggest that AI-led interviews elicit more communication signals that recruiters value (e.g., number of exchanges) and reduce those associated with lower hiring odds, such as backchannel cues and applicant-posed questions.

Our results show that the AI voice agent conducts interviews differently from human recruiters, along dimensions associated with higher offer rates. Recruiters value comprehensive interviews that cover more topics and include specific communication signals. AI-led interviews are more likely to meet these criteria, producing more comprehensive conversations with broader topic coverage and higher scores on communication signals that, in human-led interviews, predict higher job offers.

%% file: Sections_paper/6_extensions.tex
\section{Recruiter response to voice AI interviews}
\label{main_sec:results_recruiters}

In this section, we analyze how the human recruiters evaluate applicants after the interview stage. 

\subsection{Analyzing recruiters' interview scores}
\label{subsec:interview_scores_comments}

We begin by analyzing how recruiters score interviews and how they use available interview signals and the test scores. We focus our analysis on the \textit{Walk-in} mode because, in the \textit{Remote} mode, it was not mandatory for recruiters to log their interview scores in the system.\footnote{Appendix Table \ref{tab_weighting_signals_recruiters_full_sample} replicates the main result of this section using the full sample.}

\paragraph{Interview scores.} For a total of 15,303 applications (88\% of all applications in \textit{Walk-in}), we observe the interview score (1,2,3) with which a recruiter rated an applicant’s interview performance.\footnote{Availability is balanced across the \textit{Human} and \textit{AI Interviewer} conditions ($p=0.64$, two-sample test of proportions).} For a subset of 10,779 applications, we also observe their justification for the score in an open-ended text format. 

On average, applicants receive a score of 1.90 in the \textit{Human Interviewer} condition and a score of 2.01 in \textit{AI Interviewer}, a significant difference ($p<0.001$, two-sample t-test). Accordingly, recruiters rate the interviews conducted by the AI voice agent as higher than the interviews they conduct themselves. In Appendix Figure \ref{fig:main_sec_interview_score_distribution}, we plot the distribution of scores across the two conditions. We find that higher scores in \textit{AI Interviewer} are driven exclusively by a higher frequency of score 2 rather than score 1, whereas the frequency of score 3 remains similar. 

\paragraph{Interview score justification.} When providing the interview score, recruiters also submit a short justification for their score in an open-ended text format. Using natural language processing, we classify their justifications according to their sentiment. We find that 31\% of the justifications in the \textit{AI Interviewer} condition are categorized as positive sentiment, while only 24\% are so in the \textit{Human Interviewer} condition. This difference is statistically significant ($p < 0.001$, two-sample test of proportions). In contrast, the fraction of justifications that have negative sentiment is 28\% in the former and 38\% in the latter. At the same time, the fraction of neutral sentiment justifications is roughly similar across conditions (\textit{AI Interviewer}: 41\%; \textit{Human Interviewer}: 38\%). Accordingly, recruiters' comments show a positive response to the interview content produced by the AI voice agent, consistent with such interviews receiving higher interview scores.

\begin{table}[!t]
    \caption{Predicting job offer decisions of recruiters}
    \label{tab_weighting_signals_recruiters}
        \begin{threeparttable}[c]
            \input{Tables/tab_weighting_signals_recruiters}
            \begin{tablenotes}
                \item \footnotesize \textit{Notes:} 
                The table shows OLS estimates predicting job offer decisions of recruiters using standardized test scores and interview scores. The dependent variable is a dummy equal to one if an application led to a job offer. Test scores are standardized. ``\textit{AI Interviewer}'' is a dummy equal to one if the application was in the \textit{AI Interviewer} condition, and zero if the application was in the \textit{Human Interviewer} condition. Controls include an applicant's gender, source of application, pre-treatment engagement score, and whether they have applied before to any of the firm's job postings. Fixed effects include week, recruiter, application site, and job posting fixed effects. An observation is an application. Standard errors in parentheses are clustered at the applicant level. Significance levels: $^{*}\,p<0.1$,$^{**}\,p<0.05$, $^{***}\,p<0.01$.
            \end{tablenotes}
        \end{threeparttable}
\end{table}

\subsection{Recruiter weights on signals}

To determine whether to extend an offer to an applicant, recruiters have three signals about an applicant's quality available to them: (i) interview performance, (ii) standardized quantitative test score, and (iii) standardized language test score. We are interested in how much weight recruiters put on each when making decisions and, in particular, whether the weights differ across treatments. 

\paragraph{Results.} In Table \ref{tab_weighting_signals_recruiters}, we regress the recruiters' decisions about whether to extend a job offer to an applicant on interview, analytical, and language scores. For comparability, we standardize all three variables. We find that all three signals have a significant positive influence on the likelihood that a recruiter extends a job offer to an applicant in the \textit{Human Interviewer} condition. In terms of magnitude, we find that almost equal weight is put on the language and interview score, while less weight is put on the analytical score. Specifically, a one standard deviation increase in interview score is associated with an 8.9 percentage point higher likelihood of offer, keeping performance in language and analytical tests constant, controlling for baseline characteristics and including fixed effects. In contrast, a one standard deviation increase in the language and analytical score is associated with an increase of 10.8 and 3.5 percentage points, respectively. 

Importantly, we find a significant interaction when comparing the influence of each signal in \textit{Human Interviewer} relative to \textit{AI Interviewer}. The language score is significantly more predictive of offers in \textit{AI Interviewer}, while the interview score is significantly less predictive. This suggests that recruiters place more weight on the independent quality signal coming from the language test than on the interview when the signal from the interview comes from the AI and not from themselves conducting the interview. 

\paragraph{Mechanism.} To further investigate whether the interaction effect reflects recruiters' differential weighting of signals rather than other influences, we use their responses from the recruiter survey. Specifically, we use their responses to the question of how important they consider the interview and test scores in making offer decisions. As documented previously, there is a sufficient degree of heterogeneity in their responses. We exploit this heterogeneity by repeating the analysis of Table \ref{tab_weighting_signals_recruiters} separately for the sample of recruiters who state in the recruiter survey that the interview score is equally or less important than the test scores in determining their offer decisions, and recruiters who state that interview scores are more important. We would expect the previously documented interaction effect -- namely, that interview scores are less predictive of offer decisions in the \textit{AI Interviewer} condition relative to the \textit{Human Interviewer} condition -- to be stronger among recruiters who consider interview performance more important. We indeed find that the interaction of treatment with interview score is more negative (and only statistically significant) for recruiters who consider the interview as more important than the standardized scores. For details, see Appendix Table \ref{tab_weighting_signals_recruiters_heterogeneity}. This suggests that recruiters indeed weigh the signals from AI-led interviews differently.

\subsection{Recruiter heterogeneity in offer decisions}

We investigate heterogeneity in the offer decisions of recruiters along two dimensions. First, how much do offer rates differ between recruiters (level difference)? Second, how much do offer rates differ between treatments across recruiters (slope difference)? To answer these questions, we calculate the average offer rate of each recruiter, separately for the \textit{Human Interviewer} and \textit{AI Interviewer} conditions. We focus on recruiters who were assigned to review at least 25 applications in \textit{Human Interviewer} and at least 25 applications in \textit{AI Interviewer}. This leaves us with 61 recruiters out of 131 in the full sample.

We find sizable variation in average offer rates among recruiters in both the \textit{Walk-in} and \textit{Remote} mode. For instance, some recruiters have an average offer rate that is more than twice as high as other recruiters, see Appendix Figure \ref{fig:scatter_plot_recruiter_percentage_both_walkin_remote} for a scatter plot. These differences could be caused by recruiter-specific traits, regional differences in the application pool, or characteristics of the job opening for the respective client firm. At the same time, offer rates are highly correlated across treatments (Full sample: $\rho=0.87$, $p<0.001$, \textit{Walk-in:} $\rho=0.83$, $p<0.001$, \textit{Remote:} $\rho=0.47$, $p=0.007$). That is, recruiters who have, on average, a high offer rate when reviewing human interviews also have a high offer rate when reviewing interviews from the AI voice agent. 

In terms of aggregate offer behavior, we find that 69\% of recruiters have a higher average offer rate when reviewing interviews conducted by the AI voice agent relative to reviewing human interviews, while 31\% of recruiters have the reverse. Accordingly, our main effect of AI applications receiving higher offer rates is not driven by a small number of recruiters differentiating strongly between AI and human interviews. Instead, it appears to be driven by the majority of recruiters.\footnote{In Appendix \ref{app_sec:recrutier_over_time}, we further investigate whether recruiters' experience with AI-led interviews influences their offer decisions.}

\section{Applicant responses to voice AI interviews}
\label{main_sec:results_applicants}

In this section, we examine applicants' attitudes towards AI-led interviews in greater detail. Specifically, we analyze their decisions to accept job offers, their perceived interview experience, and their choices in the \textit{Choice of interviewer} condition. 

\subsection{Applicants' responses to job offers}
\label{subsec:applicants_reponse_job_offers}

After completing an AI-led interview, applicants may update their understanding of the hiring company or the prospective job, which in turn influences their decision about whether to accept a job offer. For instance, the use of AI voice agents as interviewers may affect applicants' perceptions of match quality. 

In total, of the 6,319 job offers made, applicants accepted 5,854, implying a 92.64\% acceptance rate. In \textit{Human Interviewer}, the acceptance rate is 93.64\%. The rate is slightly lower in \textit{AI Interviewer}, at 92.14\%. However, we cannot reject the equality of acceptance rates between the two treatments at the 5\% level ($p=0.086$, two-sample proportion test). In the \textit{Choice of interviewer} condition, 92.73\% of applicants who chose the human interviewer accepted an offer, compared to 92.57\% of those who chose the AI interviewer. We find no significant differences in acceptance rates when comparing the \textit{Human Interviewer} condition with applicants who selected the human interviewer, or the \textit{AI Interviewer} condition with those who selected the AI voice agent (Human: $p=0.55$; AI: $p=0.67$; two-sample test of proportions). Similarly, we find no significant difference in acceptance rates between applicants who chose the human interviewer and those who chose the AI voice agent ($p=0.92$, two-sample test of proportions). Accordingly, among applicants who receive job offers, there appears to be no negative reaction to the AI voice agent, as evidenced by no significant increase in offer refusals.

\vspace{-0.25cm}
\subsection{Applicants' interview experience}

Next, we analyze applicants' attitudes and perceptions of their interview experiences. These were measured using the customer experience survey administered to applicants after the interview.

\paragraph{Net promoter score (NPS).} We start by comparing the NPS -- how applicants rate their propensity to recommend the recruiting firm to a friend -- which is the central feedback metric used by the firm and the industry more generally. We find that the average rating for applicants in the \textit{AI Interviewer} condition is 8.97 on a 1-10 scale, whereas it is 8.84 in the \textit{Human Interviewer} condition. Accordingly, AI voice agents yield a small and insignificant increase in the NPS score ($p=0.25$, t-test). 

\paragraph{Perceived recruiter quality.} 

Applicants rated recruiter quality similarly in the \textit{AI Interviewer} and \textit{Human Interviewer} conditions. In particular, there were no significant differences in how knowledgeable applicants thought the recruiter was when giving information and answering questions about the firm ($p = 0.58$, t-test) or the role ($p = 0.11$, t-test), and in the extent to which applicants felt their time was valued by the recruiter ($p = 0.24$, t-test). The only notable difference is in the perceived relevance of the interview questions, with applicants rating the AI voice agent as asking slightly more relevant questions ($p = 0.044$, t-test). The index of perceived recruiter quality, which averages the individual items, also shows no difference between treatments ($p = 0.92$, t-test).

\paragraph{Perceived interview quality and fairness.} We find that applicants rate interviews conducted with the AI voice agent as slightly less stressful and more comfortable, but the differences are very small and far from significant at conventional levels (stressful: $p=0.65$, comfortable: $p=0.69$, t-test). In terms of naturalness, applicants rate the interview experience with the AI voice agent as significantly less natural ($p=0.014$, t-test). This results in a higher index of perceived interview quality for human-led interviews than for AI-led interviews ($p=0.076$, t-test). When asked about the follow-up flow and the frequency of feedback, the applicants in both treatments rate their experience similarly (follow-up flow: $p=0.78$, frequency of feedback: $p=0.25$, t-test). These results suggest that the AI voice agent can deliver interview-quality speech that applicants perceive as similar to that of humans. When asked to rate the fairness of their interview, applicants rate it similarly across both treatments ($p=0.68$, t-test). 

\paragraph{Open-ended feedback.} At the end of the survey, the applicants were invited to share additional feedback about their interview experience in an open-ended text response. In total, 9.60\% of applicants in the \textit{Human Interviewer} and \textit{AI Interviewer} conditions provided a response. We use sentiment analysis to classify whether applicants express negative, positive, or neutral sentiment in their text.\footnote{The analysis was conducted on the whole sample, with the instructions blind to treatment status.} We find that of those applicants who provide a response, in \textit{AI Interviewer} 71\% of the responses have a positive, 14\% a negative, and the remaining 14\% a neutral sentiment. In contrast, in the \textit{Human Interviewer} condition, 52\%, 30\%, and 19\% of responses exhibit positive, negative, and neutral sentiment, respectively. Accordingly, the likelihood that applicants express a positive sentiment is higher in the \textit{AI Interviewer} condition ($p=0.005$, two-sample test of proportion).\footnote{We also use \textit{gemini-2.5-flash} to categorize responses into 13 distinct categories (see Appendix Table \ref{tab_classification_survey_responses} for details). In total, 45\% of responses in \textit{AI Interviewer} and 19\% in \textit{Human Interviewer} mention a comfortable and positive interview experience. In total, 10\% (13\% ) of responses in \textit{AI Interviewer} (\textit{Human Interviewer}) mention problems with the audio, questions in general, or the AI voice agent specifically. The categorization results are thus in line with the sentiment analyses.} 

\medskip

Overall, we conclude that applicants generally rate their experience similarly between human recruiters and AI voice agents. 

\paragraph{Robustness.} A concern when interpreting applicants' survey responses is that their responses may be biased by demand effects. Although the survey is administered by a separate unit to which recruiters do not have access, and this was communicated to applicants in the survey, applicants may still believe that they can influence the process through their responses. For example, this may lead them to provide overly positive answers. Although such misreporting would bias the level of survey responses, it is less likely to influence our relative treatment comparison, as it would require an interaction between misreporting and treatment. To empirically assess the extent of bias, we randomly assigned applicants to either receive the survey invitation immediately after their interview or after a final decision on their application. This allows us to test whether applicants' responses differ. We find no evidence that applicants strategically change their responses, as we cannot reject the null hypothesis of equal means for our survey variables. See Appendix Table \ref{tab_balance_check_timing_survey} for details. 

\vspace{-0.25cm}
\subsection{Applicants' choices of interviewers} 
\label{subsec:applicant_persp_choice_of_interviewer}

\paragraph{Choices.}

In the \textit{Walk-in} mode, out of the 3,469 applicants that were randomized into the \textit{Choice of Interviewer} treatment, 3,420 (98.59\%) made a choice between the human recruiter and the AI voice agent. Of those, 2,370 (69.30\%) chose the AI voice agent as interviewer. In the \textit{Remote} mode, out of the 9,927 applicants that were randomized into the \textit{Choice of Interviewer} treatment, 9,659 (97.30\%) made a choice between the human interviewer and the AI voice agent. The remaining applicants did not respond to the text interview invitation or the follow-up call. Of those making a choice, 7,885 (81.63\%) chose the AI voice agent as the interviewer. Accordingly, in both modes, most applicants prefer to interview with an AI voice agent instead of a human recruiter. Aggregating across modes, we have that 78.41\% of applicants choose the AI voice agent. Choice patterns are relatively stable throughout the experiment, as Figure~\ref{fig:choice_of_recruiter_over_time} in the Appendix shows.

\paragraph{Predicting choices.} What explains the high fraction of applicants who prefer AI over humans? A primary reason may be convenience: Interviews with the AI voice agent can be scheduled at the applicant's preferred time, including immediately after receiving the interview invitation.\footnote{Indeed, we find that the time between application and interview is significantly shorter for AI-led interviews.} A second reason may be applicants' attitudes towards AI. Our survey evidence indicates that the sample generally perceives AI as having a major impact on the workplace and that this impact will generally be positive. Among all survey respondents, 48\% believe that AI will have a major impact on themselves personally, 34\% believe that the impact is minor, and 18\% believe that AI will have no impact. A total of 47\% of the survey respondents think that the impact of AI on them in the workplace will be positive, 34\% think that the positive and negative impact will be roughly balanced, and only 19\% think AI will have a negative impact on them.\footnote{Perhaps unsurprisingly, recruiters are more pessimistic about the impact of AI on them. We asked the same questions in the recruiter survey and found that 68\% of recruiters believe AI will have a significant personal impact, whereas only 12\% believe the impact will be generally positive.} We obtain similar results when looking at how the respondents assess the impact of AI on workers generally.

Importantly, for those survey takers in the \textit{Choice of Interviewer} condition, among those believing in a positive impact, 77\% choose the AI voice agent, among those believing in a balanced impact, 72\% choose the AI voice agent, and among those believing in a negative impact, 65\% choose the AI voice agent. Regressing the interview choice on the survey item, we find that the item predicts the choices, albeit only significantly so once controls and fixed effects are added. For detailed regression results, see Appendix Table \ref{tab_applicant_level_predicting_recruiter_choices}.\footnote{One might worry that applicants' perception of the directional impact of AI on themselves is affected by their experience with the AI interviewer. Comparing applicants' responses in \textit{AI Interviewer} with those in \textit{Human Interviewer}, we find no difference ($p=0.58$, t-test).}

\paragraph{Sorting.} Next, we analyze sorting effects: whether applicants differ in their quality when they can choose a recruiter compared to when they are assigned one. We use applicants' test scores as our measure of applicant quality, as they provide a signal about quality independent of interview performance. Three potential sorting effects could be present. First, positive AI sorting means that high-quality applicants opt for the AI interviewer, whereas lower-quality applicants opt for the human interviewer. Second, negative AI sorting means the reverse pattern. Third,  choice and quality may be uncorrelated. 

We start by analyzing the association between applicants' choices in \textit{Choice of Interviewer} and their test scores. Applicants who selected the AI voice agent scored an average of 3.14 (out of 6) on the language test and 47.54 (out of 100) on the analytical test. In contrast, applicants who chose the human interviewer obtained higher test scores of 3.37 and 49.77 on the language and analytical tests, respectively. These differences are statistically significant ($p<0.001$ and $p=0.001$, respectively, t-tests). The differences are robust to the inclusion of controls and fixed effects; for details, see Appendix Table \ref{tab_applicant_level_selection_test_score_only_choice}.
Similarly, we reject the equality of distributions (language: $p<0.001$, analytical: $p=0.010$, Kolmogorov–Smirnov test). For the distributions, see Appendix Figure \ref{fig:AM_test_scores_condition_on_interviewer_choice}. This pattern is consistent with models of choice-as-signal in screening \parencite{JabarianReshidiChoice}.

As a next step, we analyze whether there exist sorting patterns conditional on receiving the same interviewer. That is, we compare applicants who got assigned the AI voice agent in \textit{AI Interviewer} with those who chose it in \textit{Choice of Interviewer}, and applicants who got assigned the human recruiter in \textit{Human Interviewer} with those who chose the human recruiter in \textit{Choice of Interviewer}. This type of analysis is robust to a potential confounding factor: it may be that experiencing the AI voice agent, rather than the human recruiter, affects test performance. For example, applicants may increase or decrease their effort after interacting with the AI voice agent. More plausibly, the influence could arise from differential attrition, as applicants are more or less likely to drop out of the recruitment process after completing the interview with the AI voice agent and before taking the test.

We regress on test scores a dummy that is equal to one if the applicant has chosen the respective interviewer and zero if the applicant got assigned the interviewer. We find that applicants who choose AI have lower test scores than those assigned to it, whereas applicants who choose the human interviewer have higher test scores. However, three of the four differences are no longer statistically significant after controlling for and including fixed effects. Accordingly, the sorting effects are slightly weaker when comparing performance in the choice condition with that in the assigned conditions. For details, see the Appendix Table \ref{tab_applicant_level_selection_test_score}. Together, our results suggest the presence of negative AI sorting.

%% file: Tables/tab_weighting_signals_recruiters.tex
\begin{tabular}{@{\extracolsep{5pt}}lcc} 
\\[-1.8ex]\hline 
\hline \\[-1.8ex] 
 & \multicolumn{2}{c}{\textit{Dependent variable:} } \\ 
  & \multicolumn{2}{c}{\textit{Job Offer Made}} \\ 
\cline{2-3} 
\\[-1.8ex] & (1) & (2)\\ 
\hline \\[-1.8ex] 
 Interview score (std.)  & 0.091$^{***}$ & 0.100$^{***}$ \\ 
  & (0.010) & (0.010) \\[0.2cm]
 Language test score (std.)  & 0.108$^{***}$ & 0.115$^{***}$ \\ 
  & (0.010) & (0.010) \\[0.2cm]
 Analytical test score (std.)   & 0.034$^{***}$ & 0.028$^{***}$ \\ 
  & (0.011) & (0.010) \\[0.2cm]
 Interview score (std.) $\times$ \textit{AI Interviewer} & $-$0.047$^{***}$ & $-$0.029$^{**}$ \\ 
  & (0.012) & (0.012) \\[0.2cm]
Language test score (std.) $\times$ \textit{AI Interviewer} & 0.028$^{**}$ & 0.022$^{*}$ \\ 
  & (0.012) & (0.011) \\[0.2cm]
 Analytical test score (std.) $\times$ \textit{AI Interviewer} & $-$0.003 & 0.001 \\ 
  & (0.012) & (0.012) \\[0.2cm]
\hline \\[-1.8ex] 
Mean DV in Human Interviewer & 0.38 & 0.38 \\ 
Controls and fixed effects & -- & Yes \\ 
Observations & 9,965 & 9,864 \\ 
R$^{2}$ & 0.118 & 0.218 \\ 
\hline 
\hline \\[-1.8ex] 
\end{tabular} 

%% file: Sections_paper/7_conclusion.tex
\vspace{-0.5cm}
\section{Conclusion}
\label{main_sec:conclusion}

Most empirical research on the impacts of AI in firms so far has studied AI as a complement to human work, augmenting humans by providing them with information. This paper studies a distinct channel: AI agents replacing humans in collecting information from humans through live spoken conversation. Using a large-scale natural field experiment in recruitment, we show that conducting job interviews with AI voice agents rather than human recruiters improves match quality: we observe an increase in offer and retention rates without changing the composition of exits or worker productivity. Analyzing interview transcripts, our results suggest that these gains arise from a mechanism we label \textit{controlled variance}: AI agents conduct job interviews in a more standardized way -- with substantially less dispersion across interviews -- while maintaining sufficient within-interview adaptiveness, which improves the information collected. Hence, AI agents appear to navigate the trade-off between standardization and adaptiveness inherent in many information-collection settings more effectively. 

Our study thus informs the optimal deployment of AI as an autonomous technology that collects signals from humans through live conversations. Our treatment analysis shows the feasibility of using AI for this task. Modern AI technologies are not only technically capable of performing this task, but they may also perform the task better than humans in certain dimensions. Our mechanism results then point to the important role of standardization versus adaptiveness as levers to consider and optimally configure when adopting AI for information-collection tasks. Our setting and results suggest two key conditions to consider when deciding to adopt AI technology for collecting information about decision-relevant attributes from humans. 

The first condition is what we call \textit{screening-attribute fit}: can the relevant attributes be specified sufficiently clearly ex ante and reliably elicited through a common conversational protocol? Certain skills and attributes translate relatively easily into clear screening objectives, whereas some others may be harder to specify without relying on the interviewer’s tacit knowledge, improvised intuition, or experience. For example, language proficiency, work experience, schedule availability, and employment history are easy-to-screen attributes, whereas attributes such as research interests, leadership potential, creativity, or highly specialized expertise may be harder to screen. In our setting, the firm hires for entry-level positions that require a relatively well-defined bundle of attributes, including communication skills, educational and employment background, and practical job fit. Moreover, the availability of quantitative job performance metrics makes it possible to quantify hiring success and adjust the screening process accordingly. 

The second condition is what we call \textit{screening-environment fit}: does variation in how the relevant attributes are elicited generate useful information, or does it primarily introduce noise into the decisions that are based on the elicited information? In contexts such as ours, where the firm does not seek superstar or unconventional candidates but instead assesses a relatively homogeneous applicant pool against a quality threshold, variation in screening appears to be negative. Differences in the process such as question wording, coverage, and follow-up behavior then cause the otherwise comparable applicants to experience different screening processes. This, in turn, reduces the comparability of the resulting signals without revealing additional information about match quality. For example, some candidates may end up with short interviews, while others may face interviews with only a subset of questions that dig into specific topics rather than covering all. In both cases, this leaves the human recruiter reviewing interview transcripts with incomplete or low-quality information about candidates’ skills and attributes, increasing the likelihood of a poor matching outcome. Our setting exhibits precisely these features: high-volume recruitment requires many repetitions of the same task and, absent AI, delegates interviews across many recruiters who differ substantially in how they conduct them. In contrast, in other settings, variation in elicitation may generate useful information if a firm is recruiting from a heterogeneous candidate pool and benefits from diverse human perspectives on a specific skill.

These two conditions generate a four-option decision matrix for AI adoption. First, when the relevant attributes can be specified ex ante and variation in their elicitation is likely to be harmful to information collection, AI has its clearest advantage: the firm can encode screening objectives that ensure a standardized process while allowing adaptiveness within those constraints. Second, suppose attributes can be easily specified ex ante, but elicitation variance is informative. Then, a more flexible AI screening protocol or even a human-AI hybrid system in which humans retain discretion over selected parts of the interview could be preferable. Third, assume that attributes cannot be easily specified ex ante, and that exploratory variance in screening is informative. Here, relying on standardization risks imposing a single interpretation on an inherently complex attribute. For example, imposing a fixed interpretation of the attribute "taste" may generate consistent screening while systematically excluding valuable but non-standard (or even superstar) candidates. Hence, in this setting, the combination of human expertise and live improvisation in screening may be better than AI agents at exploring a richer set of dimensions of candidates' taste. Fourth, when attributes are hard to specify ex ante and interviewer variance is nonetheless uninformative, then there is no one obvious solution between AI and human recruiters. In this case, the priority is to improve the definition and measurement of the screening objective before considering the choice between AI and human recruiters. 

In sum, given the inherent trade-off between standardization and adaptiveness, the advantage of AI screening lies neither in full standardization nor in unrestricted adaptiveness. Instead, it lies in the fact that AI extends the set of contractable tasks available to a principal, something not possible with human workers at the same level of detail before. In particular, AI makes design choices about the degree and location of conversational variance. Firms can constrain the agent to cover common topics, use consistent screening criteria, and return to the interview protocol while preserving adaptiveness for clarification and targeted follow-up. Beyond the legal and ethical concerns, achieving a perfect or close-enough equivalent allocation of discretion to human recruiters is difficult and costly: firms can provide scripts and training, but question-by-question adherence, coverage, or the circumstances under which interviewers should deviate from the protocol are very difficult to enforce and contract. AI agents render this contract feasible, making this allocation more directly programmable and auditable.

%% file: Sections_appendix/appendix_additional_figures.tex
\section{Additional figures}
\label{app_sec:add_figures}

\begin{figure}[!ht]
\centering
\resizebox{0.9\linewidth}{!}{%
\begin{tikzpicture}[
    stage/.style ={font=\sffamily\footnotesize, align=center},
    index/.style ={font=\ttfamily\scriptsize, text=gray},
    line/.style  ={gray, thick},
]

\draw[line] (0,0) -- (16,0);  %

\foreach \x [count=\i] in {0,2.25,4.5,6.75,9,11.25,13.5,15.75}{
  \node[stage] at (\x,0.8) {%
    \ifcase\i\or Profile\\creation%
      \or Screening\\process%
      \or Scheduling\\interview%
      \or Job\\interview%
      \or Analytical\\\& language\\test%
      \or Review \&\\decision%
      \or Background\\check%
      \or Employee\\onboarding%
    \fi};
  \node[index] at (\x,-0.6) {0\i};
}
\end{tikzpicture}}%
\caption{Recruitment process}
\label{fig:hiring-funnel}
\end{figure}

\begin{figure}[!ht]
    \caption{Distribution of applications by recruitment center location}
    \label{fig:map_of_applications_by_center}
    \centering
    \includegraphics[width=1\textwidth]{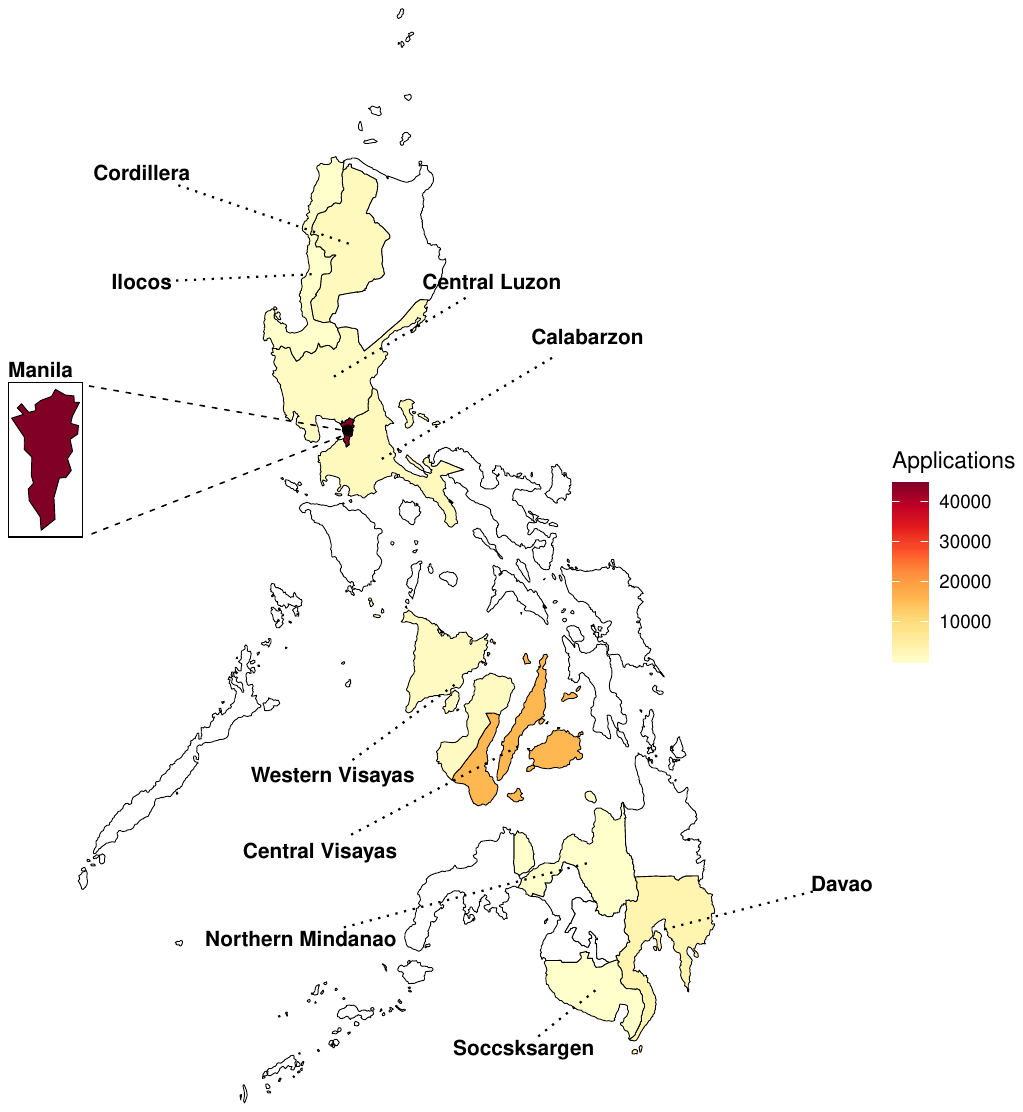}
\end{figure}

\begin{figure}[!t]
    \caption{Treatment effect on key recruiting outcomes in the conditional sample}
    \label{fig:main_sec_main_outcomes_full_figure_conditional}
    \centering
    \includegraphics[width=0.8\textwidth]{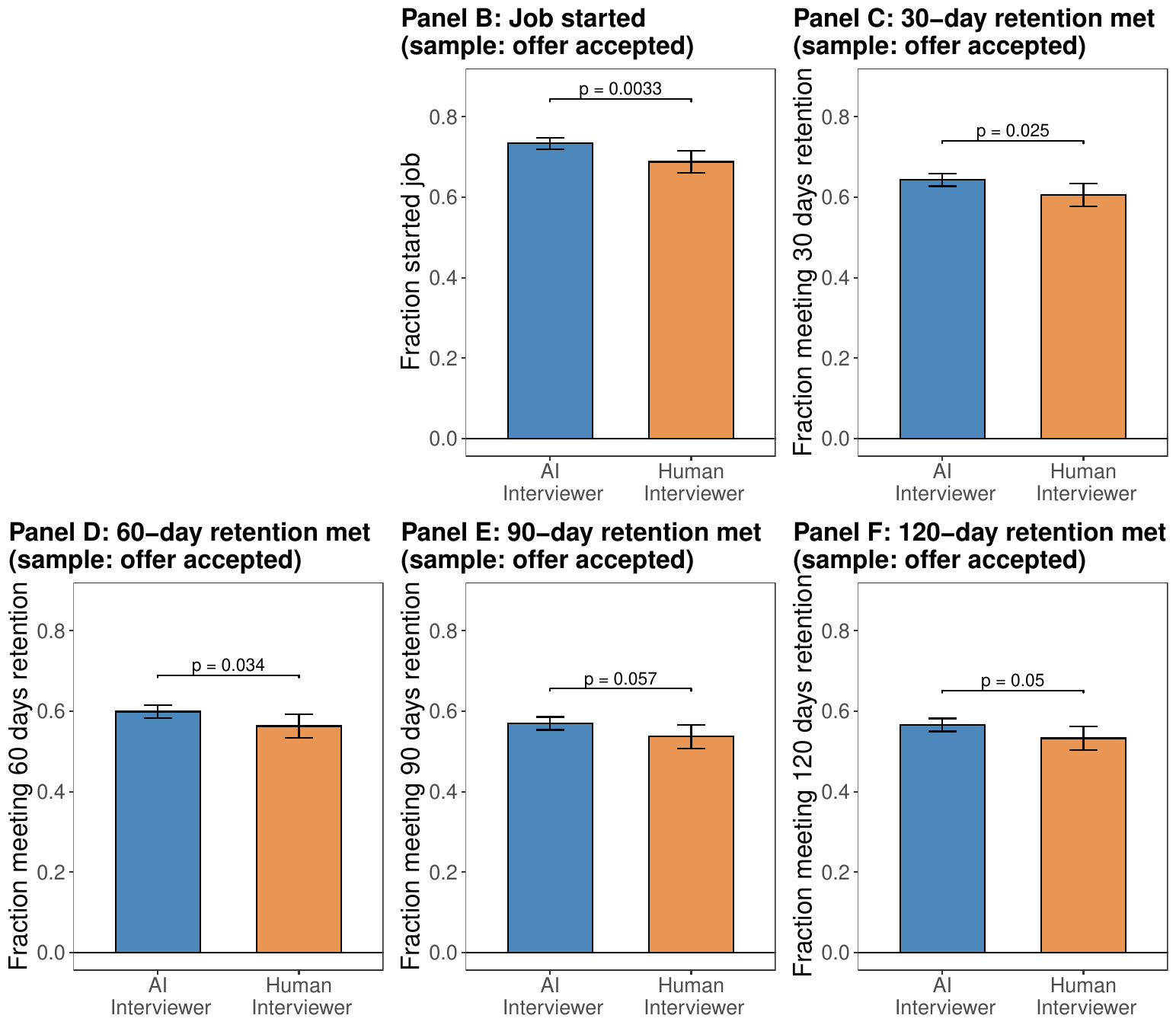}
    \figurenotes[Notes:]{The figure displays the recruiting outcomes of applicants conditional on their acceptance of a job offer. Each panel displays the fraction of applicants who realize the specific outcome. Fractions are displayed separately for the \textit{Human Interviewer} condition, in which applicants are interviewed by a human, and for the \textit{AI Interviewer} condition, in which applicants are interviewed by an AI voice agent. Bars indicate 95\% confidence intervals; p-values are calculated from a two-sample proportion test.
    }
\end{figure}

\begin{figure}[!t]
    \caption{Distribution of interviewers' language averaged per interviewer across interviews}
    \label{fig:plot_recruiter_features_average_standardization}
    \centering
    \includegraphics[width=1\textwidth]{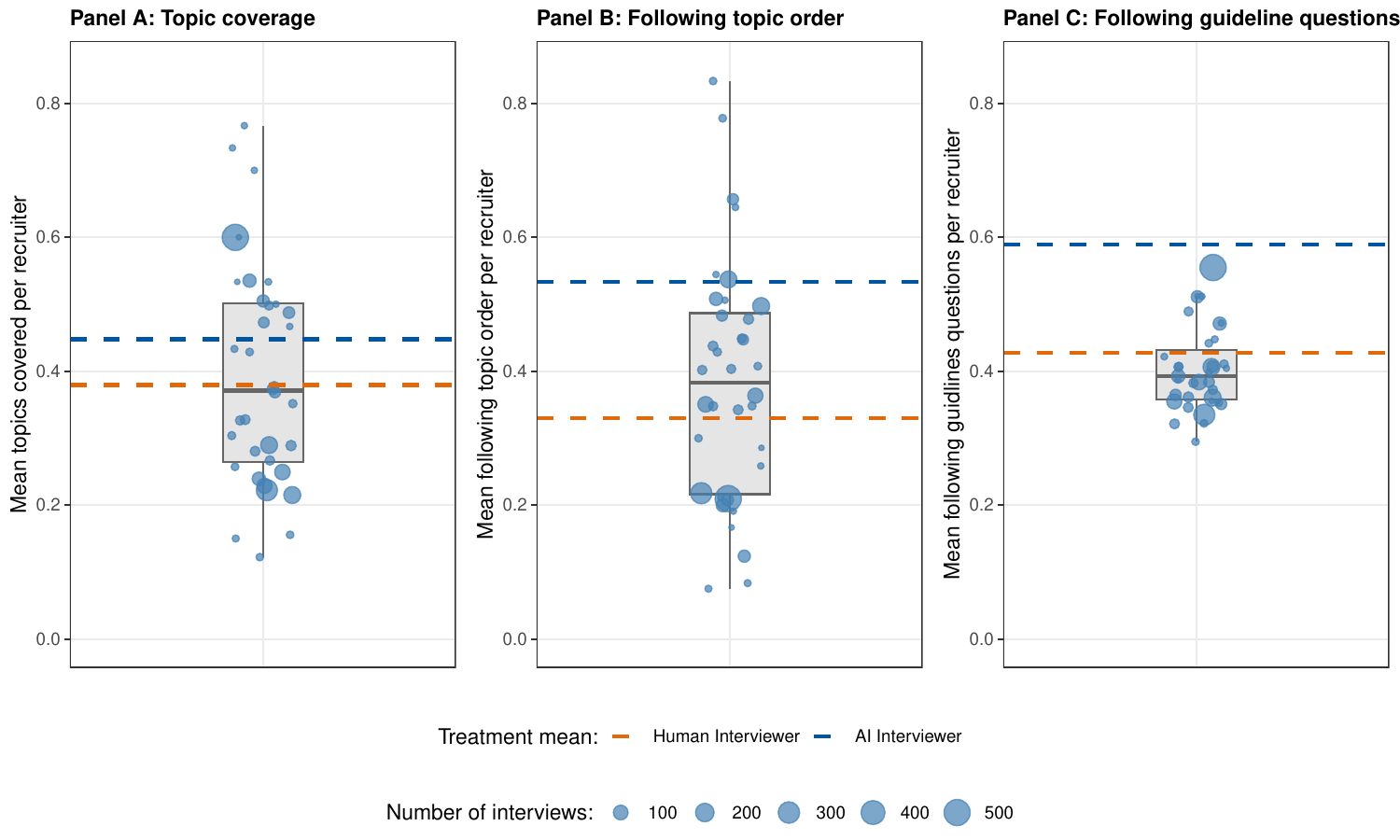}
    \figurenotes[Notes:]{
    This figure shows the distribution of interviewer-level averages of three measures of interviewer language behavior. Each panel plots one variable at the recruiter level: (A) unique topic coverage, (B) topic order correlation, and (C) topic–question similarity. Each dot corresponds to a single human recruiter, with the size of the dot proportional to the number of interviews conducted by that recruiter in the \textit{Human Interviewer} condition. The box plots summarize the distribution across recruiters: the box spans the interquartile range (25th to 75th percentile), the horizontal line indicates the median, and whiskers extend to 1.5 times the interquartile range. Dashed horizontal lines show the mean of the corresponding variable at the interview level in each treatment group. For definitions and construction of the variables, see Appendix Section~\ref{app_sec:transcript_recruiter_feature_variables}.    
    }
\end{figure}

\begin{figure}[!t]
     \caption{The distribution of hiring-relevant information pieces per treatment and question type}
    \label{fig:transcript_analysis_recrutiers_adaptiveness_distribution_information}
    \centering
    \includegraphics[width=1\textwidth]{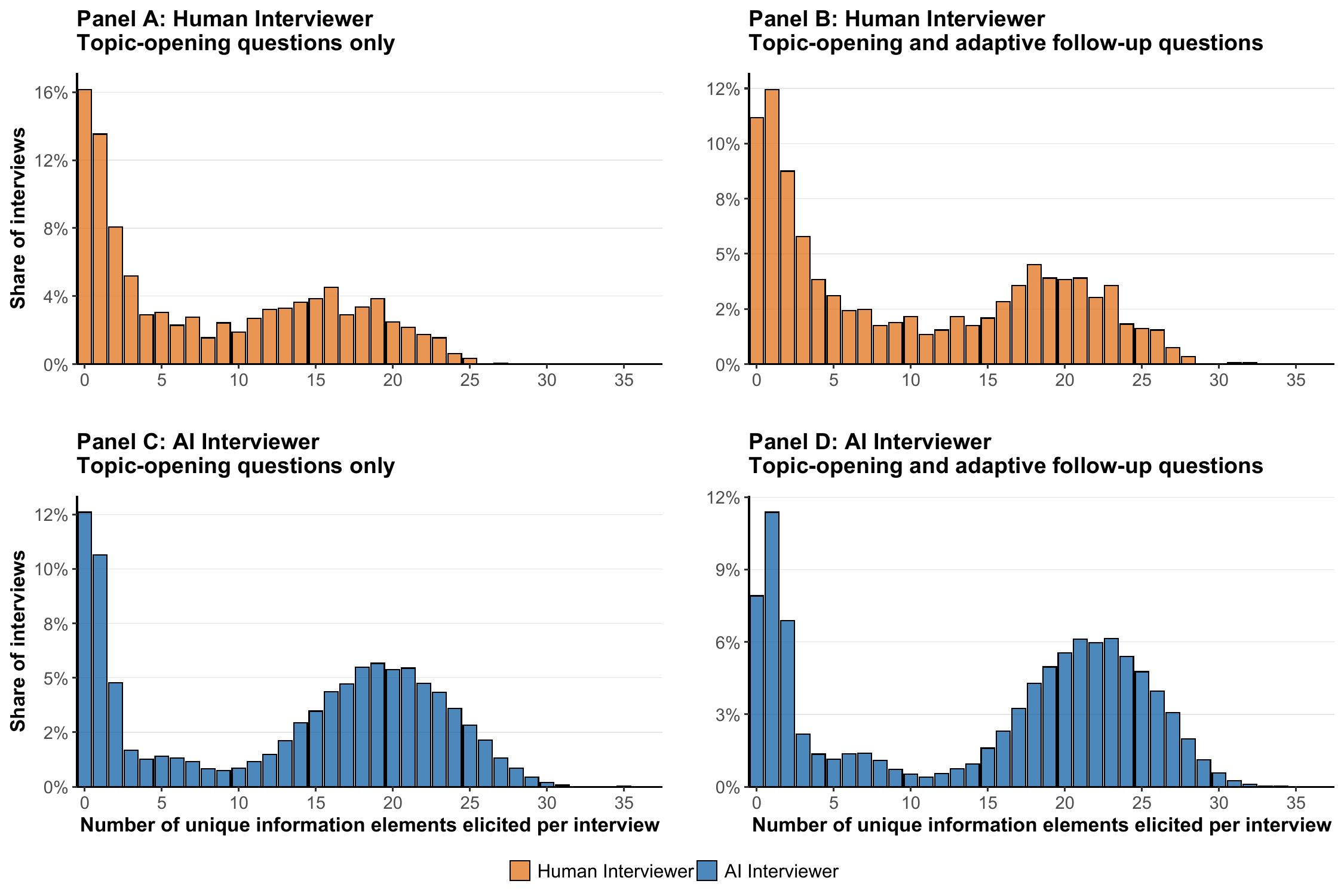}
\end{figure}

\begin{figure}[!t]
    \caption{Distribution of follow-up question rates averaged per interviewer across interviews}
    \label{fig:plot_recruiter_features_average_adaptiveness}
    \centering
    \includegraphics[width=1\textwidth]{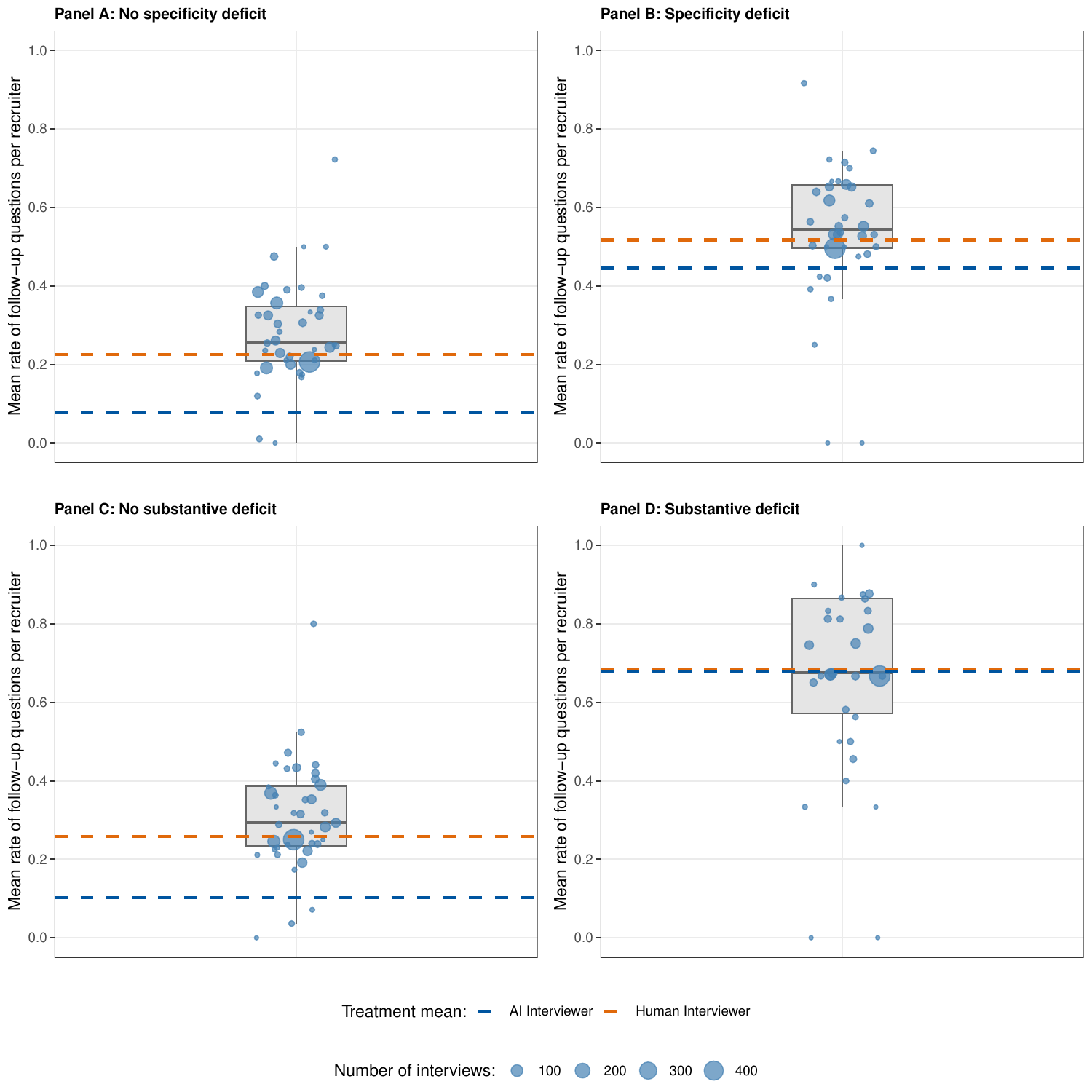}
    \figurenotes[Notes:]{
    This figure shows the distribution of recruiter-level averages of the rate of follow-up questions after applicants’ initial responses to guideline questions. Each panel plots one measure at the recruiter level: (A) no specificity deficit, (B) specificity deficit, (C) no substantive deficit, and (D) substantive deficit. Each dot corresponds to a single human recruiter, with dot size proportional to the number of eligible interviews conducted by that recruiter in the \textit{Human Interviewer} condition. The box plots summarize the distribution across recruiters: the box spans the interquartile range (25th to 75th percentile), the horizontal line indicates the median, and whiskers extend to 1.5 times the interquartile range. Dashed horizontal lines show the mean follow-up rate at the response level in each treatment group, separately for each condition. For definitions and construction of the variables, see Section~\ref{subsec:transcript_analysis_recrutiers_adaptiveness}.    
    }
\end{figure}

\begin{figure}[!t]
    \caption{Transcript variables per treatment}
    \label{fig:interview_types_treatment}
    \centering
    \includegraphics[width=0.9\textwidth]{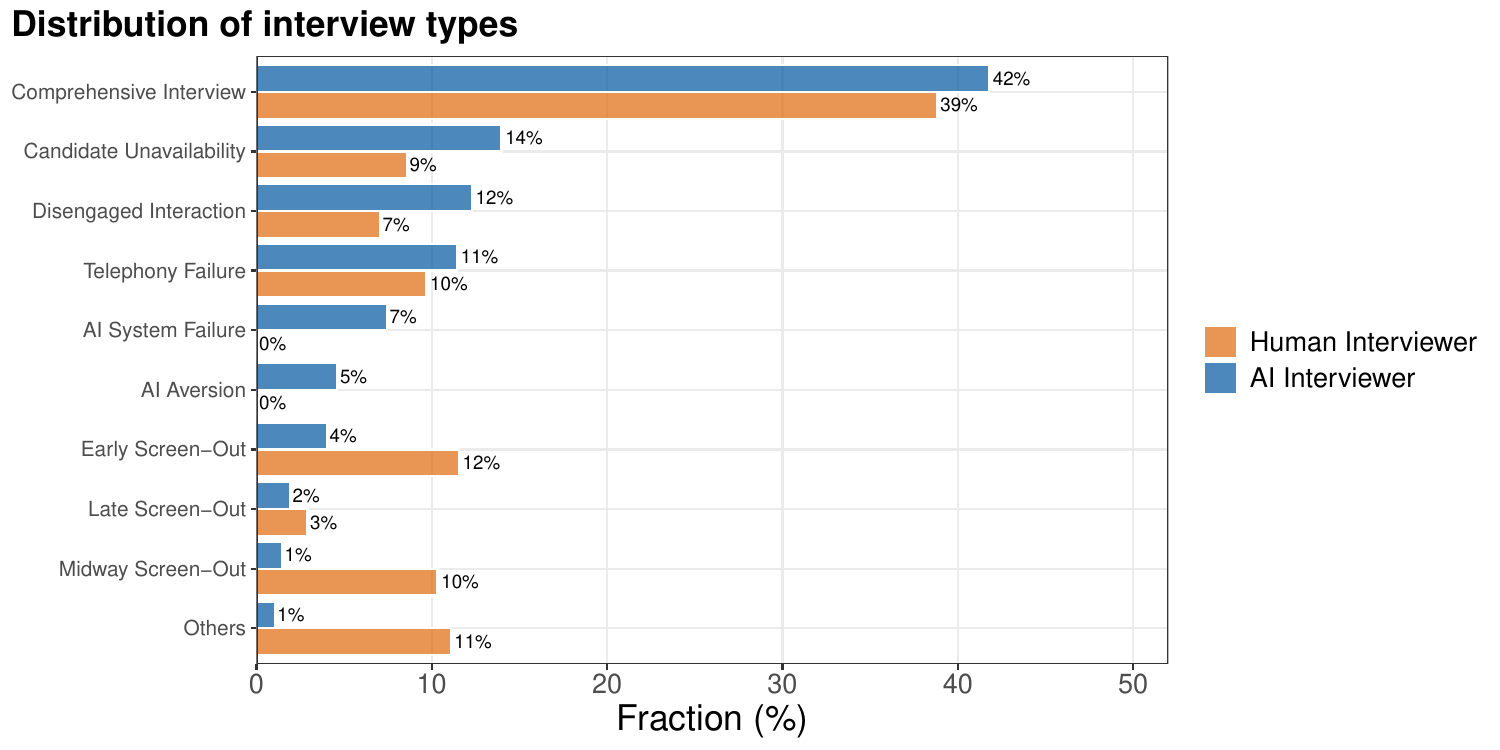}
    \figurenotes[Notes:]{The figure displays the fraction of interview transcripts that fall in each of a number of interview types. Appendix Table \ref{tab:interview_categories} provides the details about each interview type's definition. 
    }
\end{figure}

\begin{figure}[!t]
    \caption{Distribution of interview scores across treatments}
    \label{fig:main_sec_interview_score_distribution}
    \centering
    \includegraphics[width=1\textwidth]{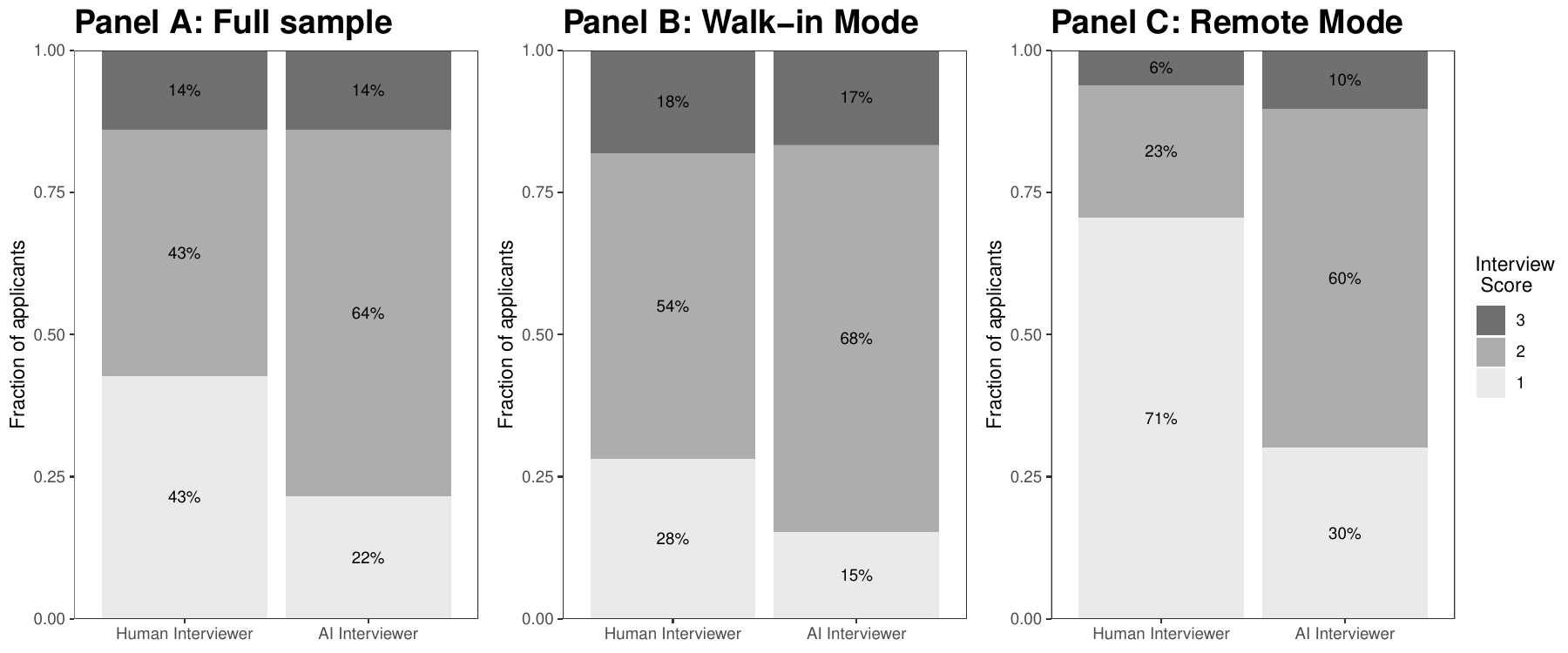}
    \figurenotes[Notes:]{The figure displays the distribution of the interview score with which recruiters assess each interview. The score is 1-poor, 2-medium, 3-good. For details on the scoring, see Appendix Table \ref{tab:score_grid}. Panel A displays the full sample results. Panel B displays the \textit{Walk-in mode} subsample, in which applicants approached the firm at one of the firm's recruitment centers. Panel C displays the \textit{Remote mode} subsample, in which applicants approached the firm online.  
    }
\end{figure}

\begin{figure}[!t]
    \caption{Average offer rate of recruiters across treatments}
    \label{fig:scatter_plot_recruiter_percentage_both_walkin_remote}
    \centering
    \includegraphics[width=1\textwidth]{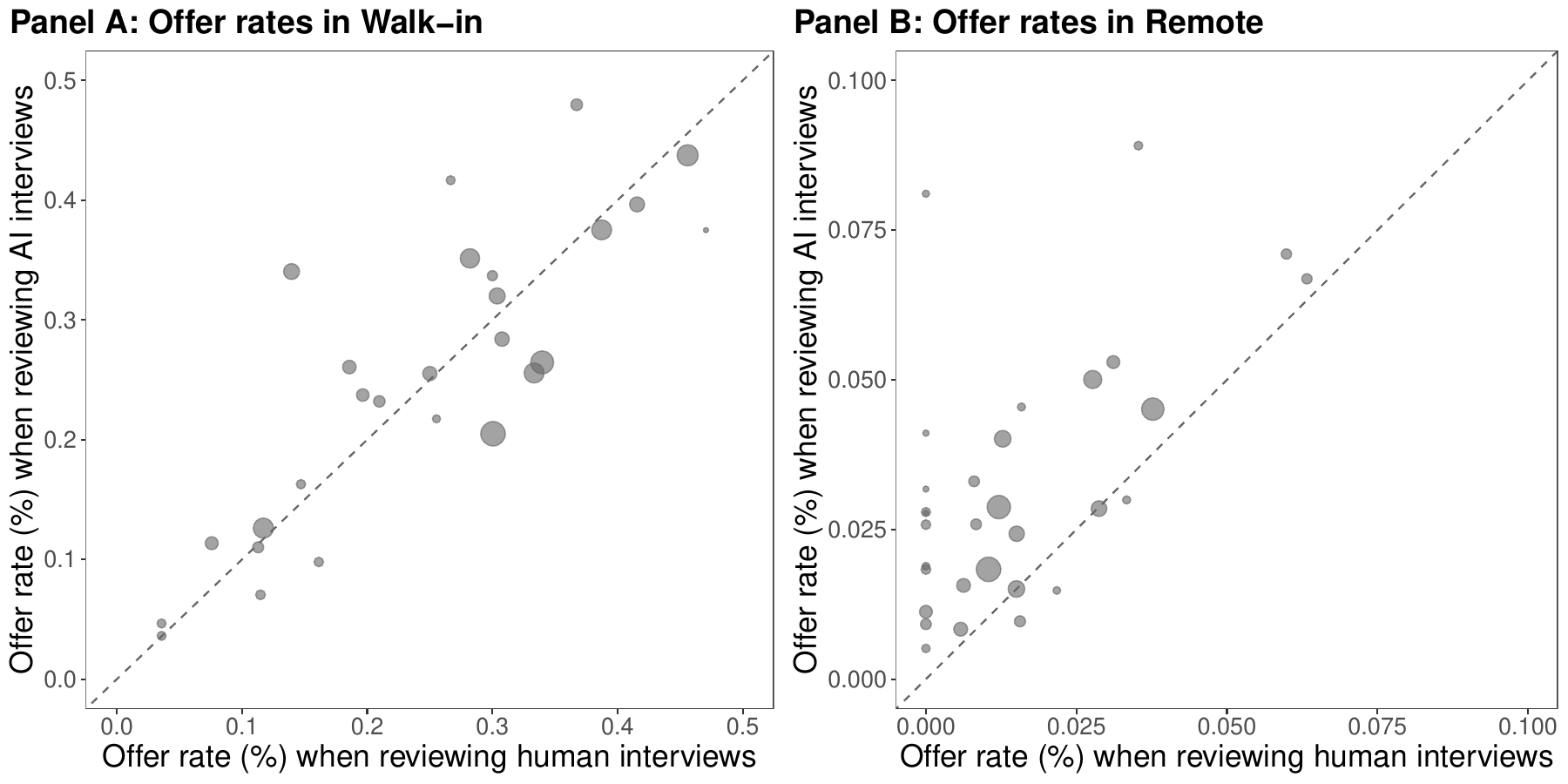}
    \figurenotes[Notes:]{The unit of observation is a recruiter. Dot size indicates the total number of interviews assigned to each recruiter for review. 
    }
\end{figure}

\begin{figure}[!t]
    \caption{Choices of interviewer in the \textit{Choice of interviewer} condition over time}
    \label{fig:choice_of_recruiter_over_time}
    \centering
    \includegraphics[width=1\textwidth]{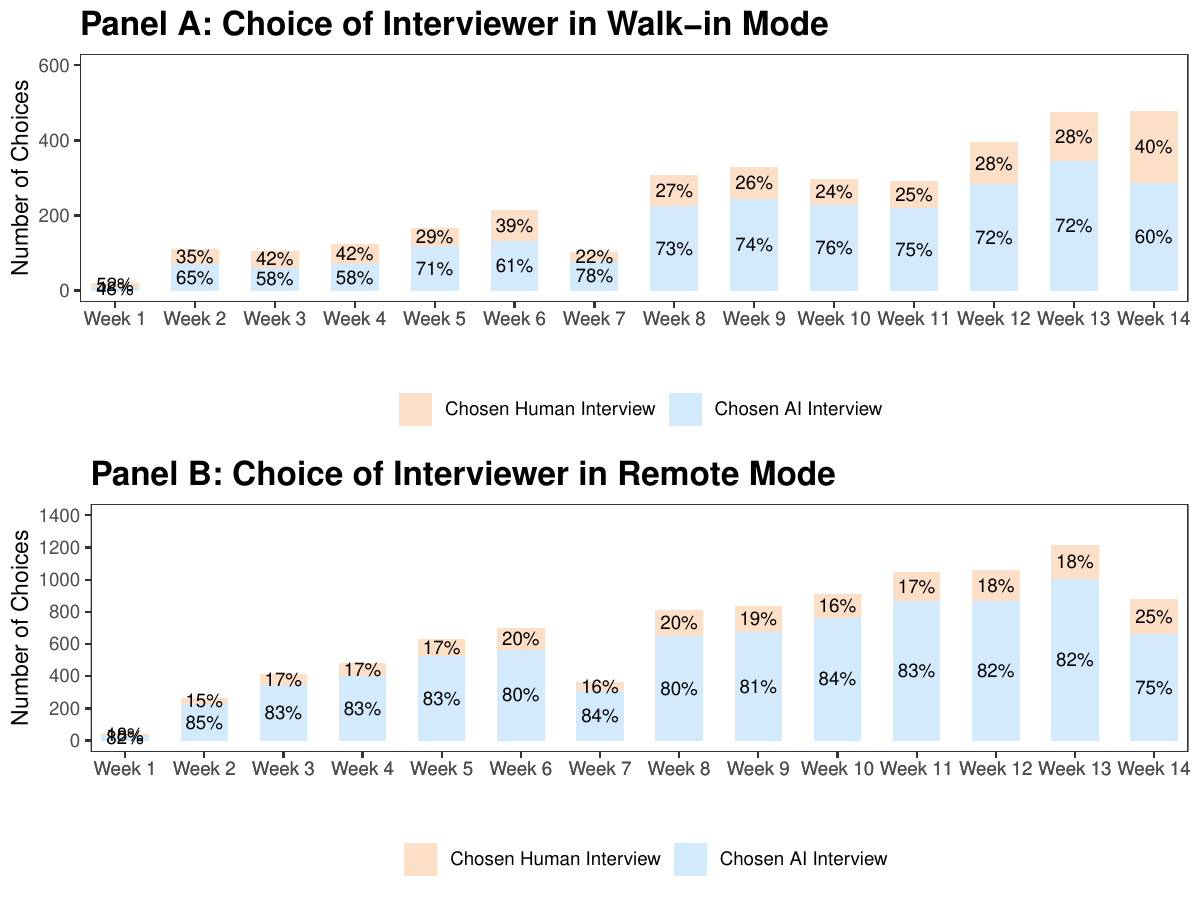}
    \figurenotes[Notes:]{The figure displays applicants' interviewer choice in the \textit{Choice of interviewer} condition over the course of the experiment. In the condition, applicants were given the choice between a human interviewer and the AI voice agent after being invited to the job interview. The experiment ran from March 7 to June 7, 2025. In the \textit{Walk-in mode} (Panel A), applicants approached the firm at one of the firm's recruitment centers. In the \textit{Remote mode} (Panel B), they approached the firm online. 
    }
\end{figure}

\begin{figure}[!t]
    \caption{Distribution of standardized test scores depending on applicants' choice of interviewer}
    \label{fig:AM_test_scores_condition_on_interviewer_choice}
    \centering
    \includegraphics[width=1\textwidth]{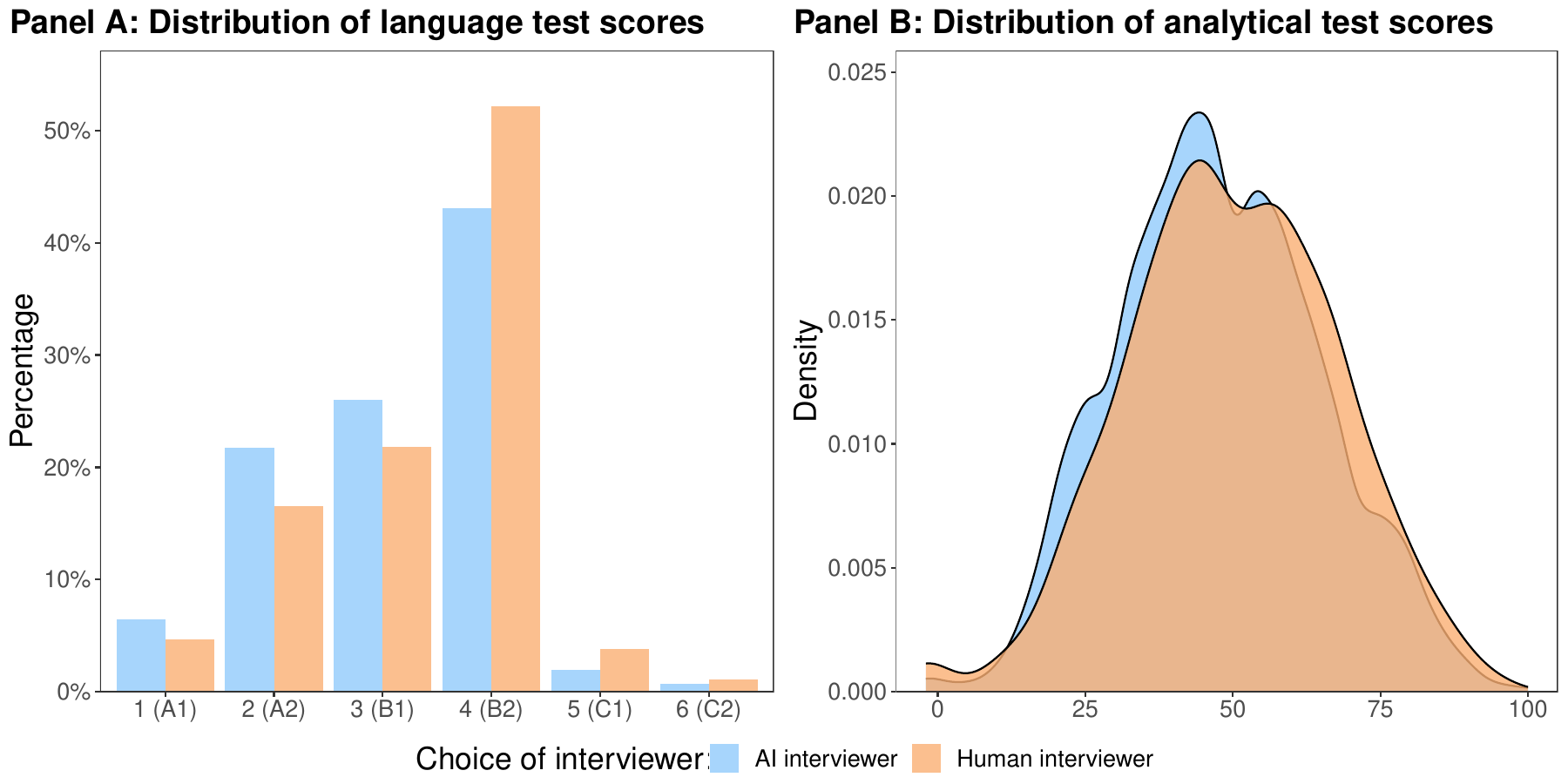}
    \figurenotes[Notes:]{
    The figure displays the distribution of test score results of applicants, split by their interviewer choice in the \textit{Choice of interviewer} condition. Panel A displays the results from the language test, which assesses applicants' writing and reading capabilities in English. Scores are based on the CEFR framework (A1 to C2). Panel B displays the results from the analytical test, which assesses in three parts applicants' attention to detail, verbal reasoning, and numerical ability. Scores are aggregated from each of the three parts and range from 0 to 100.  
    }
\end{figure}

\clearpage

%% file: Sections_appendix/appendix_additional_tables.tex
\section{Additional tables}
\label{app_sec:add_tables}

\begin{table}[htbp]
\centering
\caption{Topics covered in the interviews}
\label{tab:storyboard_full}
\renewcommand{\arraystretch}{1.2} %
\resizebox{\textwidth}{!}{%
\begin{tabular}{l p{2cm} p{7cm} p{3cm}}
\toprule
Topic & Applicability & Example of question & Key signal(s) \\
\midrule
 Introduction & All           & ``Hi [name], thanks for applying to [Employer]. Is now still a good time to chat about your experience?'' & Professionalism; applicant readiness \\[4pt]
Source verification            & All           & ``Out of curiosity, where did you see this opening advertised?'' & Channel efficacy; genuine interest \\[4pt]

 Location and commute      & On-site roles & ``Where are you based and roughly how long would the drive to our [city] site take?''& Commute feasibility; punctuality risk \\[4pt]

Motivation \& attrition risks & All           & ``What attracted you to this opportunity and how does it align with your longer-term goals?''      & Engagement; values alignment; Availability \\[4pt]
 Education verification          & All           & ``Let’s talk education—what’s the highest level you finished, and do you foresee returning to school?''    & Job readiness; up-skilling intent \\[4pt]

Compensation expectations       & All           & ``The role pays between [range]. Where do your salary expectations sit?''& Pay realism; negotiation stance \\[4pt]

 Employment history             & $\ge$1 prior job & ``Walk me through your recent call-center roles—volumes handled, key results, and why you moved on.''  & Experience depth; performance flags \\[4pt]

Re-hire eligibility check       & Former employees & ``Have you worked for [Employer] before? If so, where and who was your supervisor?''    & Prior standing; boomerang potential \\[4pt]

Availability  & All           & ``If selected, when could you start? Are you deep in any other interview processes?''    & Speed-to-hire; offer risk \\[4pt]

 Data verification  & All           & ``For our records, could you confirm the best phone, Viber, and an emergency contact?''& Contact accuracy; compliance \\[4pt]

Needs assessment   & Remote        & ``Do you have reliable internet and a laptop/PC at home for assessments?'' & Tech readiness \\[4pt]

Profiling     & All           & ``Have you ever worked for [firm] before?'' & Experience depth \\[4pt]

Further procedure    & All        &  [Explanation of further procedure]  & Tech readiness \\[4pt]

Wrap-up \& referrals            & All           & ``That’s everything from my side—any questions for me? And do you know anyone else who might thrive here?''  & Applicant curiosity; referral leads \\
\bottomrule
\end{tabular}
}
\end{table}

\begin{table}[H]
\centering
\caption{Interview performance scoring system}
\label{tab:score_grid}
\begin{tabularx}{\textwidth}{@{}c l X X X@{}}
\toprule
Score & Label & Assessment & Applicant engagement & Outcome predictions \\
\midrule
3  & Good & Clear, concise communication; strong problem-solving and critical thinking; solid grasp of role and firm & Shows keen interest; asks relevant questions & High probability of accepting offer and high job show rates and performance \\
2 & Medium & Adequate communication; basic problem-solving; satisfactory but improvable grasp of role and firm & Moderate interest; engages intermittently & Uncertain acceptance; average job show rates and performance \\
1 & Poor & Incoherent or disorganised answers; weak problem-solving; little grasp of role or firm & Low interest; few or irrelevant questions & Low acceptance likelihood; high likelihood of low job show rates and performance \\
\bottomrule
\end{tabularx}
\end{table}

\begin{table}[!ht]
    \caption{Client firm industries}
    \label{tab_industries_clients}
        \begin{threeparttable}[c]
        \begin{tabular}{lc}
        \\[-1.8ex]\hline 
        \hline \\[-1.8ex] 
         Client firm sector & Percentage (\%)  \\ 
         & ($n=41)$ \\
          \hline \\[-1.8ex]
         Technology & 22.00 \\ 
          Finance & 14.60 \\ 
          Transportation & 14.60 \\ 
          Tourism & 9.80 \\ 
          Healthcare & 7.30 \\ 
          Insurance & 7.30 \\ 
          Telecommunication & 7.30 \\ 
          Hospitality & 4.90 \\ 
          Manufacturing & 4.90 \\ 
          Retail & 4.90 \\ 
          Energy & 2.40 \\ 
        \hline
        \hline
        \end{tabular}
            \begin{tablenotes}
                \item \footnotesize \textit{Notes:} Distribution of client firm industries for which the company is hiring customer service representatives.
            \end{tablenotes}
        \end{threeparttable}
\end{table}

\clearpage

\begin{table}[!ht]
    \caption{Treatment balance tests}
    \label{tab_balance_check_main_treatment}
    \resizebox*{1\textwidth}{!}{
        \begin{threeparttable}[c]
            \input{Tables/tab_balance_check_main_treatment}
            \begin{tablenotes}
                \item \footnotesize \textit{Notes:} 
                Columns (1) - (3) display mean values of variables for the three treatments. ``Source is referral'' and ``Source is digital ad'' are binary variables equal to one if the applicant applied through a referral or a digital job advertisement, respectively. Columns (4) - (6) display p-values obtained using pairwise t-tests (variable ``Initial engagement score'') or proportion tests (all other variables).
            \end{tablenotes}
        \end{threeparttable}
       }
\end{table}

\begin{table}[H]
    \caption{Treatment effect on key recruiting outcomes among all applicants}
    \label{tab_main_sec_job_offers_treatment_main_outcomes_uncond}
    \resizebox*{1\textwidth}{!}{
        \begin{threeparttable}[c]
            \input{Tables/tab_main_sec_job_offers_treatment_main_outcomes_uncond}
            \begin{tablenotes}
                \item \footnotesize \textit{Notes:} 
                The table shows OLS estimates analyzing the treatment effect of receiving an AI voice agent instead of a human recruiter in an interview on several recruitment outcome variables. %
                Controls include an applicant's gender, source of application, pre-treatment engagement score, and whether they have applied before to any of the firm's job postings. Fixed effects include week, recruiter, application site, and job posting fixed effects. An observation is an application. Standard errors in parentheses are either clustered at the applicant level (``App.`` in row ``Clustering'') or recruiter level (``Rec''). Significance levels: $^{*}\,p<0.1$,$^{**}\,p<0.05$, $^{***}\,p<0.01$.
            \end{tablenotes}
        \end{threeparttable}
       }
\end{table}

\begin{table}[H]
    \caption{Treatment effect on key recruiting outcomes among applicants who accepted an offer}
    \label{tab_main_sec_job_offers_treatment_main_outcomes_cond}
    \resizebox*{1\textwidth}{!}{
        \begin{threeparttable}[c]
            \input{Tables/tab_main_sec_job_offers_treatment_main_outcomes_cond}
            \begin{tablenotes}
                \item \footnotesize \textit{Notes:} 
                The table shows OLS estimates analyzing the treatment effect of receiving an AI voice agent instead of a human recruiter in an interview on several recruitment outcome variables. The sample is restricted to those applicants who accepted a job offer. Controls include an applicant's gender, source of application, pre-treatment engagement score, and whether they have applied before to any of the firm's job postings. Fixed effects include week, recruiter, application site, and job posting fixed effects. An observation is an application. Standard errors in parentheses are either clustered at the applicant level (``App.`` in row ``Clustering'') or recruiter level (``Rec''). Significance levels: $^{*}\,p<0.1$,$^{**}\,p<0.05$, $^{***}\,p<0.01$.
            \end{tablenotes}
        \end{threeparttable}
       }
\end{table}

\begin{table}[!t]
    \caption{Predicting job offer decisions with interview variables}
    \label{tab_interview_vars_predicting_outcomes}
    \resizebox*{1\textwidth}{!}{
        \begin{threeparttable}[c]
            \input{Tables/tab_interview_vars_predicting_outcomes}

            \begin{tablenotes}
                \item \footnotesize \textit{Notes:} 
               The table shows OLS estimates analyzing the predictive power of interview variables on job offer decisions in the \textit{Human Interviewer} condition. The dependent variable is an indicator equal to one if an application leads to a job offer, and zero otherwise. ``Interview score by recruiter'' is the 1,2,3 score that recruiters assign to applicants' interview performance, with higher values indicating higher performance. ``Sentiment of interview text assessment by recruiter'' is a 1, 0, -1 coded variable indicating whether the sentiment of the text with which recruiters describe applicants' interview performance is positive, neutral or negative, respectively. In columns (2) and (4), we additionally include control variables, fixed effects, and applicants' test scores in the standardized language and analytical test. Controls include an applicant's gender, source of application, pre-treatment engagement score, and whether they have applied before to any of the firm's job postings. Fixed effects include week, recruiter, application site, and job posting fixed effects. An observation is an application. Standard errors in parentheses are clustered at the applicant level. Significance levels: $^{*}\,p<0.1$,$^{**}\,p<0.05$, $^{***}\,p<0.01$.
            \end{tablenotes}
        \end{threeparttable}
      }
\end{table}

\begin{table}[!ht]
    \caption{Treatment differences in the distance between average and actual recruiter question}
    \label{tab_transcript_recruiter_analysis_topic_question_similarity}
    \resizebox*{1\textwidth}{!}{
        \begin{threeparttable}[c]
            \input{Tables/tab_transcript_recruiter_analysis_topic_question_similarity}
            \begin{tablenotes}
                \item \footnotesize \textit{Notes:} P-values are obtained from pairwise t-tests and adjusted for multiple hypothesis testing using the Bonferroni correction. %
            \end{tablenotes}
        \end{threeparttable}
    }
\end{table}

\begin{table}[!t]
    \caption{Information elicited during interviews and applicant outcomes}
    \label{tab_information_elements_predicting_outcomes}
    \resizebox*{1\textwidth}{!}{ 
        \begin{threeparttable}[c]
            \input{Tables/tab_information_elements_predicting_outcomes}
            \begin{tablenotes}
                \item \footnotesize \textit{Notes:} The table shows OLS estimates predicting applicant outcomes with interview-process measures. Panel A includes the interview-average specificity deficit and substantive response deficit across applicant responses. Panel B includes the number of unique information elements elicited during the interviews. ``Received job offer'' is an indicator equal to one if an applicant received a job offer. ``30-day retention'' is an indicator equal to one if the applicant met the 30-day retention threshold. All columns include an applicant's gender, source of application, pre-treatment engagement score, and whether the applicant previously applied to any firm job posting, as well as week, recruiter, application site, and job posting fixed effects. Treatment is not included as a regressor. An observation is an application. Standard errors in parentheses are clustered at the applicant level in Columns (1) and (3), and at the recruiter level in Columns (2) and (4), in each panel. Significance levels: $^{*}\,p<0.1$, $^{**}\,p<0.05$, $^{***}\,p<0.01$.
            \end{tablenotes}
        \end{threeparttable}
    }
\end{table}

\begin{table}[!ht]
    \caption{Predicting job offer decisions of recruiters in the full sample}
    \label{tab_weighting_signals_recruiters_full_sample}
        \begin{threeparttable}[c]
            \input{Tables/tab_weighting_signals_recruiters_full_sample}
            \begin{tablenotes}
                \item \footnotesize \textit{Notes:} 
                The table shows OLS estimates predicting job offer decisions of recruiters using standardized test scores and interview scores. The dependent variable is a dummy equal to one if an application led to a job offer. Test scores are standardized. ``\textit{AI Interviewer}'' is a dummy equal to one if the application was in the \textit{AI Interviewer} condition, and zero if the application was in the \textit{Human Interviewer} condition. Controls include an applicant's gender, source of application, pre-treatment engagement score, and whether they have applied before to any of the firm's job postings. Fixed effects include week, recruiter, application site, and job posting fixed effects. An observation is an application. Standard errors in parentheses are clustered at the applicant level. Significance levels: $^{*}\,p<0.1$,$^{**}\,p<0.05$, $^{***}\,p<0.01$.
            \end{tablenotes}
        \end{threeparttable}
\end{table}

\begin{table}[!t]
    \caption{Heterogeneity in predicting job offer decisions of recruiters}
    \label{tab_weighting_signals_recruiters_heterogeneity}
        \begin{threeparttable}[c]
            \input{Tables/tab_weighting_signals_recruiters_heterogeneity}
            \begin{tablenotes}
                \item \footnotesize \textit{Notes:} 
                The table shows OLS estimates predicting job offer decisions of recruiters using standardized test scores and interview scores. The dependent variable is an indicator variable equal to one if an application led to a job offer. Test scores are standardized. ``\textit{AI Interviewer}'' is an indicator variable equal to one if the application was in the \textit{AI Interviewer} condition, and zero if the application was in the \textit{Human Interviewer} condition. We split the sample by recruiters who state in the recruiter survey that interview scores are equally or less important than the test scores for offer decisions (columns (1)-(2)) and recruiters who state that interview scores are more important (columns (3)-(4)). Controls include an applicant's gender, source of application, pre-treatment engagement score, and whether they have applied before to any of the firm's job postings. Fixed effects include week, recruiter, application site, and job posting fixed effects. An observation is an application. Standard errors in parentheses are clustered at the applicant level. Significance levels: $^{*}\,p<0.1$,$^{**}\,p<0.05$, $^{***}\,p<0.01$.
            \end{tablenotes}
        \end{threeparttable}
\end{table}
\begin{table}[!ht]
\centering
\caption{Categories of open-ended survey responses}
\label{tab_classification_survey_responses}
\renewcommand{\arraystretch}{1.2}
\resizebox{\textwidth}{!}{%
\begin{tabular}{p{4cm} p{6cm} p{8cm}}
\\[-1.8ex]\hline 
\hline \\[-1.8ex] 
\textbf{Category} & \textbf{Definition} & \textbf{Example response} \\
\midrule
Comfortable and positive experience & Applicant mentions that they had a comfortable and positive experience. & ``Yes, thank you, I just want to say I really appreciated how welcoming and professional the interview felt. The questions were thoughtful, and it gave me a clear picture of the role and firm culture. It made me even more interested in being part of the team. Thanks again for the opportunity!'' \\[8pt]
Short uninformative response & Applicant gave a short answer, expressing that they have nothing to add. & ``Nothing else.'' \\[8pt]
Problems with audio or questions &  Applicant mentions a problematic interview experience due to misunderstanding of answers, frequent interruptions, or imperfect audio quality. & ``The line is getting cut a lot and I haven't answered the question yet but the interviewer is already asking another question.'' \\[8pt]
Clear audio and instructions & Applicant mentions that instructions and speaker audio were clear. & ``I love how AI speaks clearly and how he/she ask about specific questions that need to be clear.'' \\[8pt]
Excitement to join firm & Applicant mentions their interest and excitement in joining the firm. & ``It's great and i hope I can join the team''\\[8pt]
Problems with AI voice agent  & Applicant mentions problems with the AI voice agent, such as a lack of understanding of applicant input, clarity or emotional cues. & ``AI is a useful tool for interviews and makes the process easier. However, it's different from talking to a real person. Sometimes, when I reply, the AI doesn’t fully understand me, which can lead to misunderstandings. That makes communication a bit challenging at times.''\\[8pt]
Feeling nervous & Applicant mentions that they were nervous for their first job interview. & ``It's good, I didn't really give my best answers to the interview since I was nervous because its my first time interview but for first timer it was good.'' \\[8pt]
Feeling unfairly judged & Applicant feels or expresses concern about being unfairly judged by the interviewer & ``Aside of not having a bpo experiences, there were questions that I felt and think may impact my application. I just hope my application be considered nor not discriminate of how I speak in English''\\[8pt]
Prefer human interviewer & Applicant prefers to be interviewed by a human interviewer. & ``Sometimes, you would need to have human interaction on these types of interviews.''\\[4pt]
Discussed follow-up & Applicant discusses follow-up procedures. & ``From the day i passed my final interview until now i didnt received any notification from you.'' \\[8pt]
Other & Applicant responses that could not be classified into the other categories.  & ``What can you advice of the people got hired in BPO''  \\
\hline 
\hline \\[-1.8ex] 
\end{tabular}%
}
\end{table}

\begin{table}[!ht]
    \caption{Timing of survey invitation balance tests}
    \label{tab_balance_check_timing_survey}
        \begin{threeparttable}[c]
            \input{Tables/tab_balance_check_timing_survey}
            \begin{tablenotes}
                \item \footnotesize \textit{Notes:} 
                Columns (1) and (2) display mean values of the survey variables depending on the time in the recruitment process when the survey was sent to applicants. Column (3) displays p-values obtained using a t-test.
            \end{tablenotes}
        \end{threeparttable}
\end{table}

\begin{table}[!ht]
    \caption{Predicting applicants' interviewer choices}
    \label{tab_applicant_level_predicting_recruiter_choices}
        \begin{threeparttable}[c]
            \input{Tables/tab_applicant_level_predicting_recruiter_choices}
            \begin{tablenotes}
                \item \footnotesize \textit{Notes:} The table shows OLS estimates predicting applicants' interviewer choices in the \textit{Choice of interviewer} treatment using their survey responses. The outcome variable is an indicator variable equal to one if an applicant chose the AI interviewer and zero otherwise. Higher values of ``Perceived impact of AI on applicant'' indicate a more positive impact of AI on applicants themselves. %
                Controls include an applicant's gender, source of application, and pre-treatment engagement score. Fixed effects include week, recruiter, application site, and job posting fixed effects. An observation is an applicant. Standard errors in parentheses are clustered at the applicant level. Significance levels: $^{*}\,p<0.1$,$^{**}\,p<0.05$, $^{***}\,p<0.01$.
            \end{tablenotes}
        \end{threeparttable}
\end{table}

\begin{table}[!ht]
    \caption{Differences in test scores depending on applicants' choice of interviewer}
    \label{tab_applicant_level_selection_test_score_only_choice}
        \begin{threeparttable}[c]
            \input{Tables/tab_applicant_level_selection_test_score_only_choice}
            \begin{tablenotes}
                \item \footnotesize \textit{Notes:} 
                The table shows OLS estimates predicting applicants' test scores using their interviewer choice. ``Has chosen AI'' is an indicator variable equal to one if an applicant in the \textit{Choice of Interviewer} condition chose the AI voice agent and zero if they chose the human interviewer. Controls include an applicant's gender, source of application, pre-treatment engagement score, and whether they have applied before to any of the firm's job postings. Fixed effects include week, recruiter, application site, and job posting fixed effects. An observation is an application. Standard errors in parentheses are clustered at the applicant level. Significance levels: $^{*}\,p<0.1$,$^{**}\,p<0.05$, $^{***}\,p<0.01$.
            \end{tablenotes}
        \end{threeparttable}
\end{table}

\begin{table}[!ht]
    \caption{Differences in test scores when applicants choose instead of being assigned either the human or AI interviewer}
    \label{tab_applicant_level_selection_test_score}
        \begin{threeparttable}[c]
            \input{Tables/tab_applicant_level_selection_test_score}
            \begin{tablenotes}
                \item \footnotesize \textit{Notes:} 
                The table shows OLS estimates predicting applicants' test scores using their interviewer choice. In Panel A, ``Has chosen AI'' is an indicator variable equal to one if an application was in the \textit{Choice of Interviewer} condition and the applicant chose the AI voice agent and zero if the application was in the \textit{AI interviewer} condition instead, where applicants were interviewed by the AI voice agent without a choice.  In Panel B, ``Has chosen human'' is an indicator variable equal to one if an application was in the \textit{Choice of Interviewer} condition and the applicant chose the human interviewer and zero if the application was in the \textit{Human interviewer} condition instead, where applicants were interviewed by a human recruiter without a choice. Controls include an applicant's gender, source of application, pre-treatment engagement score, and whether they have applied before to any of the firm's job postings. Fixed effects include week, recruiter, application site, and job posting fixed effects. An observation is an application. Standard errors in parentheses are clustered at the applicant level. Significance levels: $^{*}\,p<0.1$,$^{**}\,p<0.05$, $^{***}\,p<0.01$.
            \end{tablenotes}
        \end{threeparttable}
\end{table}

\begin{table}[!ht]
    \caption{Treatment differences on transcript data }
    \label{tab_transcript_analyses_treatment_on_category}
        \begin{threeparttable}[c]
            \input{Tables/tab_transcript_analyses_treatment_on_category}
            \begin{tablenotes}
                \item \footnotesize \textit{Notes:} The table shows OLS estimates analyzing treatment differences between receiving an AI voice agent instead of a human recruiter in an interview on the type of interview. The dependent variable is an indicator variable equal to one if the interview is classified as \textit{Comprehensive interview} and zero otherwise. \textit{Comprehensive interview} means it opens and closes organically and covers at least eight canonical topics. Controls include an applicant’s gender, source of application, pre-treatment engagement score, and whether they have applied before to any of the firm’s job postings. Fixed effects include week, recruiter, application site, and job posting fixed effects. An observation is an application. Standard errors in parentheses are clustered at the applicant level. Significance levels: $^{*}\,p<0.1$,$^{**}\,p<0.05$, $^{***}\,p<0.01$.
            \end{tablenotes}
        \end{threeparttable}
\end{table}

\clearpage

%% file: Tables/tab_balance_check_main_treatment.tex
\begin{tabular}{lcccccc}
\\[-1.8ex]\hline 
\hline \\[-1.8ex] 
Variable & \textit{Human} & \textit{AI} & \textit{Choice of} & $H_0$: (1) = (2) & $H_0$: (1) = (3)  & $H_0$: (2) = (3) \\ 
& \textit{Interviewer} & \textit{Interviewer} & \textit{Interviewer} & p-value & p-value & p-value \\ 
& (1) & (2) & (3) & (4)  & (5) & (6)  \\
\textbf{Panel A: Full sample} \\ \\[-1.8ex]
\hline \\[-1.8ex]
Gender (Women=1) & 0.60 & 0.60 & 0.61 & 0.21 & 0.50 & 0.04 \\ 
  Source is referral & 0.19 & 0.19 & 0.19 & 0.96 & 0.38 & 0.26 \\ 
  Source is digital ad & 0.59 & 0.59 & 0.59 & 0.47 & 0.85 & 0.34 \\ 
  Mode is Walk-in & 0.26 & 0.26 & 0.26 & 0.45 & 0.66 & 0.85 \\ 
  Initial engagement score & 37.06 & 37.12 & 37.21 & 0.80 & 0.62 & 0.73 \\     \\[-1.8ex] 
  \hline \\[-1.8ex] 
   Observations & 13,557 & 40,103 & 13,396 &  &  &  \\  \\[-1.8ex]
 \textbf{Panel B: \textit{Walk-in mode}} \\ \\[-1.8ex]
\hline \\[-1.8ex]
Gender (Women=1) & 0.56 & 0.56 & 0.56 & 1.00 & 0.84 & 0.86 \\ 
  Source is referral & 0.12 & 0.11 & 0.11 & 0.54 & 0.62 & 0.36 \\ 
  Source is digital ad & 0.05 & 0.06 & 0.06 & 0.74 & 0.70 & 0.55 \\   \\[-1.8ex] 
  \hline \\[-1.8ex] 
 Observations & 3,478 & 10,421 & 3,469 &  &  &  \\    \\[-1.8ex]
 \textbf{Panel C: \textit{Remote mode}} \\ \\[-1.8ex]
\hline \\[-1.8ex]
Gender (Women=1) & 0.62 & 0.61 & 0.62 & 0.15 & 0.48 & 0.02 \\ 
  Source is referral & 0.22 & 0.22 & 0.22 & 0.66 & 0.63 & 0.30 \\ 
  Source is digital ad & 0.77 & 0.77 & 0.78 & 0.66 & 0.66 & 0.32 \\ 
  Initial engagement score & 50.48 & 50.42 & 50.48 & 0.75 & 0.99 & 0.73 \\    \\[-1.8ex] 
  \hline \\[-1.8ex] 
 Observations & 10,079 & 29,682 & 9,927 &  &  &  \\ 
\hline
\hline
\end{tabular}

%% file: Tables/tab_main_sec_job_offers_treatment_main_outcomes_uncond.tex
\begin{tabular}{@{\extracolsep{5pt}}lccccccccc} 
\\[-1.8ex]\hline 
\hline \\[-1.8ex] 
\multicolumn{5}{@{}l}{\textbf{Panel A:}} \\
& \multicolumn{9}{c}{\textit{Dependent variable:}} \\ 
\cline{2-10} 
\\[-1.8ex] 
& \multicolumn{3}{c}{Received job offer} & \multicolumn{3}{c}{Started job} & \multicolumn{3}{c}{Employed after one month} \\ 
\cline{2-4}  \cline{5-7} \cline{8-10} 
\\[-1.8ex] & (1) & (2) & (3) & (4) & (5) & (6) & (7) & (8) & (9)\\ 
\hline \\[-1.8ex] 
\textit{AI Interviewer}  & 0.0104$^{***}$ & 0.0101$^{***}$ & 0.0101$^{***}$ & 0.0106$^{***}$ & 0.0102$^{***}$ & 0.0102$^{***}$ & 0.0091$^{***}$ & 0.0088$^{***}$ & 0.0088$^{***}$ \\ 
  & (0.0028) & (0.0026) & (0.0037) & (0.0023) & (0.0023) & (0.0032) & (0.0022) & (0.0021) & (0.0031) \\[0.2cm]
\hline \\[-1.8ex] 
Mean DV \textit{Human Interviewer} & 0.0870 & 0.0870 & 0.0870 & 0.0565 & 0.0571 & 0.0571 & 0.0498 & 0.0504 & 0.0504 \\ 
Controls and fixed effects & -- & Yes & Yes & -- & Yes & Yes & -- & Yes & Yes  \\ 
Clustering & App. & App. & Rec. & App. & App. & Rec. & App. & App. & Rec.  \\ 
Observations & 53,660 & 52,367 & 52,367 & 53,660 & 52,367 & 52,367 & 53,660 & 52,367 & 52,367 \\ 
R$^{2}$ & 0.0002 & 0.1683 & 0.1683 & 0.0004 & 0.1098 & 0.1098 & 0.0003 & 0.0984 & 0.0984 \\  
 \hline \\[-1.8ex] 
\multicolumn{5}{@{}l}{\textbf{Panel B:}} \\
& \multicolumn{9}{c}{\textit{Dependent variable:}} \\ 
\cline{2-10} 
\\[-1.8ex] 
& \multicolumn{3}{c}{Employed after two months} & \multicolumn{3}{c}{Employed after three months} & \multicolumn{3}{c}{Employed after four months} \\ 
\cline{2-4}  \cline{5-7} \cline{8-10} 
\\[-1.8ex] & (1) & (2) & (3) & (4) & (5) & (6) & (7) & (8) & (9)\\ 
\hline \\[-1.8ex] 
\textit{AI Interviewer} & 0.0087$^{***}$ & 0.0081$^{***}$ & 0.0081$^{***}$ & 0.0081$^{***}$ & 0.0076$^{***}$ & 0.0076$^{***}$ & 0.0082$^{***}$ & 0.0077$^{***}$ & 0.0077$^{***}$ \\ 
  & (0.0021) & (0.0021) & (0.0030) & (0.0021) & (0.0020) & (0.0028) & (0.0021) & (0.0020) & (0.0026) \\[0.2cm]
\hline \\[-1.8ex] 
Mean DV \textit{Human Interviewer} & 0.0463 & 0.0469 & 0.0469 & 0.0442 & 0.0448 & 0.0448 & 0.0438 & 0.0444 & 0.0444 \\ 
Controls and fixed effects & -- & Yes & Yes & -- & Yes & Yes & -- & Yes & Yes  \\ 
Clustering & App. & App. & Rec. & App. & App. & Rec. & App. & App. & Rec.  \\ 
Observations & 53,660 & 52,367 & 52,367 & 53,660 & 52,367 & 52,367 & 53,660 & 52,367 & 52,367 \\ 
R$^{2}$ & 0.0003 & 0.0893 & 0.0893 & 0.0003 & 0.0843 & 0.0843 & 0.0003 & 0.0836 & 0.0836 \\ 
\hline 
\hline \\[-1.8ex] 
\end{tabular}

%% file: Tables/tab_main_sec_job_offers_treatment_main_outcomes_cond.tex
\begin{tabular}{@{\extracolsep{5pt}}lccccccccc} 
\\[-1.8ex]\hline 
\hline \\[-1.8ex] 
\multicolumn{5}{@{}l}{\textbf{Panel A:}} \\
& \multicolumn{9}{c}{\textit{Dependent variable:}} \\ 
\cline{2-10} 
\\[-1.8ex] 
& & & & \multicolumn{3}{c}{Started job} & \multicolumn{3}{c}{Employed after one month} \\ 
  \cline{5-7} \cline{8-10} 
\\[-1.8ex] &  &  &  & (4) & (5) & (6) & (7) & (8) & (9)\\ 
\hline \\[-1.8ex] 
\textit{AI Interviewer} &  &  & & 0.0452$^{***}$ & 0.0364$^{**}$ & 0.0364$^{*}$ & 0.0372$^{**}$ & 0.0351$^{**}$ & 0.0351$^{*}$ \\
 &  &  &  & (0.0158) & (0.0156) & (0.0219) & (0.0167) & (0.0166) & (0.0194) \\[0.2cm]
\hline \\[-1.8ex] 
Mean DV \textit{Human Interviewer} &  &  & & 0.6884 & 0.6962 & 0.6962 & 0.6060 & 0.6132 & 0.6132 \\ 
Controls and fixed effects &  &  &  & -- & Yes & Yes & -- & Yes & Yes  \\ 
Clustering &  &  &  & App. & App. & Rec. & App. & App. & Rec.  \\ 
Observations &  &  & & 4,708 & 4,575 & 4,575 & 4,708 & 4,575 & 4,575 \\ 
R$^{2}$ &  &  & & 0.0018 & 0.0700 & 0.0700 & 0.0011 & 0.0611 & 0.0611 \\ 
 \hline \\[-1.8ex] 
\multicolumn{5}{@{}l}{\textbf{Panel B:}} \\
& \multicolumn{9}{c}{\textit{Dependent variable:}} \\ 
\cline{2-10} 
\\[-1.8ex] 
& \multicolumn{3}{c}{Employed after two months} & \multicolumn{3}{c}{Employed after three months} & \multicolumn{3}{c}{Employed after four months} \\ 
\cline{2-4}  \cline{5-7} \cline{8-10} 
\\[-1.8ex] & (1) & (2) & (3) & (4) & (5) & (6) & (7) & (8) & (9)\\ 
\hline \\[-1.8ex] 
\textit{AI Interviewer}  & 0.0359$^{**}$ & 0.0326$^{*}$ & 0.0326 & 0.0325$^{*}$ & 0.0287$^{*}$ & 0.0287 & 0.0334$^{*}$ & 0.0297$^{*}$ & 0.0297$^{*}$ \\ 
  & (0.0170) & (0.0170) & (0.0201) & (0.0171) & (0.0171) & (0.0183) & (0.0171) & (0.0172) & (0.0172) \\[0.2cm]
\hline \\[-1.8ex] 
Mean DV \textit{Human Interviewer} & 0.5634 & 0.5708 & 0.5708 & 0.5371 & 0.5443 & 0.5443 & 0.5326 & 0.5396 & 0.5396\\ 
Controls and fixed effects & -- & Yes & Yes & -- & Yes & Yes & -- & Yes & Yes  \\ 
Clustering & App. & App. & Rec. & App. & App. & Rec. & App. & App. & Rec.  \\ 
Observations & 4,708 & 4,575 & 4,575 & 4,708 & 4,575 & 4,575 & 4,708 & 4,575 & 4,575 \\ 
R$^{2}$ & 0.0010 & 0.0525 & 0.0525 & 0.0008 & 0.0518 & 0.0518 & 0.0008 & 0.0522 & 0.0522 \\ 
\hline 
\hline \\[-1.8ex] 
\end{tabular}

%% file: Tables/tab_interview_vars_predicting_outcomes.tex
\begin{tabular}{@{\extracolsep{5pt}}lcccccc} 
\\[-1.8ex]\hline 
\hline \\[-1.8ex] 
&  \multicolumn{6}{c}{\textit{Dependent variable:} Received job offer} \\ 
 &  \multicolumn{6}{c}{(within \textit{Human Interviewer} condition)}\\
 \cline{2-7} \\[-1.8ex] & (1) & (2) & (3) & (4) & (5) & (6)\\ 
\hline \\[-1.8ex] 
Interview score by recruiter & 0.249$^{***}$ & 0.193$^{***}$ & 0.147$^{***}$ &  &  &  \\ 
  & (0.008) & (0.016) & (0.018) &  &  &   \\[0.2cm]
Sentiment of interview justification by recruiter &  &  &  & 0.194$^{***}$ & 0.157$^{***}$ & 0.095$^{***}$ \\ 
  &  &  &  & (0.009) & (0.014) & (0.015)  \\[0.2cm]
\hline \\[-1.8ex] 
Test completed & -- & Yes & Yes & -- & Yes & Yes \\ 
Test scores included & -- & -- & Yes & -- & -- & Yes \\ 
Controls and fixed effects & -- & -- & Yes & -- & -- & Yes \\ 
Observations & 4,661 & 2,477 & 2,477 & 3,430 & 1,880 & 1,880 \\ 
R$^{2}$ & 0.179 & 0.049 & 0.237 & 0.139 & 0.059 & 0.248 \\ 
\hline 
\hline \\[-1.8ex] 
\end{tabular}

%% file: Tables/tab_transcript_recruiter_analysis_topic_question_similarity.tex
\begin{tabular}[t]{lcccc}
\toprule
Topic & \% of interviews & \textit{AI Interviewer} & \textit{Human Interviewer} & $H_0$: (2) = (3)\\ 
&covered&distance score&distance score& p-value \\ 
& (1) & (2) & (3) & (4)\\ 
\hline \\[-1.8ex] 
Introduction & 99.5\% & 0.103 & 0.217 & $p<0.001$\\
Source verification & 2.1\% & 0.486 & 0.364 & $p<0.001$\\
Location and commute & 65.7\% & 0.166 & 0.288 & $p<0.001$\\
\addlinespace
Motivation \& attrition risks – Motivation and goals & 58.4\% & 0.091 & 0.332 & $p<0.001$\\
Motivation \& attrition risks – Shifting schedule & 56.9\% & 0.046 & 0.189 & $p<0.001$\\
Motivation \& attrition risks – Age verification & 55.8\% & 0.041 & 0.330 & $p<0.001$\\
Motivation \& attrition risks – Training & 51.0\% & 0.027 & 0.248 & $p<0.001$\\
Motivation \& attrition risks – Planned vacation & 52.2\% & 0.031 & 0.101 & $p<0.001$\\
Motivation \& attrition risks – Other & 55.4\% & 0.118 & 0.341 & $p<0.001$\\
\addlinespace
Education verification & 53.8\% & 0.089 & 0.187 & $p<0.001$\\
Compensation expectations & 53.2\% & 0.035 & 0.208 & $p<0.001$\\
Employment history & 54.3\% & 0.141 & 0.272 & $p<0.001$\\
Re-hire eligibility check & 64.2\% & 0.041 & 0.270 & $p<0.001$\\
Availability & 52.2\% & 0.099 & 0.276 & $p<0.001$\\
Data verification & 51.7\% & 0.095 & 0.254 & $p<0.001$\\
\addlinespace
Needs assessment & 0.50\% & 0.462 & 0.360 & $p<0.05$\\
Profiling & 8.3\% & 0.283 & 0.445 & $p<0.001$\\
Further procedure & 49.4\% & 0.064 & 0.326 & $p<0.001$\\
Wrap-up \& referrals – Referral & 0.3\% & 0.412 & 0.341 & $p>0.10$\\
Wrap-up \& referrals – Closing & 65.6\% & 0.191 & 0.301 & $p<0.001$\\
\bottomrule
\end{tabular}

%% file: Tables/tab_information_elements_predicting_outcomes.tex
\begin{tabular}{@{\extracolsep{5pt}}lcccc}
\\[-1.8ex]\hline
\hline \\[-1.8ex]
\multicolumn{5}{@{}l}{\textbf{Panel A: Response deficits}} \\
& \multicolumn{4}{c}{\textit{Dependent variable:}} \\
\cline{2-5} \\[-1.8ex]
& \multicolumn{2}{c}{Received job offer} & \multicolumn{2}{c}{30-day retention} \\
\cline{2-3} \cline{4-5}
\\[-1.8ex] & (1) & (2) & (3) & (4) \\
\hline \\[-1.8ex]
Specificity deficit (interview average) & -0.0425$^{***}$ & -0.0425$^{*}$ & -0.0209$^{***}$ & -0.0209 \\
 & (0.0081) & (0.0223) & (0.0066) & (0.0135) \\[0.2cm]
Substantive response deficit (interview average) & -0.1084$^{***}$ & -0.1084$^{***}$ & -0.0681$^{***}$ & -0.0681$^{***}$ \\
 & (0.0079) & (0.0165) & (0.0064) & (0.0112) \\[0.2cm]
\hline \\[-1.8ex]
Mean dependent variable & 0.1476 & 0.1476 & 0.0902 & 0.0902 \\
Pre-treatment controls & Yes & Yes & Yes & Yes \\
Fixed effects & Yes & Yes & Yes & Yes \\
Clustering & App. & Rec. & App. & Rec. \\
Observations & 25,136 & 25,136 & 25,136 & 25,136 \\
R$^{2}$ & 0.1892 & 0.1892 & 0.1214 & 0.1214 \\
\hline \\[-1.8ex]
\multicolumn{5}{@{}l}{\textbf{Panel B: Information collection}} \\
& \multicolumn{4}{c}{\textit{Dependent variable:}} \\
\cline{2-5} \\[-1.8ex]
& \multicolumn{2}{c}{Received job offer} & \multicolumn{2}{c}{30-day retention} \\
\cline{2-3} \cline{4-5}
\\[-1.8ex] & (1) & (2) & (3) & (4) \\
\hline \\[-1.8ex]
Unique information elements elicited & 0.0064$^{***}$ & 0.0064$^{***}$ & 0.0040$^{***}$ & 0.0040$^{***}$ \\
 & (0.0002) & (0.0006) & (0.0002) & (0.0005) \\[0.2cm]
\hline \\[-1.8ex]
Mean dependent variable & 0.1571 & 0.1571 & 0.0961 & 0.0961 \\
Pre-treatment controls & Yes & Yes & Yes & Yes \\
Fixed effects & Yes & Yes & Yes & Yes \\
Clustering & App. & Rec. & App. & Rec. \\
Observations & 23,396 & 23,396 & 23,396 & 23,396 \\
R$^{2}$ & 0.1975 & 0.1975 & 0.1259 & 0.1259 \\
\hline
\hline \\[-1.8ex]
\end{tabular}

%% file: Tables/tab_weighting_signals_recruiters_full_sample.tex
\begin{tabular}{@{\extracolsep{5pt}}lcc} 
\\[-1.8ex]\hline 
\hline \\[-1.8ex] 
 & \multicolumn{2}{c}{\textit{Dependent variable:} } \\ 
  & \multicolumn{2}{c}{\textit{Job Offer Made}} \\ 
\cline{2-3} 
\\[-1.8ex] & (1) & (2)\\ 
\hline \\[-1.8ex] 
 Interview score (std.) & 0.077$^{***}$ & 0.088$^{***}$ \\ 
  & (0.010) & (0.010) \\ 
  & & \\ 
 Language test score (std.)& 0.109$^{***}$ & 0.111$^{***}$ \\ 
  & (0.010) & (0.009) \\ 
  & & \\ 
 Analytical test score (std.)& 0.033$^{***}$ & 0.029$^{***}$ \\ 
  & (0.010) & (0.010) \\ 
  & & \\ 
 Interview score (std.) $\times$ \textit{AI Interviewer} & $-$0.038$^{***}$ & $-$0.026$^{**}$ \\ 
  & (0.011) & (0.011) \\ 
  & & \\ 
Language test score (std.) $\times$ \textit{AI Interviewer}& 0.015 & 0.015 \\ 
  & (0.011) & (0.010) \\ 
  & & \\ 
 Analytical test score (std.) $\times$ \textit{AI Interviewer} & $-$0.003 & 0.001 \\ 
  & (0.011) & (0.011) \\ 
  & & \\ 
\hline \\[-1.8ex] 
Mean DV in \textit{Human Interviewer}  & 0.38 & 0.37 \\ 
Controls and fixed effects & -- & Yes \\ 
Observations & 12,934 & 12,732 \\ 
R$^{2}$ & 0.102 & 0.211 \\ 
\hline 
\hline \\[-1.8ex] 
\end{tabular} 

%% file: Tables/tab_weighting_signals_recruiters_heterogeneity.tex
\begin{tabular}{@{\extracolsep{5pt}}lcccc} 
\\[-1.8ex]\hline 
\hline \\[-1.8ex] 
 & \multicolumn{4}{c}{\textit{Dependent variable:} Job Offer Made} \\[0.2cm]  
  & \multicolumn{2}{c}{Recruiters who consider}  & \multicolumn{2}{c}{Recruiters who consider} \\
  & \multicolumn{2}{c}{interview $\leq$ test (survey)}  & \multicolumn{2}{c}{interview $>$ test (survey)} \\  
\cline{2-3} \cline{4-5} 
\\[-1.8ex] & (1) & (2) & (3) & (4)\\ 
\hline \\[-1.8ex] 
 Interview score (std.)  & 0.087$^{***}$ & 0.117$^{***}$ & 0.142$^{***}$ & 0.120$^{***}$ \\ 
  & (0.016) & (0.015) & (0.026) & (0.028) \\ 
  & & \\ 
 Language test score (std.) & 0.067$^{***}$ & 0.077$^{***}$ & 0.107$^{***}$ & 0.114$^{***}$ \\ 
  & (0.015) & (0.015) & (0.029) & (0.030) \\ 
  & & \\ 
 Analytical test score (std.)  & 0.027$^{*}$ & 0.012 & 0.026 & 0.016 \\ 
  & (0.016) & (0.015) & (0.029) & (0.029) \\ 
  & & \\   
 Interview score (std.) $\times$ \textit{AI Interviewer} & $-$0.021 & $-$0.024 & $-$0.086$^{***}$ & $-$0.059$^{*}$ \\ 
  & (0.018) & (0.017) & (0.029) & (0.031) \\ 
  & & \\ 
Language test score (std.) $\times$ \textit{AI Interviewer} & 0.042$^{**}$ & 0.036$^{**}$ & 0.045 & 0.038 \\ 
  & (0.017) & (0.016) & (0.031) & (0.032) \\ 
  & & \\ 
 Analytical test score (std.) $\times$ \textit{AI Interviewer}& $-$0.001 & 0.006 & 0.026 & 0.031 \\ 
  & (0.018) & (0.017) & (0.032) & (0.031) \\ 
  & & \\ 
\hline \\[-1.8ex] 
Mean DV in Human Interviewer & 0.32 & 0.32 & 0.42 & 0.41 \\ 
Controls and fixed effects & -- & Yes & -- & Yes \\ 
Observations & 4,574 & 4,538 & 2,147 & 2,091 \\ 
R$^{2}$ & 0.101 & 0.243 & 0.152 & 0.242 \\ 
\hline 
\hline \\[-1.8ex] 
\end{tabular} 

%% file: Tables/tab_balance_check_timing_survey.tex
\begin{tabular}{lccc}
\\[-1.8ex]\hline 
\hline \\[-1.8ex] 
 & \multicolumn{2}{c}{\textit{Survey fielded}}& $H_0$: (1) = (2) \\ 
\cline{2-3} 
 Variable& Post interview & Post recruitment  & p-value  \\ 
& (1) & (2) & (3)  \\
\hline \\[-1.8ex]
Net promoter score & 8.98 & 8.91 & 0.37 \\ 
  Perceived impact of AI index (direction) & 0.28 & 0.26 & 0.77 \\ 
  Perceived impact of AI index (magnitude) & 2.37 & 2.35 & 0.63 \\ 
  Perceived recruiter quality index & 3.79 & 3.76 & 0.23 \\ 
  Perceived interview quality index & 3.77 & 3.75 & 0.73 \\ 
  Knowledge of AI & 2.48 & 2.35 & 0.10 \\ 
  Usage of AI & 6.74 & 6.45 & 0.19 \\ 
  \hline \\[-1.8ex] 
  Observations & 1,844 & 920 &  \\ 
\hline
\hline
\end{tabular}

%% file: Tables/tab_applicant_level_predicting_recruiter_choices.tex
\begin{tabular}{@{\extracolsep{5pt}}lcc} 
\\[-1.8ex]\hline 
\hline \\[-1.8ex] 
 & \multicolumn{2}{c}{\textit{Dependent variable:}} \\ 
\cline{2-3} 
\\[-1.8ex] & \multicolumn{2}{c}{Choice of Interviewer (AI = 1)} \\ 
\\[-1.8ex] & (1) & (2)\\ 
\hline \\[-1.8ex] 
 Perceived impact of AI on applicant (direction) & 0.058 & 0.108$^{**}$ \\ 
  & (0.040) & (0.043) \\ 
\hline \\[-1.8ex] 
Controls and fixed effects & -- & Yes \\
Observations & 186 & 177 \\ 
R$^{2}$ & 0.011 & 0.241 \\ 
\hline 
\hline \\[-1.8ex] 
\end{tabular} 

%% file: Tables/tab_applicant_level_selection_test_score_only_choice.tex
\begin{tabular}{@{\extracolsep{5pt}}lcccc} 
\\[-1.8ex]\hline 
\hline \\[-1.8ex] 
 & \multicolumn{4}{c}{\textit{Dependent variable:}} \\ 
\\[-1.8ex] & \multicolumn{2}{c}{Language test score (1-6)} & \multicolumn{2}{c}{Analytical test score (0-100)} \\ 
\cline{2-3}  \cline{4-5} 
\\[-1.8ex] & (1) & (2) & (3) & (4) \\ 
\hline \\[-1.8ex] 
Has chosen AI  & $-$0.227$^{***}$ & $-$0.170$^{***}$ & $-$2.228$^{***}$ & $-$1.665$^{**}$ \\ 
  & (0.040) & (0.044) & (0.689) & (0.775) \\ 
\hline \\[-1.8ex] 
Mean DV when human is chosen & 3.37 & 3.37 & 49.77 & 49.82 \\ 
Controls and fixed effects & -- & Yes & -- & Yes \\ 
Observations & 3,377 & 3,318 & 3,435 & 3,374 \\ 
R$^{2}$ & 0.009 & 0.046 & 0.003 & 0.036 \\ 
\hline 
\hline \\[-1.8ex] 
\end{tabular} 

%% file: Tables/tab_applicant_level_selection_test_score.tex
\begin{tabular}{@{\extracolsep{5pt}}lcccc} 
\\[-1.8ex]\hline 
\hline \\[-1.8ex] 
\multicolumn{5}{@{}l}{\textbf{Panel A: Choice of AI versus assigned AI interviewer}} \\
& \multicolumn{4}{c}{\textit{Dependent variable:}} \\ 

\\[-1.8ex] & \multicolumn{2}{c}{Language test score (1-6)} & \multicolumn{2}{c}{Analytical test score (0-100)} \\ 
\cline{2-3}  \cline{4-5} 
\\[-1.8ex] & (1) & (2) & (3) & (4) \\ 
\hline \\[-1.8ex] 
 Has chosen AI & $-$0.040$^{*}$ & $-$0.005 & $-$1.179$^{***}$ & $-$0.686$^{*}$ \\ 
  & (0.022) & (0.022) & (0.374) & (0.376) \\ 
\hline \\[-1.8ex] 
Mean DV in assigned AI& 3.18 & 3.18 & 48.72 & 48.62 \\ 
Controls and fixed effects & -- & Yes & -- & Yes \\ 
Observations & 13,857 & 13,601 & 14,120 & 13,861 \\ 
R$^{2}$ & 0.000 & 0.033 & 0.001 & 0.025 \\ 
\hline 
\hline \\[-1.8ex] 
\multicolumn{5}{@{}l}{\textbf{Panel B: Choice of human versus assigned human interviewer}} \\
 & \multicolumn{4}{c}{\textit{Dependent variable:}} \\ 

\\[-1.8ex] & \multicolumn{2}{c}{Language test score (1-6)} & \multicolumn{2}{c}{Analytical test score (0-100)} \\ 
\cline{2-3}  \cline{4-5} 
\\[-1.8ex] & (1) & (2) & (3) & (4)\\ 
\hline \\[-1.8ex] 
Has chosen human  & 0.099$^{**}$ & 0.036 & 0.598 & $-$0.091 \\ 
  & (0.039) & (0.040) & (0.683) & (0.713) \\ 
\hline \\[-1.8ex] 
Mean DV in assigned human & 3.27 & 3.26 & 49.17 & 49.11 \\ 
Controls and fixed effects & -- & Yes & -- & Yes \\ 
Observations & 3,672 & 3,583 & 3,740 & 3,649 \\ 
R$^{2}$ & 0.002 & 0.059 & 0.000 & 0.044 \\ 
\hline 
\hline \\[-1.8ex] 
\end{tabular} 

%% file: Tables/tab_transcript_analyses_treatment_on_category.tex
\begin{tabular}{@{\extracolsep{5pt}}lcc} 
\\[-1.8ex]\hline 
\hline \\[-1.8ex] 
 & \multicolumn{2}{c}{\textit{Dependent variable:}} \\ 
\cline{2-3} 
\\[-1.8ex] & \multicolumn{2}{c}{Interview is comprehensive} \\ 
\\[-1.8ex] & (1) & (2)  \\ 
\hline \\[-1.8ex] 
 \textit{Direct AI Interview }& 0.0299$^{***}$ & 0.0482$^{***}$ \\ 
  & (0.0116) & (0.0099)   \\[0.2cm]
\hline \\[-1.8ex] 
Mean DV in Human Interviewer & 0.3879 & 0.3880 \\ 
Controls and fixed effects & -- & Yes \\ 
Observations & 29,221 & 28,785 \\ 
R$^{2}$ & 0.0002 & 0.2086 \\ 
\hline 
\hline \\[-1.8ex] 
\end{tabular}

%% file: Sections_appendix/appendix_recruitmentprocess.tex
\section{Recruitment process details}
\label{app_sec:recruitment}

\subsection{Job description wording}
\label{app_sec:jobdescriptions}
\subsubsection*{Customer Expert}

\noindent
Our \textit{Customer Service Representatives} and \textit{Technical Support Representatives} are vital members of our company. You will field customer inquiries and find innovative ways to respond. You will have the chance to work in a highly collaborative and engaging environment that provides dynamic work experience with different cultures, as well as unlimited opportunities to grow your potential and develop your career.

\vspace{0.5em}

\noindent
As a \textit{Customer Service Representative} / \textit{Technical Support Representative}, your responsibilities will include:
\begin{itemize}
    \item Handling and carefully responding to all customer inquiries via inbound calls and email
    \item Providing excellent customer service through active listening
    \item Working with confidential customer information in a secure manner
    \item Aiming to resolve issues on the first call by being proactive
    \item Appropriately and adequately communicating with customers
\end{itemize}

\vspace{0.5em}

\noindent\textbf{Working hours:} \\
Monday to Friday – 8:00 PM to 5:00 AM PH Time / 10:00 PM to 7:00 AM PH Time

\vspace{0.5em}

\noindent\textbf{Background Requirements:}
\begin{itemize}
    \item NBI clearance
    \item Birth certificate
    \item Fit-to-work clearance
\end{itemize}

\vspace{0.5em}

\noindent\textbf{Compensation:} Depends on job, between Php~16{,}000 and Php~25{,}000

\vspace{0.5em}

\noindent\textbf{Minimal Requirements:}
\begin{itemize}
    \item SHS Grad / HS Grad
    \item Average communication skills
\end{itemize}

\subsection{Engagement-score algorithm (non-proprietary summary)}
\label{app_sec:engagement_score_algorithm}

Each applicant receives a base score that depends on how their profile entered the system (self-application vs.\ recruiter-added). The algorithm adds points for (i) valid phone and e-mail information, (ii) successful delivery of SMS/e-mail messages, (iii) message openings, and (iv) positive “Yes” responses; it subtracts points for explicit “No” responses. Missing or invalid contact details do not affect the score. Once the cumulative score exceeds a low, account-specific threshold, the applicant is queued for interview scheduling. Applicants who do not reach the threshold are not contacted. Applicants who are queued for interview scheduling are randomized into one of the three experimental treatments: \textit{Human interviewer}, \textit{AI interviewer}, or \textit{Choice of interviewer}.

\subsection{Invitation text}
\label{app_sec:invitation_texts}

\subsubsection{Treatments \textit{Human interviewer} and \textit{AI interviewer}}

\textbf{Email:} If a recruiter decides to interview an applicant, the following invitation text is sent via email to them. The text is identical across the \textit{Human interviewer} and \textit{AI interviewer} conditions. \medskip

{\itshape
\noindent [Subject line:] Interview Invitation: Schedule Your interview for the [Name Position] \smallskip

\noindent Hi [First name] [Last name], \smallskip

\noindent I hope you're doing well! My name is [Name recruiter], from [Name recruiting partner], the recruiting partner of [Name recruiting firm]. We've had a chance to review your application for the [Name Position]. We currently have an immediate need to fill this position. Not all roles require [Position-specific] experience, providing opportunities for various backgrounds and skill sets. If you are interested, please click on the button below.\smallskip
}

\noindent [Button ``Get Interview Call'']\smallskip

\noindent [Clicking the button will redirect applicants to the firm's website, where they can schedule their interview.]\\

\noindent \textbf{Phone:} At the same time as the email is sent, the following invitation text is sent via text message to applicants' phones. Again, the text is identical across the \textit{Human interviewer} and \textit{AI interviewer} conditions. \medskip

{\itshape
\noindent Subject:  Interview Invitation: Schedule Your Interview for the [Name position] \smallskip

\noindent Hi [First name] [Last name],\smallskip

\noindent We are reaching out to you regarding your application for the [Name position]. We'd like you to schedule your interview. We've sent the interview invitation to your email.
}

\subsubsection{Treatment \textit{Choice interviewer}}

Applicants in the \textit{Choice interviewer} treatment receive the same email and phone message as in the other two treatments. The only difference is that clicking the button ``Get Interview Call'' in the email will redirect them to an interview scheduling preference page. The text on the page is as follows:\medskip 

{\itshape
\noindent Interview Scheduling Preference\smallskip 

\noindent Congratulations! You have been shortlisted for an interview. Please select your preferred interviewer:

\begin{itemize}
    \item AI Interviewer: The call can be scheduled at your convenience.
    \item Human Interviewer: You'll need to schedule the interview based on the human recruiter's availability. 
\end{itemize}
}

\noindent [After selecting the interviewer, applicants can schedule their interview exactly as in the other two treatments.] 

\subsection{Details on the interview process}

\subsubsection{How human recruiters conduct interviews}

Human recruiters are provided with a structured interview script that ensures consistency while allowing flexibility in addressing individual candidate profiles. The interview begins with a standardized introduction in which the recruiter confirms the candidate's identity and explains the purpose of the call. The recruiter then proceeds with scenario-based questions tailored to the candidate's background. For example, if the candidate has gaps in their employment history, the recruiter asks about the reasons for those gaps and how the candidate maintained their motivation during that period. Similarly, if a candidate has frequently changed jobs, the recruiter asks about the reasons for these transitions.

In addition to these customized questions, the script includes general questions applicable to all candidates. These questions explore the candidate's recent employment, their strategies for handling stress, and their expectations regarding salary. For candidates with specific backgrounds, such as financial accounting experience or technical support, the recruiter conducts mock calls or role-playing exercises to assess their practical skills in handling customer inquiries. For example, a candidate with a background in financial accounts might be asked to role-play a scenario where a customer inquires about a declined payment or a loan application process.

Although the script provides a comprehensive structure, recruiters are allowed to deviate from it as long as the essential questions are addressed. This flexibility allows recruiters to adapt their questioning to better suit the flow of the conversation and to probe deeper into areas of interest or concern. 

\subsubsection{How AI voice agents conduct interviews}

The AI voice agent is instructed to follow the same structure as human recruiters. At the beginning of the interview, the AI voice agent uses the following standardized text as the introduction:  \medskip

{\itshape
\noindent \textbf{AI voice agent}: Hi [Applicant name]. This is Anna, [firm name]'s AI recruiter and I am calling about the [job] role you applied to recently. Do you have a couple of minutes to chat about your application?   \smallskip

\noindent \textbf{Applicant:} [Example response: Hi, yes I have time.] \smallskip

\noindent \textbf{AI voice agent:} Great! Since I am an AI recruiter, as I ask you questions, if you are not clear on my question, please feel free to ask me for clarification. Does that work for you?  \smallskip

\noindent\textbf{Applicant:} [Example response: Yes, that works for me.]   \smallskip

\noindent \textbf{AI voice agent:}  Ok, I also want to let you know a human recruiter will review the recording from our discussion today and will make the final decision on your application for employment. The questions I will be asking you are the same questions my human counterpart would ask. Does that sound OK?
}

\clearpage
\subsection{Full interview examples}
\label{app_sec:interview_full_examples}

\input{Sections_appendix/appendix_example_interviews}

%% file: Sections_appendix/appendix_example_interviews.tex
The following transcripts present two sample interviews, one from the \textit{Human Interviewer} and one from the \textit{AI Interviewer} condition. As noted in the main text, the content of interviews varies across application sites and job profiles. These examples are therefore provided for illustrative purposes and are not necessarily representative of the interviews conducted in the experiment. To protect applicant confidentiality, the transcripts have been edited to preserve anonymity and prevent any possibility of direct or indirect identification.

\subsection{Human Interviewer example}

\textbf{Interviewer:} Hello. \newline
\textbf{Candidate:} Hello, good morning. \newline
\textbf{Interviewer:} Hi, this is \textcolor{lightblue}{[Anonymized]} from \textcolor{lightblue}{[Anonymized]} Recruitment. I’ll be conducting the interview and I’ll be asking you personal details and some common interview questions. Then I'll be giving you feedback prior to your position, okay? \textcolor{lightblue}{[Topic 1: Introduction]}\newline
\textbf{Candidate:} Yes, of course. \newline
\textbf{Interviewer:} Now, I just want to remind you that this phone call is recorded for quality assurance purposes. \newline
\textbf{Candidate:} Yes. \newline
\textbf{Interviewer:} Okay. So can you state your complete name, your middle name, your age, and your birthplace? \newline
\textbf{Candidate:} My name is \textcolor{lightblue}{[Anonymized]} and I’m \textcolor{lightblue}{[Anonymized]}.\newline
\textbf{Interviewer:} I’m sorry, how old are you? \newline
\textbf{Candidate:} \textcolor{lightblue}{[Anonymized]}.\newline
\textbf{Interviewer:} \textcolor{lightblue}{[Anonymized]}? \newline
\textbf{Candidate:} Yeah, correct. \newline
\textbf{Interviewer:} Can you turn down on your phone maybe? \newline
\textbf{Candidate:} Yes. \newline
\textbf{Interviewer:} Have you ever worked for \textcolor{lightblue}{[Anonymized]} in the past? \newline
\textbf{Candidate:} No. This is my second job in \textcolor{lightblue}{[Anonymized]}. \newline
\textbf{Interviewer:} And where are you currently residing? \textcolor{lightblue}{[Topic 2: Location Verification]}\newline
\textbf{Candidate:} In \textcolor{lightblue}{[Anonymized]}. \newline
\textbf{Interviewer:} If you are ever hired, will there be any issues commuting or traveling from your place that you work? \newline
\textbf{Candidate:} No, there’s no issues at all because right now we do have a direct train from \textcolor{lightblue}{[Anonymized]} going to your place. \newline
\textbf{Interviewer:} May I know what will be your mode of transportation or how long will it take you to travel? \textcolor{lightblue}{[Topic 2: Mode of Transportation]}\newline
\textbf{Candidate:} Yeah, just a few. I think it will be around an hour. \newline
\textbf{Interviewer:} Can you state your highest educational attainment? \textcolor{lightblue}{[Topic 3: Educational Background]}\newline
\textbf{Candidate:} I did \textcolor{lightblue}{[Anonymized]}. \newline
\textbf{Interviewer:} Hello, I’m sorry. I’m sorry I can’t hear you clearly. Can you turn the muffled sound if you can? Can you transfer for a better reception or maybe using your headset? \newline
\textbf{Candidate:} Yeah, I’m currently using my headset right now. \newline
\textbf{Interviewer:} If you can, can we just mute your phone so that you can hear me clearly? \newline
\textbf{Candidate:} Okay, ill just remove my headset. Sorry. Hello? \newline
\textbf{Interviewer:} Hello, can you hear me clearly? \newline
\textbf{Candidate:} Okay, that’s better. Thank you so much. First, I graduated \textcolor{lightblue}{[Anonymized]}. Then I took up my masters degree in \textcolor{lightblue}{[Anonymized]} but unfortunately I just finished \textcolor{lightblue}{[Anonymized]} because there were some circumstances that I have to face here in the Philippines. Thats why I go back there. \newline
\textbf{Interviewer:} Okay. May I know what’s your masters degree? \newline
\textbf{Candidate:} Masters degree in \textcolor{lightblue}{[Anonymized]}, majoring in \textcolor{lightblue}{[Anonymized]}.\newline
\textbf{Interviewer:} May I ask you if you have plans to pursue your studies? \textcolor{lightblue}{[Topic 4: Candidate’s Commitment and Attrition Risks -  Future Study Plan]}\newline
\textbf{Candidate:} Right now, no because I need to focus first with \textcolor{lightblue}{[Anonymized]}. Thats the reason why I came back here in the Philippines. \newline
\textbf{Interviewer:} Thank you. Are you currently employed? \newline
\textbf{Candidate:} Sorry? \newline
\textbf{Interviewer:} Are you currently employed? \newline
\textbf{Candidate:} No. I’m not employed right now. \newline
\textbf{Interviewer:} Do you have any pending applications from the other companies? \textcolor{lightblue}{[Topic 4: Candidate’s Commitment and Attrition Risks -  Other Applications]}\newline
\textbf{Candidate:} To be honest, I just submitted my application and fortunately this is the first company who called me for an interview. Thats why I’m here right now. \newline
\textbf{Interviewer:} Thank you. How about any previous work experience? All in all? Any foreign? \textcolor{lightblue}{[Topic 5: Past Employment History]}\newline
\textbf{Candidate:} First, my first three years I worked in \textcolor{lightblue}{[Anonymized]} as \textcolor{lightblue}{[Anonymized]}. Then after that, I went abroad to \textcolor{lightblue}{[Anonymized]}. I worked as a \textcolor{lightblue}{[Anonymized]} for \textcolor{lightblue}{[Anonymized]} years. Then I go back here in the Philippines to work as \textcolor{lightblue}{[Anonymized]} for \textcolor{lightblue}{[Anonymized]}. Then I went to \textcolor{lightblue}{[Anonymized]} and worked as an \textcolor{lightblue}{[Anonymized]}.\newline
\textbf{Interviewer:} So the most recent job that you had is in \textcolor{lightblue}{[Anonymized]}, right? \newline
\textbf{Candidate:} Yeah, correct. \newline
\textbf{Interviewer:} Can you please tell me more about it? \newline
\textbf{Candidate:} As a \textcolor{lightblue}{[Anonymized]}, I’m the one responsible for \textcolor{lightblue}{[Anonymized]}. All the problems related with \textcolor{lightblue}{[Anonymized]}, they put in \textcolor{lightblue}{[Anonymized]}. Then every morning I have to check \textcolor{lightblue}{[Anonymized]} and check all the priorities which need to handle first. Also, I have to give feedback to \textcolor{lightblue}{[Anonymized]}. And I also receive some \textcolor{lightblue}{[Anonymized]} from the customers regarding the problems or issues with their equipment or any computer related problems. And especially, I need to resolve all those issues to reduce the downtime on their side. Basically, that’s all my job. \newline
\textbf{Interviewer:} And may I know what’s the reason why you left that job? \newline
\textbf{Candidate:} I left my job. Thats what I’m saying because I have to face some problems here in the Philippines because \textcolor{lightblue}{[Anonymized]}, they have some issues. I have to weigh in which is better for the future. But I will not sacrifice my \textcolor{lightblue}{[Anonymized]}. So I’d rather choose \textcolor{lightblue}{[Anonymized]} rather than \textcolor{lightblue}{[Anonymized]}. \newline
\textbf{Interviewer:} May I ask if you have any certificates of employment from the previous job? \newline
\textbf{Candidate:} Yeah, I do. \newline
\textbf{Interviewer:} So what made you decide to apply? \newline
\textbf{Candidate:} Yeah, because I submitted applications and fortunately this is the first company who called me for an interview. Also, I think since I'm challenging, like trying something like \textcolor{lightblue}{[Anonymized]} which is very in demand in this country. And I also want to improve my comm skills as well because working in \textcolor{lightblue}{[Anonymized]} will improve your comm skills. \newline
\textbf{Interviewer:} May I know what’s your salary expectation? \textcolor{lightblue}{[Topic 6: Compensation]}\newline
\textbf{Candidate:} To be honest, I would like to. I dont have a specific since I dont have any background experience in \textcolor{lightblue}{[Anonymized]}. So I would like to assume that from the range between \textcolor{lightblue}{[Anonymized]} to \textcolor{lightblue}{[Anonymized]}\newline
\textbf{Interviewer:} \textcolor{lightblue}{[Anonymized]} to \textcolor{lightblue}{[Anonymized]}, right? \newline
\textbf{Candidate:} Yes, yes. \newline
\textbf{Interviewer:} The range that we can offer you if ever you’re going to get hired will be from \textcolor{lightblue}{[Anonymized]} to \textcolor{lightblue}{[Anonymized]}  depending on what account are you going to be with. Are you okay with it? \newline
\textbf{Candidate:} Yes. \newline
\textbf{Interviewer:} Okay. So I'm sure you have \textcolor{lightblue}{[Anonymized]}  experience. \newline
\textbf{Candidate:} Yeah, \textcolor{lightblue}{[Anonymized]}. The customers will call \textcolor{lightblue}{[Anonymized]} to help to resolve their problem or some issues. For example, for instance, they have an internet issue and the client or the customer will call \textcolor{lightblue}{[Anonymized]} and the job of \textcolor{lightblue}{[Anonymized]} is to resolve, to find the root cause and to resolve the issues of \textcolor{lightblue}{[Anonymized]}. \newline
\textbf{Interviewer:} And how would you define quality \textcolor{lightblue}{[Anonymized]}? \newline
\textbf{Candidate:} Oh yeah. First, you must be polite to the service and you must also handle the average customer and you should be able to resolve all the \textcolor{lightblue}{[Anonymized]} issues and concerns. Especially with less supervision from \textcolor{lightblue}{[Anonymized]}. You must be able to handle that on your own. \newline
\textbf{Interviewer:} So if ever youre going to encounter irate or difficult \textcolor{lightblue}{[Anonymized]}, how do you think you can handle them effectively? \newline
\textbf{Candidate:} Yeah, I always stay calm. I already encountered with irate \textcolor{lightblue}{[Anonymized]} in \textcolor{lightblue}{[Anonymized]}. I stay calm and explain to them everything in \textcolor{lightblue}{[Anonymized]}: I explain everything clearly and I try to focus on the problem and solution to resolve the issues of \textcolor{lightblue}{[Anonymized]}. \newline
\textbf{Interviewer:} What are skills do you think you have that can be a great asset for the job position? \newline
\textbf{Candidate:} First, I’m a hardworking and dynamic self-disciplined person and I can also work as an individual or a team player and I’m good at handling \textcolor{lightblue}{[Anonymized]}. \newline
\textbf{Interviewer:} If ever you’re going to encounter conflict between your peers or your colleagues, how are you going to address them? \newline
\textbf{Candidate:} I will have to help them personally and privately, not in front of the audience, and help them and try to resolve the issues. \newline
\textbf{Interviewer:} Do you have any \textcolor{lightblue}{[Anonymized]} or at least most of the time I mentioned? \newline
\textbf{Candidate:} No. \textcolor{lightblue}{[Anonymized]}\newline
\textbf{Interviewer:} Do you have any other certifications aside from your work? \newline
\textbf{Candidate:} No, not at all. \newline
\textbf{Interviewer:} Do you have any travel plans within the next six months, international or local? \textcolor{lightblue}{[Topic 7: Candidate’s Commitment and Attrition Risks - Planned Vacation]}\newline
\textbf{Candidate:} No, I don’t have one. \newline
\textbf{Interviewer:} Have you ever had the willingness to work on-site? \newline
\textbf{Candidate:} Yeah, of course. \newline
\textbf{Interviewer:} You can work in shifting schedules, in extended hours, holidays, weekends, weekdays? \textcolor{lightblue}{[Topic 7: Candidate’s Commitment and Attrition Risks - Working during Holidays]}\newline
\textbf{Candidate:} Yeah, no problem at all. Because when I worked in \textcolor{lightblue}{[Anonymized]} for \textcolor{lightblue}{[Anonymized]}, I worked in various shifts for at least \textcolor{lightblue}{[Anonymized]}. \newline
\textbf{Candidate:} So working in various shifts is not a problem at all. \newline
\textbf{Interviewer:} Can you start as soon as possible if ever youre going to get hired? \newline
\textbf{Candidate:} Yeah, absolutely. And I'm willing to start. \newline
\textbf{Interviewer:} And may I know how long are you planning to work here with us? \textcolor{lightblue}{[Topic 7: Candidate’s Commitment and Attrition Risks - Long-term Intention]}\newline
\textbf{Candidate:} Sorry? \newline
\textbf{Interviewer:} How long are you planning to work here with us? \newline
\textbf{Candidate:} As long as I'm enjoying my job in our company, Ill pursue my goal. Because if I'm going to start here as a regular \textcolor{lightblue}{[Anonymized]}, thats not my goal. \newline
\textbf{Candidate:} My goal is to become a higher position, which I can use my masters degree in handling people, something like that. \newline
\textbf{Interviewer:} Many of you already have your government documents like \textcolor{lightblue}{[Anonymized]} to help you by any chance? \textcolor{lightblue}{[Topic 8: Data Verification]}\newline
\textbf{Candidate:} Right now I dont have \textcolor{lightblue}{[Anonymized]} yet, but I do have \textcolor{lightblue}{[Anonymized]}. \newline
\textbf{Interviewer:} And may I know what valid IDs do you have with you? \newline
\textbf{Candidate:} I do have my \textcolor{lightblue}{[Anonymized]}.\newline
\textbf{Interviewer:} And are you vaccinated? \newline
\textbf{Candidate:} \textcolor{lightblue}{[Anonymized]}\newline
\textbf{Interviewer:} Do you have any medical issues that we need to know? \newline
\textbf{Candidate:} \textcolor{lightblue}{[Anonymized]}.\newline
\textbf{Interviewer:} Okay, I think that concludes the end of our interview. The next step is the assessment. Please wait for the contact officer to call you, okay? \textcolor{lightblue}{[Topic 9: Conclusion]}\newline
\textbf{Candidate:} Okay, thank you so much. 

\subsection{AI Interviewer example}

\textbf{Interviewer:} Hi \textcolor{lightblue}{[Anonymized]}. This is Anna, \textcolor{lightblue}{[Anonymized]} AI recruiter and I am calling about the \textcolor{lightblue}{[Anonymized]} role you applied to recently. Do you have fifteen minutes to talk about your application? \textcolor{lightblue}{[Topic 1: Introduction + Candidate Availability]} \newline
\textbf{Candidate:} Yes.  \newline
\textbf{Interviewer:} Great! Since I am an AI recruiter, as I ask you questions, if you are not clear on my question, please feel free to ask me for clarification. Does that work for you? \newline
\textbf{Candidate:} Yes.  \newline
\textbf{Interviewer:} Ok, I also want to let you know a human recruiter will review the recording from our discussion today and will make the final decision on your application for employment. The questions I will be asking you are the same questions my human counterpart would ask. Does that sound OK? \newline
\textbf{Candidate:} Yes.  \newline
\textbf{Interviewer:} Let's get started. Have you ever worked for \textcolor{lightblue}{[Anonymized]} in the past?  \newline
\textbf{Candidate:} No. This is my first time working in, um, \textcolor{lightblue}{[Anonymized]}. \newline
\textbf{Interviewer:} That's okay, we get a lot of new applicants. Can you tell me a little bit about yourself and why you're interested in this role?  \newline
\textbf{Candidate:} Firstly, good afternoon. I'm \textcolor{lightblue}{[Anonymized]} from \textcolor{lightblue}{[Anonymized]} and I am an \textcolor{lightblue}{[Anonymized]}  graduate and have also completed \textcolor{lightblue}{[Anonymized]} semester of \textcolor{lightblue}{[Anonymized]} at \textcolor{lightblue}{[Anonymized]}  in \textcolor{lightblue}{[Anonymized]}. My year of experience in \textcolor{lightblue}{[Anonymized]} has equipped me with strong customer service and skills, and I'm confident in my ability to make a significant contribution to your team.  \newline
\textbf{Interviewer:} That's great to hear about your experience and education. This position is an on-site position and is located at \textcolor{lightblue}{[Anonymized]}. Can you share with me where you reside? \textcolor{lightblue}{[Topic 2: Location Verification]} \newline
\textbf{Candidate:} I'm residing in \textcolor{lightblue}{[Anonymized]}. Is there any available position in \textcolor{lightblue}{[Anonymized]}?   \newline
\textbf{Interviewer:} Our office is actually located in \textcolor{lightblue}{[Anonymized]}. How far a drive would this be for you?  \newline
\textbf{Candidate:} I think, um, it's one and a half hour from \textcolor{lightblue}{[Anonymized]}. But it would be better if there's any open position in \textcolor{lightblue}{[Anonymized]} .  \newline
\textbf{Interviewer:} Ok, how about transportation to and from work. Can you share with me the mode of transportation you would use to commute to our office daily? \textcolor{lightblue}{[Topic 2: Mode of Transportation]}  \newline
\textbf{Candidate:} Oh, from \textcolor{lightblue}{[Anonymized]}, I'd estimate my commute to \textcolor{lightblue}{[Anonymized]}  would be around \textcolor{lightblue}{[Anonymized]}  minutes to \textcolor{lightblue}{[Anonymized]}  depending on traffic. But to ensure punctuality, uh, also plan to commute at least \textcolor{lightblue}{[Anonymized]} before my shift, starts since it's a one ride trip only, and I'm confident about reaching the time reaching the office on time.  \newline
\textbf{Interviewer:} The minimum age requirement for this position is 18 years of age, are you at least 18 years of age? \textcolor{lightblue}{[Topic 3: Age Requirement]} \newline
\textbf{Candidate:} I'm \textcolor{lightblue}{[Anonymized]}  years old.  \newline
\textbf{Interviewer:} Most of the opportunities we have are on a shifting schedule where you could work during holidays and/or weekends. Have you had previous employment that has required you to work a similar schedule? \textcolor{lightblue}{[Topic 4: Candidate’s Commitment and Attrition Risks - Working during Holidays]} \newline
\textbf{Candidate:} Yes.  \newline
\textbf{Interviewer:} In addition to a shifting schedule, we sometimes require overtime. Would you be able to accommodate extended hours when needed? \textcolor{lightblue}{[Topic 4: Candidate’s Commitment and Attrition Risks - Working Overtime]} \newline
\textbf{Candidate:} Yes. I will.  \newline
\textbf{Interviewer:} I'd like to ask you some questions about your career goals and motivations. When you read our job advertisement, what attracted you to the position at \textcolor{lightblue}{[Anonymized]} \textcolor{lightblue}{[Topic 4: Candidate’s Commitment and Attrition Risks - Motivations]} \newline
\textbf{Candidate:} Actually, my previous coworker referred me to because as per him, there's open roles for a \textcolor{lightblue}{[Anonymized]}  account and as I have \textcolor{lightblue}{[Anonymized]} experience and I think I will be fit for this role. That's why I applied for this job.  \newline
\textbf{Interviewer:} That's interesting. As you consider your career goals, how long could you envision working for \textcolor{lightblue}{[Anonymized]} and performing the duties of this job? \textcolor{lightblue}{[Topic 4: Candidate’s Commitment and Attrition Risks - Long-term Intentions]}\newline
\textbf{Candidate:} I think I see myself um, still working to your company \textcolor{lightblue}{[Anonymized]}  to \textcolor{lightblue}{[Anonymized]}  years. If there's a good position or good environment, I think I I'm able to work as long as I work on my previous job.  \newline
\textbf{Interviewer:} As you consider making a job change, what is important to you as you evaluate job opportunities? \newline
\textbf{Candidate:} Right now, my goal is to work at your company since I as I check on my multimedia reviews on \textcolor{lightblue}{[Anonymized]} that your company's team good. That that's why I want to focus on working with company as well.  \newline
\textbf{Interviewer:} Thank you for your answers, I appreciate the insight.  \newline
\textbf{Candidate:} That's why I think I didn't need to find another job. I'm just interested in the job position that you have open. \newline
\textbf{Interviewer:} I'd like to shift back to a few more questions about the position we are considering you for. Our training requires mandatory attendance. If selected for the position, would you be able to attend all training?  \newline
\textbf{Candidate:} Yes. I will. Mm-mm.  \newline
\textbf{Interviewer:} Do you have any planned vacations over the next 6 months? \textcolor{lightblue}{[Topic 4: Candidate’s Commitment and Attrition Risks - Planned Vacations]} \newline
\textbf{Candidate:} No. I don't have any plans. My, um, go for this year or this month is work. Focus on work.  \newline
\textbf{Interviewer:} Are there any activities that may affect your attendance or any prior commitments that you have to attend to?  \newline
\textbf{Candidate:} For now, there aren’t any. \newline
\textbf{Interviewer:} Let's take a few minutes to review your education background. What is your highest level of education? \textcolor{lightblue}{[Topic 5: Educational Background]} \newline
\textbf{Candidate:} I am a \textcolor{lightblue}{[Anonymized]} graduate and have also completed two semester of \textcolor{lightblue}{[Anonymized]}  at \textcolor{lightblue}{[Anonymized]}. \newline
\textbf{Interviewer:} That's interesting. Are there any activities that may affect your attendance or any prior commitments that you have to attend to?  \newline
\textbf{Candidate:} For now, I think there are no activities in the future that affect my attendance because I will now talk as to work, and I have no any vacation pending.  \newline
\textbf{Interviewer:} Let's take a few minutes to review your education background. You've already told me about your highest level of education completed, which is \textcolor{lightblue}{[Anonymized]}  and having completed \textcolor{lightblue}{[Anonymized]}  at \textcolor{lightblue}{[Anonymized]}. I just want to confirm, is that correct?  \newline
\textbf{Candidate:} Yes. \newline
\textbf{Candidate:} I'm an \textcolor{lightblue}{[Anonymized]} and a \textcolor{lightblue}{[Anonymized]}, um, completer, uh, for \textcolor{lightblue}{[Anonymized]}. I think it's correct.  \newline
\textbf{Interviewer:} Since you have completed \textcolor{lightblue}{[Anonymized]}, do you plan to return to \textcolor{lightblue}{[Anonymized]}?  \newline
\textbf{Candidate:} For now, I don't think so. I just my goal for now is to get a job. Permanent job, to sustain my bill needs.  \newline
\textbf{Interviewer:} If we can shift our discussion to compensation expectations, that would be helpful. What is your expected salary? \textcolor{lightblue}{[Topic 6: Compensation]} \newline
\textbf{Candidate:} I consider my ex considering my experience in customer facing enroll, I was hoping for a salary closer to \textcolor{lightblue}{[Anonymized]} to \textcolor{lightblue}{[Anonymized]}.  \newline
\textbf{Interviewer:} The roles that we have open right now are between the salary package range of \textcolor{lightblue}{[Anonymized]} to \textcolor{lightblue}{[Anonymized]} per Month, depending on your qualifications, you could be offered a salary package between these amounts. Would you be okay with this range?  \newline
\textbf{Candidate:} Is there any closer for \textcolor{lightblue}{[Anonymized]} to \textcolor{lightblue}{[Anonymized]}? May I know the specific or may I know the account first?  \newline
\textbf{Interviewer:} The roles that we have open right now are between the salary package range of \textcolor{lightblue}{[Anonymized]} to \textcolor{lightblue}{[Anonymized]}, depending on your qualifications, you could be offered a salary package between these amounts. Would you be okay with this range?  \newline
\textbf{Candidate:} Yes. I would be okay.  \newline
\textbf{Interviewer:} Thank you again for the information you've shared. I have a few more questions to go over, so let's continue. I'd like to talk about your \textcolor{lightblue}{[Anonymized]} experience. can you share with me the volume of \textcolor{lightblue}{[Anonymized]} on a daily basis? \textcolor{lightblue}{[Topic 7: Past Employment History]} \newline
\textbf{Candidate:} My previous job, I have been in \textcolor{lightblue}{[Anonymized]} near \textcolor{lightblue}{[Anonymized]}  and I received \textcolor{lightblue}{[Anonymized]} well, we received \textcolor{lightblue}{[Anonymized]} about \textcolor{lightblue}{[Anonymized]}. That's our goal for the day, and I get \textcolor{lightblue}{[Anonymized]} that's passing the \textcolor{lightblue}{[Anonymized]} for a day.  \newline
\textbf{Interviewer:} That sounds like a significant vo-  \newline
\textbf{Candidate:} yes. Continue.  \newline
\textbf{Interviewer:} Were these \textcolor{lightblue}{[Anonymized]} solving technical issues or were you responding to \textcolor{lightblue}{[Anonymized]}?  \newline
\textbf{Candidate:} We conduct troubleshooting and also we create \textcolor{lightblue}{[Anonymized]}, and we talk to a customer and explain their needs, what their needs, and also like that.  \newline
\textbf{Interviewer:} Now, I'd like to ask about your previous work experience. Have you worked with other companies before this role?  \newline
\textbf{Candidate:} My previous job, I ended my journey there \textcolor{lightblue}{[Anonymized]}.I think. Yeah. That's my last job.  \newline
\textbf{Interviewer:} So, it sounds like you have some work experience, specifically from your last job. \newline
\textbf{Candidate:} Yes.  \newline
\textbf{Interviewer:} Now, could you please provide the name of the most recent company you worked for? \newline
\textbf{Candidate:} I've been for \textcolor{lightblue}{[Anonymized]} for a year, and my first job was \textcolor{lightblue}{[Anonymized]}. Then the second one, is I've I have \textcolor{lightblue}{[Anonymized]}. \newline
\textbf{Interviewer:} Now, can you tell me what are your employment dates for those companies you mentioned. \newline
\textbf{Candidate:} My employment date like, probably this job I've been there for \textcolor{lightblue}{[Anonymized]} and ended \textcolor{lightblue}{[Anonymized]}. It's almost \textcolor{lightblue}{[Anonymized]}  \newline
\textbf{Interviewer:} So, your employment dates are from \textcolor{lightblue}{[Anonymized]} to \textcolor{lightblue}{[Anonymized]}. That's helpful to know.  \newline
\textbf{Candidate:} Mhmm. That's correct.  \newline
\textbf{Interviewer:} Now. Can you tell methe location of the company we are discussing. Where is this company located?  \newline
\textbf{Candidate:} This company must my previous job was located at \textcolor{lightblue}{[Anonymized]}.  \newline
\textbf{Interviewer:} So your previous job was located at \textcolor{lightblue}{[Anonymized]}. Now, I'd like to ask, do you have any subsequent work experience with other companies? \newline
 \textbf{Candidate:} I'm also I also work as health care account at \textcolor{lightblue}{[Anonymized]}. Before I apply for at \textcolor{lightblue}{[Anonymized]}, then I applied for this.  \newline
\textbf{Interviewer:} That's interesting. Now, could you please provide the name of the second most recent company you worked for?  \newline
\textbf{Candidate:} Second reason, company I work for. I work for \textcolor{lightblue}{[Anonymized]}. I'm \textcolor{lightblue}{[Anonymized]} there. I've been there almost \textcolor{lightblue}{[Anonymized]}.  \newline
\textbf{Interviewer:} I'd like to know more about your work history. What are your employment dates for these previous roles, if you have them available?  \newline
\textbf{Candidate:} For a while, let me double check here because I don't I don't remember. I think it's to be sure, let's me let me double check here on my resume for a while.  \newline
\textbf{Interviewer:} Where is this company located?  \newline
\textbf{Candidate:} Okay. Before I applying for a \textcolor{lightblue}{[Anonymized]} as a \textcolor{lightblue}{[Anonymized]}  for \textcolor{lightblue}{[Anonymized]} account. Before that, I've been in a \textcolor{lightblue}{[Anonymized]}  as a \textcolor{lightblue}{[Anonymized]}. Duration was \textcolor{lightblue}{[Anonymized]}. It's located at \textcolor{lightblue}{[Anonymized]}. \newline
 \textbf{Interviewer:} So, it seems like you have some experience in customer service. Do you have any subsequent work experience with other companies besides \textcolor{lightblue}{[Anonymized]} and \textcolor{lightblue}{[Anonymized]}?  \newline
\textbf{Candidate:} I also work at \textcolor{lightblue}{[Anonymized]}. I've been there for duration is \textcolor{lightblue}{[Anonymized]} to \textcolor{lightblue}{[Anonymized]} as a \textcolor{lightblue}{[Anonymized]}. All my job all my previous job experience is \textcolor{lightblue}{[Anonymized]} roles.  \newline
\textbf{Interviewer:} Could you please provide the name of the third most recent company you worked for?  \newline
\textbf{Candidate:} My third is the \textcolor{lightblue}{[Anonymized]}. It's a \textcolor{lightblue}{[Anonymized]}, and my position there a \textcolor{lightblue}{[Anonymized]}. The duration is \textcolor{lightblue}{[Anonymized]} until \textcolor{lightblue}{[Anonymized]}. This company was located at \textcolor{lightblue}{[Anonymized]}.  \newline
\textbf{Interviewer:} You mentioned you worked at \textcolor{lightblue}{[Anonymized]} as a \textcolor{lightblue}{[Anonymized]}, can you tell me why you left that company? \newline
\textbf{Candidate:} I left at \textcolor{lightblue}{[Anonymized]} because I'm sick. \newline
\textbf{Interviewer:} If you are offered the job, when would you be able to start? \textcolor{lightblue}{[Topic 8: Availability]} \newline
\textbf{Candidate:} I can for now because since I don't have any work, I can work  \newline
\textbf{Interviewer:} Do you have any pending applications this this week in \textcolor{lightblue}{[Anonymized]} \newline
\textbf{Candidate:} This is my first application before I ended my contract on my previous job.  \newline
\textbf{Interviewer:} Do you have any pending applications with other companies?  \newline
\textbf{Candidate:} I don't have any pending applications.  \newline
\textbf{Interviewer:} What is your contact number? \textcolor{lightblue}{[Topic 9: Data Verification]} \newline
\textbf{Candidate:} Okay. My contact number and my cell phone number \textcolor{lightblue}{[Anonymized]}  \newline
\textbf{Interviewer:} Could y \newline
ou please share your \textcolor{lightblue}{[Anonymized]} number with me?  \newline
\textbf{Candidate:} Okay. Same as I said earlier, my phone number through \textcolor{lightblue}{[Anonymized]} is \textcolor{lightblue}{[Anonymized]} \textbf{Interviewer:} May I know your Marital/Civil Status, please?  \newline
\textbf{Candidate:} \textcolor{lightblue}{[Anonymized]}.  \newline
\textbf{Interviewer:} \textcolor{lightblue}{[Anonymized]} do you have any questions you would like for me to document and have reviewed by the recruiting team?  \newline
\textbf{Candidate:} Yes. I want to ask if there's any available position for your local account in tour site at \textcolor{lightblue}{[Anonymized]} because I was prepared my by my previous by my previous coworker and I really want to join their team because I I believe, like, I have a role because or I fitted the job this job because I have a \textcolor{lightblue}{[Anonymized]}. As per my referral, your current employee, that's health care account for your local account or bill care account is now open for hiring.  \newline
\textbf{Interviewer:} I have noted your question, and I will make sure to document it, after this call, someone from our team will get back to you regarding the available position for \textcolor{lightblue}{[Anonymized]}, and we will also verify the information about \textcolor{lightblue}{[Anonymized]} being open for hiring.  \newline
\textbf{Candidate:} Okay. Thank you. That's all.  \newline
\textbf{Interviewer:} Thank you \textcolor{lightblue}{[Anonymized]}. This concludes the questions I have for you today. Following this interview, our recruiting team will be reviewing the summary of our conversation to determine next steps regarding your application for employment. Due to the high demand and urgency to fill our openings, we encourage you to complete an online assessment. By taking this assessment right away, you'll demonstrate your commitment to this opportunity. \textcolor{lightblue}{[Topic 10: Conclusion + Assessment Test]} \newline
\textbf{Candidate:} Mhmm.  \newline
\textbf{Interviewer:} Great! We'll see you on the next hiring process. Good luck!

%% file: Sections_appendix/appendix_interview_transcript_variables.tex
\section{Interview transcript variables}
\label{app_sec:transcript_variables}

\paragraph{Data preparation.} The firm shared with us raw verbatim transcripts. In total, we have transcripts available for 34,109 applications, which is a subset of all interviews conducted. We used LLMs to transform them into structured and anonymized transcripts to be able to use the text as data. Specifically, we preprocessed the raw transcripts using prompt methods based on few-shot examples, roles, and instructions with the \textit{gemini-2.0-flash} model, following the techniques described in \textcite{brown2020language}. First, we created speaker tags for each candidate and recruiter and then applied this classification to each turn within a job interview. Second, we removed personally identifiable information (PII), such as name, address, phone number, and email address, from the speakers or the companies mentioned during the job interview. See Appendix Sections \ref{app:speakerprompt} and \ref{app:anonprompt} for the prompts.

\subsection{Interview type variables}
\label{app_sec:transcript_topics_types}

\paragraph{Setup.} We performed an LLM-driven natural language processing analysis on anonymized structured transcripts using few-shot, role-based, and instruction-based prompt engineering techniques to categorize interviews into mutually exclusive types. We combine multiple prompting strategies to achieve a reliable and accurate transcript classification. Specifically, we combine role-based prompting, chain-of-thought (CoT) prompting, and in-context examples (zero- or few-shot). First, role-based prompting appears at the start of the prompt (``You are an interview expert''), which focuses the model on domain knowledge of recruitment and interviewing. Second, CoT prompting provides a structured decision hierarchy, reducing misclassifications when labels overlap \parencite{vatsal2024survey, wei2022chain}. Finally, we included in-context examples to help the model recognize category cues. The prompt can be found in 
Appendix Section \ref{app:intclassprompt}. In Table \ref{tab:interview_categories}, we display the classification of interview types, which was developed in collaboration with the company to achieve a comprehensive coverage of the different types of interviews.

\paragraph{Results.} We find that the largest share of interviews is classified as \emph{comprehensive interviews}, characterized by an organic opening and closing, high applicant engagement, and at least eight distinct covered topics. In total, 42\% of interviews conducted by the AI voice agent and 39\% of interviews conducted by a human recruiter are classified in this way.\footnote{This difference is significant at the 1\% level and robust to the inclusion of controls and fixed effects (see Appendix Table \ref{tab_transcript_analyses_treatment_on_category}).} In total, 7\% of AI and 25\% of human interviews are \emph{Screen-outs}, in which the interview is ended by the recruiter because the applicant does not meet certain requirements. Instances of \emph{Disengaged interactions}, where applicants are unresponsive, distracted, or disinterested, account for 12\% and 9\% of interviews, respectively, whereas instances of applicants being unavailable account for 14\% and 9\% of interviews. Lastly, we identify two AI-specific categories. In 7\% of AI-led interviews, the interview is aborted due to a technical failure of the AI voice agent, and in 5\% of interviews, the applicant explicitly expresses the unwillingness to continue speaking with the AI voice agent. As a summary, Appendix Figure \ref{fig:interview_types_treatment} plots the distribution of the interview types.

\begin{table}[H]
\renewcommand{\arraystretch}{1.5}
\begin{tabular}{
    l%
    p{7cm}     %
    p{1cm}     %
    l%
}
\hline
\textbf{Category} & \textbf{Definition} & \textbf{Duration} & \textbf{Topic Coverage} %
\\
\hline
Comprehensive Interview &
Interview has a natural opening and closure, a high-quality engagement, and contains $\geq$ eight expected topics.  &
Average & High ($\geq8$) 
\\[4pt]
AI Aversion &
The candidate explicitly expresses an unwillingness to continue speaking with an AI recruiter. &
Short & Low 
\\[4pt]
Early Screen-Out &
Interview ends early because the candidate is immediately disqualified based on a non-negotiable requirement related to the job (e.g., location). &
Short & Very Low (0–2)
\\[4pt]
Midway Screen-Out &
Interview ends after some initial engagement due to a mismatch discovered during the conversation (e.g., conflicting school plans). &
Medium & Moderate (3–7)  
\\[4pt]
Late Screen-Out &
Interview proceeds nearly to completion but the candidate fails a final, critical criterion (e.g., rehire status). &
Long & High (8+)  
\\[4pt]
Telephony Failure &
Interview ends due to issues with cellular network, signal loss, or VOIP instability. &
Varies & Varies 
\\[4pt]
AI System Failure &
LLM/voice agent stalls, crashes, or fails to respond in the interview. &
Varies & Varies %
\\[4pt]
Disengaged Interaction &
Applicant is disinterested, unresponsive, and/or distracted during the interview, and the interview has poor continuity.  &
Varies & Low-Moderate ($<8$) %
\\[4pt]
Candidate Unavailability &
Applicant states they are unable to talk, and the interview ends for this reason.  &
Short & Very Low (0–2)  \\[4pt]
Others &
Interview type that does not fit in any other category.  &
Varies & Varies \\[4pt]

\hline
\end{tabular}
\caption{Transcript type classification}
\label{tab:interview_categories}
\end{table}

\subsection{Standardization variables}
\label{app_sec:transcript_recruiter_feature_variables}

We use the following five variables to capture features of recruiters' language and behavior in the interviews. To construct the variables, we used LLM-based prompting with \textit{gemini-2.5-flash}, a transformer model, and regex NLP methods.

\paragraph{Topic coverage.} We use a controlled-vocabulary prompt to categorize how many of the maximum number of 14 topics that recruiters can cover in a given transcript are substantively covered. A topic is counted as 'covered' only when (i) the interviewer explicitly probes the theme, and (ii) the candidate offers a nontrivial reply (at least three content words) that contains at least one keyword from a topic-specific lexicon supplied to the model. To reduce semantic drift and ensure consistency across transcripts, we use the firm's predefined topic labels. We also add a final, ''other'' topic to capture off-topic conversations, hence there are 15 in total. Appendix Table \ref{tab:storyboard_full} displays the full list of possible topics, and Appendix Section \ref{app:topicprompt} the prompt. Topic coverage is then a percentage, defined as the ratio of topics covered divided by the maximum number of 15 topics. 

\paragraph{Following topic order.} This variable measures the consistency of the order of topics in a given interview transcript compared to the firm's guideline topic order. See Table \ref{tab:storyboard_full} for the template order and description of topics. For each transcript, we extract the sequence in which topics appear and remove repeated occurrences so that each topic is represented once. Next, we compute the Kendall rank correlation coefficient ($\tau$) between the transcript's topic sequence and the guideline's topic sequence. Kendall's $\tau$ measures the degree of similarity in the ordering between the two sequences, ranging from -1 to 1. Higher $\tau$ values thus indicate stronger similarity between actual and guideline topic order. 

\paragraph{Following guideline questions.} To measure the extent to which recruiters follow the guideline questions, we calculate the semantic similarity between recruiter questions asked in each interview and the template questions supplied by the firm as guidelines. To do so, we first match each question to one of the 14 topics (Table \ref{tab:storyboard_full}) and convert the text into a 384-dimension embedding using the all-MiniLM-L6-v2 sentence transformers model.\footnote{The model all-MiniLM-L6-v2 is a sentence embedding model that is fine-tuned to perform well on sentence similarity tasks. The model is obtained from \url{https://huggingface.co/sentence-transformers/all-MiniLM-L6-v2}.} In the same way, we convert the template questions to word embeddings, averaging the embeddings in case of multiple template questions. We then compute the cosine similarity score between the recruiter's word embedding of the question and the corresponding embedding of the template questions for that particular topic. For each interview, we then average the similarity scores across all topics. Higher values indicate higher overall similarity of recruiter questions with the guideline questions.

\paragraph{Question distance score.} This variable measures the semantic variance of recruiter questions per topic. We again first match each question to one of the 14 topics (Table \ref{tab:storyboard_full}) and convert the text into a 384-dimension embedding using the all-MiniLM-L6-v2 sentence transformers model. Then, separately for each topic and condition (AI voice agent and human recruiter), we measure the ''topic centroid embedding'' by calculating the average of word embeddings of all recruiter questions within that topic per condition. We then calculate the topic distance score as 1 - cosine similarity score between the word embedding of each question and the corresponding topic centroid embedding. For each interview, we then calculate the average distance score across all questions asked for the respective topic. A higher question distance score indicates greater semantic variation in the phrasing of the questions relative to the average recruiter question.

\subsection{Applicant variables}
\label{app_sec:transcript_applicant_feature_variables}

We use the following eight variables to capture linguistic features of applicants. The relevant phrases and markers were captured using spaCy and regex NLP methods.

\paragraph{Vocabulary richness score.} This variable is defined as the number of unique words divided by the square root of the total number of words spoken by the applicant. Higher scores indicate greater vocabulary diversity and linguistic sophistication.

\paragraph{Syntactic complexity score.} This variable captures the average number of subordinate clauses, specific details, and explanatory phrases per response used by the applicant, normalized by response length. Higher values capture greater thoroughness and nuance in responses.

\paragraph{Discourse markers frequency.} This variable captures the average number of sequential, causal, and clarifying discourse markers per minute (e.g., “first”, “because”, “specifically”) in the applicant’s responses. Higher values indicate more frequent use of discourse markers.

\paragraph{Filler words frequency.} This variable represents the average number of basic and conversational fillers per minute (e.g., “uh”, “uhm”, “like”, “you know”) in the applicant’s responses. Higher values indicate more frequent use of filler words.

\paragraph{Backchannel cue frequency.} This variable measures the average number of verbal backchannel cues per minute — short verbal cues supplied by the applicant to indicate attention or agreement (e.g., “sure”, “got it”, “mhm”, “okay”, “yeah”, “yes”). Higher values indicate more frequent use of backchannel cues.

\paragraph{Number of exchanges interviewer–applicant.} This variable counts the total number of conversational exchanges, where an exchange is defined as a sequence in which the interviewer speaks, followed by the applicant.

\paragraph{Number of questions by applicant.} This is the total number of questions asked by the applicant during the conversation.

\paragraph{Linguistic style match index.} This index captures the similarity of linguistic style between the applicant and interviewer. It is constructed as the average similarity score across nine function word categories:  
(1) personal pronouns (e.g., “I”, “you”),  
(2) impersonal pronouns (e.g., “this”, “it”),  
(3) articles (e.g., “a”, “the”),  
(4) auxiliary verbs (e.g., “am”, “have”),  
(5) high-frequency adverbs (e.g., “very”, “well”),  
(6) prepositions (e.g., “in”, “around”),  
(7) conjunctions (e.g., “but”, “while”),  
(8) negations (e.g., “not”, “no”), and  
(9) quantifiers (e.g., “many”, “few”).  
Each category score is defined as  
\[
1 - \frac{|\text{rate}_{\text{interviewer}} - \text{rate}_{\text{applicant}}|}{\text{rate}_{\text{interviewer}} + \text{rate}_{\text{applicant}} + 0.0001},
\]  
where rates represent the percentage usage in each speaker’s text. Higher index values indicate greater linguistic style similarity between the interviewer and applicant.

%% file: Sections_appendix/appendix_recruiter_experience.tex
\section{Recruiter behavior over time}
\label{app_sec:recrutier_over_time}

Here, we investigate whether recruiters' experience with or learning from handling AI and human interviews matters for their decision-making. Note that the firm piloted the roll-out of the AI-led interviews in the weeks prior to the start of the experiment. Hence, most recruiters have been exposed to and worked with AI-led interviews. Accordingly, the process was not entirely new to them. Nevertheless, our recruiters differ substantially in the number of interviews they conduct over the course of the experiment. This allows us to analyze the correlation between the number of interviews recruiters are assigned to review and their propensity to extend offers. 

We start by investigating whether a general experience effect exists, i.e., whether the total number of interviews assigned for review is correlated with the likelihood of extending an offer. We find a significant association: regressing offer rates on the total number reveals that, on average, reviewing 10 additional interviews is associated with a -0.06 percentage point lower offer rate. For details, see Appendix Table \ref{tab_recruiter_experience_on_offer_rate_long} column (1). Note that we are not claiming that experiencing more interviews has a causal effect on offers or that it is the effect of exposure to interviews. There are likely recruiter-specific factors that influence both offer rates and the number of interviews reviewed. For instance, there are likely skill differences among recruiters in how efficiently they handle applications.

Instead, we are mainly interested in whether the association of experience with offer rates \textit{differs} between the \textit{Human Interviewer} and \textit{AI Interviewer} conditions. Accordingly, we interact the number of interviews reviewed with a treatment dummy. We find a significant and positive interaction effect ($p=0.02$, column (2) of Appendix Table \ref{tab_recruiter_experience_on_offer_rate_long}). Hence, the negative association of the number of applications with offer rates is less pronounced in the \textit{AI Interviewer} condition relative to the \textit{Human Interviewer} condition.

\begin{table}[!ht]
    \caption{Association of review experience with job offer rate among recruiters}
    \label{tab_recruiter_experience_on_offer_rate_long}
        \begin{threeparttable}[c]
            \input{Tables/tab_recruiter_experience_on_offer_rate_long}
            \begin{tablenotes}
                \item \footnotesize \textit{Notes:} 
                The table shows OLS estimates. The dependent variable is a recruiter's job offer rate (0-1), i.e., the sum of offers a recruiter gave to applicants during the experiment divided by the total sum of applications evaluated by the recruiter. ``Number of applications assigned to recruiter'' is the total number of applications a recruiter received per treatment as part of the experiment. In column (1), we pool both \textit{AI interviewer} and \textit{Human interviewer} conditions. In column (2), we add an indicator variable indicating treatment status. An observation is a recruiter-treatment unit. Standard errors in parentheses are clustered at the recruiter level. Significance levels: $^{*}\,p<0.1$,$^{**}\,p<0.05$, $^{***}\,p<0.01$.
            \end{tablenotes}
        \end{threeparttable}
\end{table}

%% file: Tables/tab_recruiter_experience_on_offer_rate_long.tex
\begin{tabular}{@{\extracolsep{5pt}}lcc} 
\\[-1.8ex]\hline 
\hline \\[-1.8ex] 
 & \multicolumn{2}{c}{\textit{Dependent variable:}} \\ 
\cline{2-3} 
\\[-1.8ex] &  \multicolumn{2}{c}{Average job offer rate (\%)}  \\ 
& (1) & (2) \\
\hline \\[-1.8ex] 
 Number of applications assigned to recruiter & $-$0.00006$^{***}$ & $-$0.00013$^{***}$ \\ 
  & (0.00001) & (0.00004) \\ 
  & & \\ 
 \textit{AI Interviewer}&  & 0.01862 \\ 
  &  & (0.02150) \\ 
  & & \\  
  Number of applications assigned to recruiter $\times$ \textit{AI Interviewer}&  & 0.00007$^{**}$ \\ 
  &  & (0.00003) \\ 
  & & \\ 
\hline \\[-1.8ex] 
Recruiters & 112 & 112 \\ 
Observations & 224 & 224 \\ 
R$^{2}$ & 0.03181 & 0.04271 \\ 
\hline 
\hline \\[-1.8ex] 
\end{tabular} 

%% file: Sections_appendix/appendix_applicant_survey_instructions.tex
\section{Applicant survey instructions}
\label{app_sec:survey}

\paragraph{Invitation text short survey.}

We want to hear about your recent candidate experience. Please take a few minutes to share your feedback. Your feedback is confidential and does not impact any employment decisions.

\paragraph{Invitation text long survey.}

We want to hear about your recent candidate experience. Please take a few minutes to share your feedback. As an appreciation for your time, we will send you a gift card in the amount of 4USD when you complete this survey. Your feedback is confidential and does not impact any employment decisions.

\subsection{Long survey wording}\label{appendix:surveylong}

\subsubsection{Procedural trust}

\begin{enumerate}
    \item Based on your experience, how likely is it that you would recommend our company to a friend or colleague as a place to apply for work? [NPS question]
    
    \item Was the recruiter knowledgeable about the company?
    \begin{itemize}
        \item Very knowledgeable
        \item Somewhat knowledgeable
        \item Slightly knowledgeable
        \item Not knowledgeable at all
    \end{itemize}

    \item Was the recruiter knowledgeable about the role you were applying for?
    \begin{itemize}
        \item Very knowledgeable
        \item Somewhat knowledgeable
        \item Slightly knowledgeable
        \item Not knowledgeable at all
    \end{itemize}

    \item Were the questions asked during your phone interview relevant to the job you applied for?
    \begin{itemize}
        \item Very relevant
        \item Somewhat relevant
        \item Slightly relevant
        \item Not relevant at all
    \end{itemize}

    \item Do you feel your time was valued during the recruitment process?
    \begin{itemize}
        \item Very much valued
        \item Somewhat valued
        \item Slightly valued
        \item Not valued at all

    \end{itemize}

    \item Did you feel that the recruiter was able to follow up appropriately based on your answers?
    \begin{itemize}
        \item Always
        \item Often
        \item Sometimes
        \item Rarely
        \item Never
    \end{itemize}
\end{enumerate}

\subsubsection{Social experience}

\begin{enumerate}
    \item How natural did the interaction with the recruiter feel?
    \begin{itemize}
        \item Very natural
        \item Somewhat natural
        \item Neutral
        \item Somewhat unnatural
        \item Very unnatural
    \end{itemize}

    \item How comfortable did you feel during the interview with the recruiter?
    \begin{itemize}
        \item Very comfortable
        \item Somewhat comfortable
        \item Neutral
        \item Somewhat uncomfortable
        \item Very uncomfortable
    \end{itemize}

    \item Did you find talking to the recruiter stressful?
    \begin{itemize}
        \item Not at all stressful
        \item Somewhat stressful
        \item Moderately stressful
        \item Very stressful
        \item Extremely stressful
    \end{itemize}

    \item  How frequently did you receive live feedback from the recruiter during your interview?
    \begin{itemize}
        \item Very frequently
        \item Somewhat frequently
        \item Occasionally
        \item Rarely
        \item Never
    \end{itemize}
\end{enumerate}

\subsubsection{Perceived Discrimination}

\begin{enumerate}
    \item Did you feel discriminated against by the recruiter because of your gender identity?
    \begin{itemize}
        \item Yes
        \item No
        \item Not sure
    \end{itemize}

    \item Do you believe the interview process was fair compared to your past interview experiences?
    \begin{itemize}
        \item Much more fair
        \item More fair
        \item About the same
        \item Less fair
        \item Much less fair
        \item This was my first ever job interview. (N/A).
    \end{itemize}
\end{enumerate}

\subsubsection{General AI awareness, knowledge and usage}

\begin{enumerate}
    \item Select the correct definition of Artificial Intelligence (AI)
    \begin{itemize}
        \item AI is the process of enhancing industrial machinery efficiency using automated control systems for mechanical and electrical improvements. [Incorrect]
        \item AI involves developing computer systems to perform tasks that usually require human intelligence, such as language understanding and pattern recognition. [Correct]
        \item AI is a method to develop software applications specifically designed to manage financial transactions and banking operations efficiently. [Incorrect]
        \item AI refers to the creation of complex spreadsheets for data analysis and business forecasting, emphasizing numerical computations. [Incorrect]
        \item I don’t know. [Incorrect]
    \end{itemize}

    \item Select the correct definition of Generative Artificial Intelligence (GenAI)
    \begin{itemize}
        \item GenAI automates genetic analysis to modify DNA sequences for medical purposes. [Incorrect]
        \item GenAI is AI that creates new content, such as text and images, by learning from existing data. [Correct]
        \item GenAI involves the creation of algorithms for solving complex mathematical problems and optimizing industrial processes. [Incorrect]
        \item GenAI is a system that focuses on generating engineering techniques to enhance crop yield and resistance to pests in agriculture. [Incorrect]
        \item I don’t know. [Incorrect]
    \end{itemize}

    \item How often do you use the following products?
    \begin{itemize}
        \item ChatGPT [Daily, Weekly, Monthly, Never, I have never heard of it, Not rated N/A]
        \item Character.AI [Daily, Weekly, Monthly, Never, I have never heard of it, Not rated N/A]
        \item QuillBot [Daily, Weekly, Monthly, Never, I have never heard of it, Not rated N/A]
        \item Midjourney [Daily, Weekly, Monthly, Never, I have never heard of it, Not rated N/A]
    \end{itemize}

    \item Thinking about customer service, which of the following uses artificial intelligence (AI)?
    \begin{itemize}
        \item A detailed Frequently Asked Questions webpage [Incorrect]
        \item An online survey sent to customers that allows them to provide feedback [Incorrect]
        \item A contact page with a form available to customers to provide feedback [Incorrect]
        \item A chatbot that immediately answers customer questions [Correct]
        \item Not sure [Incorrect]
    \end{itemize}

    \item When using email, which of the following uses artificial intelligence (AI)?
    \begin{itemize}
        \item The email service marking an email as read after the user opens it [Incorrect]
        \item The email service allowing the user to schedule an email to send at a specific time in the future [Incorrect]
        \item The email service categorizing an email as spam [Correct]
        \item The email service sorting emails by time and date [Incorrect]
        \item Not sure [Incorrect]
    \end{itemize}

    \item Thinking about online shopping, which of the following uses artificial intelligence (AI)?
    \begin{itemize}
        \item Storage of account information, such as shipping addresses [Incorrect]
        \item Records of previous purchases [Incorrect]
        \item Product recommendations based on previous purchases [Correct]
        \item Product reviews from other customers [Incorrect]
        \item Not sure [Incorrect]
    \end{itemize}
\end{enumerate}

\subsubsection{AI perception on the labor market}

[The following items were taken from \textcite{pew2022}.] \medskip
 
\noindent Over the next 20 years, how much impact do you think the use of artificial intelligence (AI) in the workplace will have on...

\begin{enumerate}
    \item Workers generally
    \begin{itemize}
        \item A major impact
        \item A minor impact
        \item No impact
        \item Not sure
        \item No answer
    \end{itemize}

    \item You, personally
    \begin{itemize}
        \item A major impact
        \item A minor impact
        \item No impact
        \item Not sure
        \item No answer
    \end{itemize}

\noindent Thinking about the use of artificial intelligence (AI) in the workplace over the next 20 years, what do you think the outcome will be for...

    \item Workers generally
    \begin{itemize}
        \item AI will help more than it hurts
        \item AI will equally help and hurt
        \item AI will hurt more than it helps
        \item Not sure
        \item No answer
    \end{itemize}

    \item You, personally
    \begin{itemize}
        \item AI will help more than it hurts
        \item AI will equally help and hurt
        \item AI will hurt more than it helps
        \item Not sure
        \item No answer
    \end{itemize}
\end{enumerate}

\subsubsection{Open-ended feedback}

\begin{enumerate}
    \item Do you have any additional feedback you’d like to share about your interview experience?
\end{enumerate}

\subsection{Short survey wording}\label{appendix:surveyshort}

\subsubsection{Procedural trust}

\begin{enumerate}

    \item Based on your experience, how likely is it that you would recommend our company to a friend or colleague as a place to apply for work? [NPS question]

    \item Was the recruiter knowledgeable about the company?
    \begin{itemize}
        \item Very knowledgeable
        \item Somewhat knowledgeable
        \item Slightly knowledgeable
        \item Not knowledgeable at all
    \end{itemize}

    \item Were the questions asked during your phone interview relevant to the job you applied for?
    \begin{itemize}
        \item Very relevant 
        \item Somewhat relevant
        \item Slightly relevant
        \item Not relevant at all

    \end{itemize}
\end{enumerate}

\subsubsection{Social experience}

\begin{enumerate}
    \item How natural did the interaction with the recruiter feel?
    \begin{itemize}
        \item Very natural
        \item Somewhat natural
        \item Neutral
        \item Somewhat unnatural
        \item Very unnatural
    \end{itemize}

    \item  How frequently did you receive live feedback from the recruiter during your interview?
    \begin{itemize}
        \item Very frequently
        \item Somewhat frequently
        \item Occasionally
        \item Rarely
        \item Never
    \end{itemize}
\end{enumerate}

\subsubsection{Perceived discrimination}

\begin{enumerate}
    \item Did you feel discriminated against by the recruiter because of your gender identity?
    \begin{itemize}
        \item Yes
        \item No
        \item Not sure
    \end{itemize}
\end{enumerate}

\subsubsection{General AI awareness, knowledge and usage}

\begin{enumerate}
    \item Select the correct definition of Generative artificial intelligence (GenAI)
    \begin{itemize}
        \item GenAI automates genetic analysis to modify DNA sequences for medical purposes. [incorrect]
        \item GenAI is AI that creates new content, such as text and images, by learning from existing data. [Correct]
        \item GenAI involves the creation of algorithms for solving complex mathematical problems and optimizing industrial processes. [incorrect]
        \item GenAI is a system that focuses on generating engineering techniques to enhance crop yield and resistance to pests in agriculture. [incorrect]
        \item I don’t know [incorrect]
    \end{itemize}

    \item How often do you use the following products?
    \begin{itemize}
        \item ChatGPT [Daily, Weekly, Monthly, Never, I have never heard of it, Not rated N/A]
        \item Character.AI [Daily, Weekly, Monthly, Never, I have never heard of it, Not rated N/A]
        \item QuillBot [Daily, Weekly, Monthly, Never, I have never heard of it, Not rated N/A]
        \item Midjourney [Daily, Weekly, Monthly, Never, I have never heard of it, Not rated N/A]
    \end{itemize}
\end{enumerate}

\subsubsection{AI perception on the labor market}

\noindent Over the next 20 years, how much impact do you think the use of artificial intelligence (AI) in the workplace will have on...

\begin{enumerate}
    \item Workers generally
    \begin{itemize}
        \item A major impact
        \item A minor impact
        \item No impact
        \item Not sure
        \item No answer
    \end{itemize}

    \item You, personally
    \begin{itemize}
        \item A major impact
        \item A minor impact
        \item No impact
        \item Not sure
        \item No answer
    \end{itemize}
\end{enumerate}

\subsubsection{Open-ended feedback}

\begin{enumerate}
    \item Do you have any additional feedback you’d like to share about your interview experience?
\end{enumerate}

%% file: Sections_appendix/appendix_recruiter_survey_instructions.tex
\section{Recruiter survey instructions}
\label{app_sec:survey_recruiter}

Welcome! Thank you for taking part in this short survey. The survey asks about your experiences and views on hiring. We are interested in hearing your thoughts and perspectives. There are no right or wrong answers - we’re simply interested in your honest opinion. The survey will only take a few minutes to complete, and your responses will remain confidential.

\begin{enumerate}
    \item Did you evaluate interviews conducted by Anna, our AI voice agent?
\begin{itemize}
  \item Yes
  \item No
\end{itemize}
\end{enumerate}

\subsection{Predicting differences between human and AI-led interviews}

For the next questions, please consider all interviews that were conducted at PSG in the last three months, i.e., from March to June 2025.

\begin{enumerate}
\item Across all interviews, do you expect AI-led interviews to be of \textbf{higher, lower, or equal quality} compared to human-led interviews?

\begin{itemize}
  \item Much higher quality
  \item Slightly higher quality
  \item About the same quality
  \item Slightly lower quality
  \item Much lower quality
\end{itemize}

\item Across all interviews, do you expect AI-interviewed candidates to \textbf{receive job offers at a higher, lower, or equal rate} compared to human-interviewed candidates?

\begin{itemize}
  \item Higher
  \item Equal
  \item Lower
\end{itemize}

\item \emph{[Shown only if “Higher” was selected in the previous question 2.]}\\
On the previous page, you indicated that you think AI-interviewed candidates receive job offers at a higher rate. \textbf{If out of 1,000 candidates who were interviewed by human recruiters, 85 got a job offer, how many of the 1,000 do you think would have gotten a job offer if they were interviewed by the AI instead?}

\item \emph{[Shown only if “Lower” was selected in the previous question 2.]}\\
On the previous page, you indicated that you think human-interviewed candidates receive job offers at a higher rate. \textbf{If out of 1,000 candidates who were interviewed by human recruiters, 85 got a job offer, how many of the 1,000 do you think would have gotten a job offer if they were interviewed by the AI instead?}

\item Across all candidates who eventually received job offers, do you expect AI-interviewed candidates to \textbf{stay employed longer, shorter, or for a similar length of time} compared to human-interviewed candidates?

\begin{itemize}
  \item AI-interviewed are \textbf{much longer} employed
  \item AI-interviewed are \textbf{slightly longer} employed
  \item AI-interviewed are \textbf{equally long} employed
  \item AI-interviewed are \textbf{slightly shorter} employed
  \item AI-interviewed are \textbf{much shorter} employed
\end{itemize}

\item Across all candidates who eventually started their job, do you expect AI-interviewed candidates to \textbf{have higher, the same, or lower on-the-job productivity} compared to human-interviewed candidates?

\begin{itemize}
  \item AI-interviewed are \textbf{much more} productive
  \item AI-interviewed are \textbf{slightly more} productive
  \item AI-interviewed are \textbf{similarly} productive
  \item AI-interviewed are \textbf{slightly less} productive
  \item AI-interviewed are \textbf{much less} productive
\end{itemize}
\end{enumerate}

\subsection{Experience evaluating AI voice agent interviews}

\noindent [Shown only if “Yes” selected for the question of whether the recruiter evaluated AI voice agent interviews]

\begin{enumerate}

\item Compared to evaluating the interviews you conducted yourself, how \textbf{easy/difficult was it for you to evaluate} AI-led interviews in terms of time and effort?

\begin{itemize}
  \item Much more difficult to evaluate AI
  \item Somewhat more difficult to evaluate AI
  \item About the same
  \item Somewhat easier to evaluate AI
  \item Much easier to evaluate AI
\end{itemize}

\item When you decide whether to make an offer to an applicant, how \textbf{important} are an applicant’s [Name of standardized test] test scores compared to their performance in the interview itself?

\begin{itemize}
  \item Interview performance is \textbf{much more important} than [Name of test] scores
  \item Interview performance is \textbf{somewhat more important} than [Name of test] scores
  \item Interview performance and [Name of test] scores are \textbf{equally important}
  \item Interview performance is \textbf{somewhat less important} than [Name of test] scores
  \item Interview performance is \textbf{much less important} than [Name of test] scores
\end{itemize}

\item Compared to the interviews you conducted yourself, how do you rate the \textbf{quality of the information} you received from AI-led interviews?

\begin{itemize}
  \item Much lower quality from AI
  \item Somewhat lower quality from AI
  \item About the same quality
  \item Somewhat better quality from AI
  \item Much better quality from AI
\end{itemize}

\item Overall, how would you rate the introduction of AI-interviews in the recruiting process?

\begin{itemize}
  \item Very negative
  \item Negative
  \item Neutral
  \item Positive
  \item Very positive
\end{itemize}

\item Compared to the interviews you conducted yourself, \textbf{how high are your standards} for applicants from AI-led interviews?

\begin{itemize}
  \item Much lower standards for applicants from AI
  \item Lower standards for applicants from AI
  \item About the same standards
  \item Higher standards for applicants from AI
  \item Much higher standards for applicants from AI
\end{itemize}

\item Please share any additional thoughts or suggestions regarding your experience with AI-led interviews, especially improvements or changes you would like to see.\\

\subsection{AI perception on the labor market}

\noindent Artificial intelligence (AI) can be used by employers to collect and analyze data, make decisions, and complete tasks. Some employers are using AI in hiring, for worker evaluations, or even to do jobs humans used to do.

\noindent Over the next 20 years, how much impact do you think the use of artificial intelligence (AI) in the workplace will have on...

    \item Workers generally
    \begin{itemize}
        \item A major impact
        \item A minor impact
        \item No impact
        \item Not sure
        \item No answer
    \end{itemize}

    \item You, personally
    \begin{itemize}
        \item A major impact
        \item A minor impact
        \item No impact
        \item Not sure
        \item No answer
    \end{itemize}

Thinking about the use of artificial intelligence (AI) in the workplace over the next 20 years, what do you think the outcome will be for...

    \item Workers generally
    \begin{itemize}
        \item AI will help more than it hurts
        \item AI will equally help and hurt
        \item AI will hurt more than it helps
        \item Not sure
        \item No answer
    \end{itemize}

    \item You, personally
    \begin{itemize}
        \item AI will help more than it hurts
        \item AI will equally help and hurt
        \item AI will hurt more than it helps
        \item Not sure
        \item No answer
    \end{itemize}
\end{enumerate}

%% file: Sections_appendix/appendix_prompts.tex
\UseRawInputEncoding
\section{Text analysis: prompt content}\label{sec:prompt_strategies}

\subsection{Speaker labeling prompt}\label{app:speakerprompt}

\begin{tcolorbox}[
    breakable,
    colback=gray!2,
    colframe=gray!40,
    boxrule=0.1pt,
    title={
      \textcolor{black}{\sffamily\textsc{Speaker labeling prompt}}
    }
]
\small \ttfamily
\begin{lstlisting}[language={}, basicstyle=\small\ttfamily,
                   columns=flexible]
You are an expert conversation analyst tasked with meticulously labeling an interview transcript. Your goal is to accurately identify each speaker and label their lines with either "Interviewer:" or "Candidate:". Do not repeat any portions of the original text. Only output the labeled transcript.

If it's unclear who is speaking, make your best judgment based on the overall context of the conversation.

Return the transcript with speaker labels in the format "Speaker: Dialogue". *Place each speaker's turn on a new line, but do not include \textbackslash n within the line itself.

\#\#\# Example

Input:

[Anonymized]: Hi, this is  [Anonymized] from Teleperformance Recruitment. Ill be conducting an interview and Ill be asking you personal details and some common interview questions in order to give you feedback on how I should proceed your application, okay? 

[Name applicant]: Okay. 

[Anonymized]: And I just want to remind you that this phone call is recorded for quality assurance purposes, okay? 
[Name applicant]: Alright, got it. 

[Anonymized]: Okay

Output:

Interviewer: Hi, this is  [Anonymized] from Teleperformance Recruitment. Ill be conducting an interview and Ill be asking you personal details and some common interview questions in order to give you feedback on how I should proceed your application, okay? 

Candidate: Okay. 

Interviewer: And I just want to remind you that this phone call is recorded for quality assurance purposes, okay? 
Candidate: Alright, got it. 

Interviewer: Okay

Here is the transcript to label: [omitted for privacy reasons].
\end{lstlisting}
\end{tcolorbox}

\subsection{Anonymization prompt}\label{app:anonprompt}
\begin{tcolorbox}[
    breakable,
    colback=gray!2,
    colframe=gray!40,
    boxrule=0.1pt,
    title={
      \textcolor{black}{\sffamily\textsc{Anonymization prompt}}
    }
]
\small \ttfamily 
\begin{lstlisting}[language={}, basicstyle=\small\ttfamily,
                   columns=flexible]
You are an expert transcript anonymizer.
Your job is to remove any personally identifiable information (PII) from the following text. That includes:
- Person names (replace with **AnonymizedNAME**)
- Email addresses (replace with **AnonymizedEMAIL**)
- Phone numbers (replace with **AnonymizedPHONE**)
- Organization names (replace with **AnonymizedORG**)
- Street address or city or location (replace with **AnonymizedADDRESS**)
- Date of Birth (replace with **AnonymizedDOB**)
Return only the cleaned text - do not add any explanation.
Text to anonymize:
"""{text}"""
\end{lstlisting}
\end{tcolorbox}

\subsection{Interview classification prompt}\label{app:intclassprompt}
The following shows the prompt used to classify interviews into types. Note that the prompt contains the additional category ``Expectation Mismatch''. For the paper, we put this category under ``Other'' given that less than 1\% of interviews have this category.

\begin{tcolorbox}[
    breakable,     
    colback=gray!2,
    colframe=gray!40,
    boxrule=0.1pt,
    title={
      \textcolor{black}{\sffamily\textsc{Interview classification prompt}}
    }
]

\small \ttfamily
\begin{lstlisting}[language={}, basicstyle=\small\ttfamily,
                   columns=flexible]
You are an interview expert. Your task is to classify the provided interview data into one of the following categories. 

Interview Transcript:
1. Interview Transcript: {transcript}

Interview Meta Data:
1. Call Duration: {duration} minutes
2. Number of Topics Covered: {topic_count}
3. Treatment: {treatment}

Instructions:
1.  Evaluate interview transcript and metadata based on order of priority, starting from Priority 1 (Interview-Stopping Events), Priority 2 (Screening Out Interviews), Priority 3 (Interview Analysis), to Priority 4 (Others).
2.  Provide your reasoning. In your explanation, cite specific phrases from the transcript and data points (like topic_count or call duration) to justify your choice.

Category Definitions Based on Order of Priority:
Priority 1 - Interview-Stopping Events
- Check for these following categories first

1.  Candidate Unavailability: The candidate explicitly states they are currently unable to talk (e.g., "I'm driving," "I'm in a meeting," "Can I call you back later?"). The call ends quickly due to this reason.
2.  AI Aversion: The candidate explicitly expresses unwillingness to continue speaking with an AI recruiter. e.g., "I'm not comfortable talking to AI," "I'd prefer to speak with a human recruiter.")
    - Note: Do not classify human interviews (Treatment that contains "Human Interview") into "AI Aversion"
3.  Telephony Failure: Issues with cellular network, signal loss, or VOIP instability. Conversation is not finished, no conclusion remark
    - Note: when there are some repeated questions, topic_count >= 8, conversation ends with concluding remark, it cannot be classified as "Telephone Failure"
4.  AI System Failure: The LLM/voice agent malfunctions (e.g., stalls, crashes, fails to respond, repeats itself endlessly).  This is a failure of the AI itself, not the connection.
    - Important Exclusion: This category does not apply if an interview recovers and ends with concluding remark. A few repeated question does not automatically mean "AI System Failure"
    - Note: Do not classify human interviews (Treatment that contains "Human Interview") into "AI System Failure"

Priority 2 - Screening Out Interviews
- If interview-stopping events do not apply, check for these following categories

5.  Early Screen-Out: The interview ends early because the candidate is immediately disqualified based on a non-negotiable requirement related to the job (e.g., salary expectations, location, visa status). Important rules include topic_count is low (0-2) and call duration is short. Recruiter states the reason for ending the call due to disqualification. 
    - Important Rule: A short duration alone does not automatically mean "Early Screen-Out." "Early Screen-Out" requires an explicit disqualification based on a non-negotiable requirement stated by the recruiter.
    - Important Rule: if a call ends without concluding remark from the recruiter, "Early Screen-Out" does not apply
6.  Midway Screen-Out: The interview ends after some initial engagement due to a mismatch discovered during the conversation (e.g., availability issues that weren't immediately apparent, conflicting school plans, a skill gap). Important rules include topic_count is moderate (3-7) and call duration is in the middle. Recruiter states the reason for ending the call due to disqualification.
    - Important Rule: if a call ends without concluding remark from the recruiter, "Midway Screen-Out" does not apply
7.  Late Screen-Out: The interview proceeds nearly to completion but the candidate fails a final, critical criterion (e.g., rehire status, a serious attitudinal concern revealed late in the interview). Important rules include topic_count is high (>= 8) and call duration is typically long. Recruiter states the reason for ending the call due to disqualification.
    - Important Rule: if a call ends without concluding remark from the recruiter, "Late Screen-Out" does not apply

Priority 3 - Interview Analysis
- If none of above categories apply, check for these following categories

8.  Disengaged Interaction: The candidate demonstrates disinterest, unresponsive, distracted, and poor continuity. This category applies when the candidate initially engaged in the conversation. 
    - Number of topics is strictly less than 8 topics (topic_count < 8)
9.  Comprehensive Interview: Natural opening and closure; topic_count at least 8 (>= 8) expected topics or more; high-quality engagement from both parties.  
    - The candidate answers questions fully and asks relevant questions.
    - There is a concluding remark. e.g. "We'll see you on the next hiring process. Good luck!"
10. Expectation Mismatch: A full interview is conducted, topic_count is high (>= 8), but the candidate has a fundamental misunderstanding of the role, the company, or the requirements.

Priority 4 - Others
- If none of above categories apply, check for the following category
11. Others: An interview is not in English or does not fit to above categories.

Expected output:
1.  Category: [Chosen Category Name]
2.  Explanation: [Your concise explanation, citing evidence from the provided data and transcript.]

"""

\end{lstlisting}
\end{tcolorbox}

\subsection{Topic coverage prompt}\label{app:topicprompt}
\begin{tcolorbox}[
    breakable,
    colback=gray!2,
    colframe=gray!40,
    boxrule=0.1pt,
    title={
      \textcolor{black}{\sffamily\textsc{Topic coverage prompt}}
    }
]
\small \ttfamily
\begin{lstlisting}[language={}, basicstyle=\small\ttfamily,
                   columns=flexible]
You are a highly skilled conversation analyst reviewing interview transcripts. Your task is to determine whether the following interview conversation covers a specific list of topics.

**Topics to Identify:**
['INTRODUCTION', 'SOURCE VERIFICATION', 'LOCATION/COMMUTE/TRANSPORTATION VERIFICATION', 'CHECKING FOR RED FLAGS/COMMITMENT/ATTRITION RISKS', 'EDUCATION VERIFICATION', 'COMPENSATION', 'SCREENING FOR EMPLOYMENT HISTORY', 'REHIRE ELIGIBILITY CHECK', 'AVAILABILITY', 'DATA VERIFICATION', 'NEEDS ASSESSMENT (in preparation for AMCAT)', 'PROFILING', 'ICIMS \& AMCAT', 'REFERRAL']

**Instructions:**

1.  Carefully analyze the provided interview transcript to determine which of the topics listed above are substantively discussed.
2.  SOURCE VERIFICATION is to understand where candidates find the opening
3.  PROFILING is to understand whether candidates have worked in the company before, their past work experiences, or skills related to the job. If no, then it will explain briefly about the company 
4.  For each topic, consider whether the conversation includes sufficient information or questions to indicate that the topic was genuinely addressed.  Brief mentions or passing references should *not* be considered as "covered."
5.  Organize your output into three distinct sections: "Topics Covered," "Topics Not Covered," and "Explanations."
6.  For the "Explanations" section, provide a concise justification (1-2 sentences) for why each topic is classified as "covered" or "not covered," referencing specific parts of the transcript if possible.

**Output Format:**

*   **Topics Covered:** [List of covered topics, e.g., ['INTRODUCTION', 'SOURCE VERIFICATION']]
*   **Topics Not Covered:** [List of topics not covered, e.g., ['AVAILABILITY', 'COMPENSATION']]
*   **Explanations:**
    *   INTRODUCTION: [Explanation of why INTRODUCTION is covered, e.g., "The interviewer and candidate exchanged greetings and discussed the purpose of the interview."]
    *   SOURCE VERIFICATION: [Explanation of why SOURCE VERIFICATION is covered, e.g., "The interviewer asked the candidate how they found the job posting."]
    *   AVAILABILITY: [Explanation of why AVAILABILITY is not covered, e.g., "The interviewer did not ask about the candidate's start date or work schedule."]
    *   ... (Continue for all topics, both covered and not covered)
**Interview Transcript:**
"""{text}"""
\end{lstlisting}
\end{tcolorbox}

\subsection{Topic order prompt}\label{app:topicorderprompt}
\begin{tcolorbox}[
    breakable,
    colback=gray!2,
    colframe=gray!40,
    boxrule=0.1pt,
    title={
      \textcolor{black}{\sffamily\textsc{Topic coverage prompt}}
    }
]
\small \ttfamily
\begin{lstlisting}[language={}, basicstyle=\small\ttfamily,
                   columns=flexible]
You are a highly skilled conversation analyst reviewing interview transcripts. 

INSTRUCTIONS:
1. Your task is to assign a unique topic id (integer, starting from 1)
2. Output the sequence of topic id as they appear in the transcript
3. If a topic reappears in later conversations, use the same topic id again and do not create a new topic id

TOPICS TO IDENTIFY:
1. INTRODUCTION
- Hi (Candidates name), my name is ______ with the Teleperformance recruiting team. I am calling to thank you for applying to the _________ job opening and was hoping you had time to speak with me about your prior work experience. Is this a good time to talk?

2. SOURCE VERIFICATION
- Before we begin, can you share with me where you found this job opportunity listed? We like to track where candidates are finding our job openings.

3. LOCATION/COMMUTE/TRANSPORTATION VERIFICATION
- Ok, let's jump into some more specific questions. Where do you reside?
- This position is an onsite position and is located at _______, how far a drive would this be for you?
- What mode of transportation would you use?

4. CHECKING FOR RED FLAGS/COMMITMENT/ATTRITION RISKS
- The minimum age requirement for this position is 18 years of age, are you at least 18 years of age or older?
- Most of the opportunities we have are on a shifting schedule where you could even work during holidays and/or weekends. Would you be okay with this type of schedule?
- Also, would you be willing to render over time if necessary?
- Can you share with me what your main motivations are to consider this job opportunity?
- What motivated you to apply to this Teleperformance job opening?
- As you consider your career goals, how long could you envision working for Teleperformance and performing the duties of this job?
- How do your short and long term goals align to the opportunities this job presents?
- The training process we offer is mandatory to attend.
- Do you have any planned vacation in the next 6mos?
- What are the activities that may affect your attendance or any prior commitments that you have to attend to?  

5. EDUCATION VERIFICATION
- What is your highest level of education completed?
- If you are a college grad, do you plan to return to school to complete a post undergraduate degree?
- Why did you not continue with (HS/College)

6. COMPENSATION
- If we can shift our discussion to compensation expecations, that would be helpful. What is your expected salary?
- The roles that we have open right now are between the salary package range of (Salary Range of available campaigns); depending on your qualifications, you could be offered a salary package between these amounts. Would you be okay with this range?

7. SCREENING FOR EMPLOYMENT HISTORY
- I'd like to talk more about your call center experience that we spoke about earlier. Can you share with me the volume of calls you managed on a daily basis? Were these calls solving technical issues or were you responding to general inquiries?  Were you ever put on a performance plan during your prior employment?                                                                                             
- Tell me about your last 3 companies, starting with the most recent. - (Get employment dates, location)
- What was your role?
- Why did you leave the company?

8. REHIRE ELIGIBILITY CHECK
- A few more questions and we are almost done. Have you ever worked for Teleperformance in the past?
- If yes, where did you work and do you recall your prior manager's name?

9. AVAILABILITY
- If you are offered the job, when would you be able to start?
- Do you have any pending applications with other companies? If yes, what stage in the interview process are you in?

10. DATA VERIFICATION
- What are your Contact Number & Alternate Contact Number (if there's any)?
- What is your Viber Account Number?
- When is your Birthdate?
- What is your Gender:
- What is your Marital/Civil Status:
- Who is your Emergency Contact Person (Name, Relationship, Number): 
- When is your date of birth?

11. NEEDS ASSESSMENT (in preparation for AMCAT)
- Do you have a reliable internet connection? Do you have a laptop/PC available at home?

12. COMPANY OVERVIEW
- Thanks for answering my questions. Based on your background, I can see that you're a good fit for the CSR/TSR position I'm working on for Teleperformance. 
- TP is one of the largest BPO Companies. Our client has over 20 sites in the PH alone. They have different accounts that offer paid training, HMO, and leadership opportunities. Are you ready to hear more about the job?

13. ICIMS & AMCAT
- As soon as you're finished with the assessment, please let me know and send me a screenshot of your completion. Take note of the time stamp after the assessment. Then I will keep in touch with you with the hour. Do you have any questions so far?

14. REFERRAL
- Btw, we have multiple openings here in TP, can you recommend anyone that has similar skills and experience such as yourself? Could you provide me their names and phone numbers?

15. GENERAL CONVERSATION / TRANSITION
- Okay, thank you for that.
- How your day has been?

OUTPUT:
{{
  "topics": [{{"id": 1, "name": "INTRODUCTION"}}, ...],
  "sequence": [1,2,1,3,...]
}}

OUTPUT INSTRUCTIONS:
1. Do not wrap the json in markdown code fences, like ```json ... ```
2. Entire output should be a valid, single, well-formed json object
3. JSON object have to be pretty-printed with an indentation of 2 spaces
4. Output should directly begin with {{ and }}

INTERVIEW TRANSCRIPT:
\"\"\"{text}\"\"\"
\end{lstlisting}
\end{tcolorbox}

\subsection{Interview conversational move classification prompt}\label{app:conversational_move_interviewer_move}

The following is the prompt used to identify and classify conversational moves in interview transcripts. The paper's \textit{interviewer question type} variable is constructed using the prompt's \texttt{category} output. \textit{Topic-opening} questions correspond to \texttt{topic\_switch} and \texttt{same\_topic\_followup}, which respectively open a new guideline topic and move to the next subtopic within the current topic. \textit{Adaptive follow-up} questions correspond to \texttt{deeper\_followup}, while off-topic remarks correspond to \texttt{off\_template}. Moves classified as \texttt{return\_to\_template} are included in the off-topic remarks category.

\begin{tcolorbox}[
    breakable,   
    colback=gray!2,
    colframe=gray!40,
    boxrule=0.1pt,
    title={
      \textcolor{black}{\sffamily\textsc{Interview conversational move classification prompt}}
    }
]
\small \ttfamily
\begin{lstlisting}[language={}, basicstyle=\small\ttfamily,
                   columns=flexible]
 Classify the interviewer's most recent move and identify what DROVE it.
  This is a TWO-DIMENSIONAL classification: (1) what kind of move
  (category), (2) why it happened (trigger). Also report semantic distance
  from the topic's template.
</task>

<context>
  IMPORTANT: These are VOICE transcripts converted via speech-to-text (ASR).
  Expect disfluencies, garbled text, and paraphrased wording.
  Human interviewers routinely PARAPHRASE template questions -- rewording
  does NOT mean they are adapting to the candidate.

  The standard screening template follows this topic order:
  {TEMPLATE_STR}
</context>

<topic>{topic}</topic>

<template_questions>
{template_qs_str}
</template_questions>

<asked_already>
{asked_already_str}
</asked_already>

<recent_history>
{recent_history}
</recent_history>

<previous_interviewer_question>
{prev_interviewer_text}
</previous_interviewer_question>

<prior_candidate_turn>
{prior_candidate_turn}
</prior_candidate_turn>

<interviewer_move>
{interviewer_move}
</interviewer_move>

<definitions>
  SEMANTICALLY EQUIVALENT: a correct answer to one question would also be
  a correct answer to the other.

  TEMPLATE PROGRESSION: the move would have been asked regardless of how
  the candidate answered the previous turn -- it is the next slot in the
  script.

  RESPONSE-CONTINGENT: the move is conditional on something specific in
  the candidate's prior turn. Test: would this move make sense if the
  candidate had said something completely different? If no -->
  response-contingent.
</definitions>

<categories>
  same_topic_followup: progressing through current topic's template,
    paraphrasing a template question, or asking a routine factual
    checklist item for the topic.

  deeper_followup: response-contingent move that goes beyond every
    template question for this topic OR is an adaptive re-ask of a
    template question driven by an inadequate response.

  topic_switch: moving to a different scripted topic in template order.

  off_template: tangent unrelated to any scripted topic.

  return_to_template: resuming the scripted flow after a deviation.
</categories>

<triggers>
  template_progression: the move would have happened regardless of the
    candidate's prior turn. Decoupled from candidate state.

  insufficiency_driven: the move is responsive to an observably inadequate
    prior response — the candidate's turn was vague, garbled, off-topic,
    incomplete, or showed confusion/misunderstanding. The interviewer is
    re-asking, clarifying, or probing because the candidate did not
    provide the information sought. Judge from the TEXT of the candidate's
    turn, not from any external score.

  substantive_engagement: the move is responsive to a SUFFICIENT prior
    response that legitimately invites engagement. Examples: negotiation
    moves in COMPENSATION after the candidate stated a range;
    counter-offers; probing a complete but interesting disclosure
    (e.g., candidate fully answered employment history and mentioned
    managing a team -- interviewer asks about team size).

  tangent_pursuit: candidate said something off-topic and the interviewer
    followed it.

  ambiguous: cannot determine the driver from available evidence.
</triggers>

<decision_procedure>
  Apply in order. First step that fires determines (category, trigger).

  Step 1 -- Mechanical retry (silence)?
    Candidate was silent or gave no substantive response (e.g., "hello?",
    silence timeout) and interviewer is repeating to get any response
    --> classify the underlying question; trigger = template_progression.

  Step 2 -- Adaptive re-ask after observably insufficient response?
    The candidate DID respond but their response was observably vague,
    incomplete, off-topic, garbled, or showed confusion/misunderstanding
    (judge from the candidate's text), and the interviewer is rephrasing
    or re-asking the same question.
    --> category = deeper_followup, trigger = insufficiency_driven. STOP.

  Step 3 -- Template progression / paraphrase / routine checklist?
    Move is the next scripted question, semantically equivalent to a
    template question, or a routine factual item for this topic.
    --> category = same_topic_followup.
    --> trigger = template_progression UNLESS the prior candidate turn
        shows observable signs of inadequacy (vague, garbled, off-topic,
        non-responsive) on information this question recovers, in which
        case trigger = insufficiency_driven.
    STOP.

  Step 4 -- Response-contingent probe within current topic?
    Move goes beyond every template question for this topic AND would
    not make sense if the candidate had answered differently.
    --> category = deeper_followup.
    --> trigger = insufficiency_driven if the prior candidate turn shows
        observable signs of inadequacy (vague, garbled, off-topic,
        confused); else substantive_engagement.
    STOP.

  Step 5 -- Resuming after deviation?
    Previous move was deeper_followup or off_template, this move returns
    to scripted flow --> return_to_template,
    trigger = template_progression. STOP.

  Step 6 -- Different scripted topic?
    --> topic_switch, trigger = template_progression. STOP.

  Step 7 -- Otherwise --> off_template.
    Trigger: if candidate raised the off-topic content first
    --> tangent_pursuit; else ambiguous.
</decision_procedure>

<semantic_distance>
  Output semantic_distance_from_template in [0.0, 1.0]:
    0.0  = move is a verbatim or near-verbatim template question.
    0.25 = paraphrase of a template question, same intent.
    0.5  = related to a template question but probes a different angle.
    0.75 = clearly within the topic but not aligned with any template
           question.
    1.0  = unrelated to any template question for this topic.
  Compute as the minimum semantic distance over all template questions
  for the current topic.
</semantic_distance>

<output_format>
  Return ONLY a JSON object -- NO tie-breaking, NO conservative defaults.
  Report what you actually see, with confidence.

  {{
    "category": "<one of the 5>",
    "category_confidence": <0.0-1.0>,
    "category_runner_up": "<second-most-likely category>",
    "category_runner_up_confidence": <0.0-1.0>,
    "trigger": "<one of the 5 triggers>",
    "trigger_confidence": <0.0-1.0>,
    "semantic_distance_from_template": <0.0-1.0>,
    "step_fired": <1-7>,
    "matched_template_question": "<template question matched, or null>",
    "reason": "<=30 words"
  }}

  If category and runner-up confidences are within 0.15 of each other,
  that ambiguity IS the signal -- record it, don't resolve it.
</output_format>"""
\end{lstlisting}
\end{tcolorbox}

\subsection{Interview applicant response classification prompt}\label{app:conversational_move_applicant_move}

The following is the prompt used first to define interview \textit{information elements} (labeled \texttt{slots} in the prompt) and then to classify applicants' responses to interviewer questions and the information collected through those responses. In the paper, \textit{specificity deficit} refers to the prompt variable \texttt{d1\_specificity\_deficit}, while \textit{substantive deficit} refers to \texttt{d3\_clarification\_or\_evasion}. Both are coded as indicators equal to one when the corresponding prompt score exceeds $0.5$ and zero otherwise. The paper's \textit{number of unique information elements elicited} is the interview-level count of distinct entries in \texttt{slots\_filled} across all applicant responses.

\begin{tcolorbox}[
    breakable,   
    colback=gray!2,
    colframe=gray!40,
    boxrule=0.1pt,
    title={
      \textcolor{black}{\sffamily\textsc{Interview applicant response classification prompt}}
    }
]
\small \ttfamily
\begin{lstlisting}[language={}, basicstyle=\small\ttfamily,
                   columns=flexible]
SLOT_SCHEMA = {
    'INTRODUCTION': {
        'candidate_name_confirmed': 'Candidate confirms their identity/name',
        'availability_to_talk': 'Candidate confirms they have time for the call',
    },
    'SOURCE VERIFICATION': {
        'application_source': 'Where the candidate found the job listing',
    },
    'LOCATION/COMMUTE/TRANSPORTATION VERIFICATION': {
        'current_address_or_city': 'Where the candidate currently lives',
        'distance_to_worksite': 'How far the candidate is from the office',
        'commute_feasibility': 'Whether commute is feasible for the candidate',
        'transportation_method': 'Mode of transportation to/from work',
    },
    'CHECKING FOR RED FLAGS/COMMITMENT/ATTRITION RISKS': {
        'minimum_age_eligibility': 'Candidate confirms they are at least 18',
        'shift_availability': 'Willingness to work shifting/holiday/weekend schedules',
        'overtime_willingness': 'Willingness to render overtime',
        'motivation': 'Main motivations for applying to this role',
        'career_goals': 'Short-term and long-term career goals, envisioned tenure',
        'vacation_plans': 'Planned vacations in the next 6 months',
        'attendance_commitments': 'Activities or commitments that may affect attendance',
    },
    'EDUCATION VERIFICATION': {
        'highest_level': 'Highest level of education completed',
        'school_name': 'Name of the school/institution',
        'course': 'Course or field of study',
        'graduation_year': 'Year of graduation or last school year attended',
        'further_study_plans': 'Plans to return to school for further study',
    },
    'COMPENSATION': {
        'acceptable_range_lower': 'Lower bound of expected salary',
        'acceptable_range_upper': 'Upper bound of expected salary',
        'current_or_last_pay': 'Current or most recent salary/pay',
        'flexibility_signals': 'Willingness to accept the offered salary range',
    },
    'SCREENING FOR EMPLOYMENT HISTORY': {
        'employer_name': 'Name of employer/company',
        'job_title': 'Job title or role',
        'start_date': 'Employment start date',
        'end_date': 'Employment end date',
        'reason_for_leaving': 'Reason for leaving the company',
        'key_responsibilities': 'Key duties, call volume, technical vs general',
        'customer_service_skills': 'How candidate handles irate customers, conflicts',
        'performance_history': 'Whether ever on a performance plan',
    },
    'REHIRE ELIGIBILITY CHECK': {
        'prior_tp_employment': 'Whether candidate has worked for Teleperformance before',
        'prior_location': 'Where candidate previously worked at TP',
        'prior_manager': 'Name of prior manager at TP',
    },
    'AVAILABILITY': {
        'start_date': 'When candidate can start if offered the job',
        'pending_applications': 'Pending applications with other companies and their stage',
    },
    'DATA VERIFICATION': {
        'contact_phone': 'Contact number',
        'alternate_phone': 'Alternate contact number',
        'viber_number': 'Viber account number',
        'date_of_birth': 'Date of birth',
        'gender': 'Gender',
        'marital_status': 'Marital/civil status',
        'emergency_contact': 'Emergency contact person, relationship, number',
    },
    'NEEDS ASSESSMENT (in preparation for AMCAT)': {
        'internet_connection': 'Whether candidate has reliable internet',
        'laptop_or_pc': 'Whether candidate has a laptop/PC at home',
    },
    'COMPANY OVERVIEW': {
        'interest_confirmed': 'Candidate confirms interest in hearing about the role',
    },
    'ICIMS & AMCAT': {
        'assessment_readiness': 'Candidate readiness to complete the assessment',
        'questions_asked': 'Any questions the candidate has',
    },
    'REFERRAL': {
        'referral_name': 'Name of referred person',
        'referral_contact': 'Phone number of referred person',
    },
    'CLOSING': {
        'closing_acknowledged': 'Candidate acknowledges closing statements',
    },
}

For a single candidate turn in a job interview, score how INSUFFICIENT
  the response is relative to what the current topic is designed to elicit.
  Insufficiency is a measure of HIRING-RELEVANT UNCERTAINTY LEFT UNRESOLVED
  -- not a measure of length, politeness, or eloquence.

  A long, fluent answer can be highly insufficient (lots of words, no
  concrete facts). A short answer can be fully sufficient ("I left in March
  2024 because I was relocating" -- three facts, all topic slots filled).
</task>

<context>
  IMPORTANT: These are VOICE transcripts converted via speech-to-text (ASR).
  Expect disfluencies (um, uh, repetitions), ASR errors (garbled/misheard
  words), and informal spoken language. Do NOT penalize for grammar,
  disfluency, or ASR artifacts. Focus ONLY on hiring-relevant content.

  For yes/no or confirmation questions, a short "yes"/"no" is fully
  sufficient (insufficiency = 0.0). For open-ended questions, the candidate
  must provide at least one piece of relevant information.
</context>

<topic>{topic}</topic>

<slot_schema>
{slot_schema_str}
</slot_schema>

<prior_turns>
{prior_turns_str}
</prior_turns>

<interviewer_question>
{interviewer_question}
</interviewer_question>

<candidate_turn>
{candidate_turn}
</candidate_turn>

<scoring_dimensions>
  Score each from 0.0 (no problem) to 1.0 (severe problem). 
  
  d1_specificity_deficit:
    Low concrete-fact density. Vague modifiers ("a while ago", "pretty
    good", "around there"), no numbers, no dates, no proper nouns where
    the slot calls for one.

  d2_filler_share:
    High proportion of filler, hedging, or backchannel tokens ("um",
    "like", "you know", "I guess", "kind of") relative to substantive
    content.

  d3_clarification_or_evasion:
    Candidate asks for clarification, deflects, says "I don't know",
    changes subject, or otherwise does not engage the question's
    substance.

  d4_contradiction_with_prior:
    The turn contradicts something the candidate said earlier in the
    interview. 0.0 if no contradiction; 1.0 if direct contradiction.

  d5_missing_slots:
    Fraction of HIRING-RELEVANT slots in <slot_schema> that the
    interviewer's question targeted but the candidate did not fill.
    Compute as: (slots_targeted_but_unfilled) / (slots_targeted).
    If no slots were targeted by this question, output 0.0.
</scoring_dimensions>

<output_format>
  Return ONLY a JSON object:
  {{
    "dimensions": {{
      "d1_specificity_deficit": <0.0-1.0>,
      "d2_filler_share": <0.0-1.0>,
      "d3_clarification_or_evasion": <0.0-1.0>,
      "d4_contradiction_with_prior": <0.0-1.0>,
      "d5_missing_slots": <0.0-1.0>
    }},
    "slots_targeted": ["<slot_name>", ...],
    "slots_filled": ["<slot_name>", ...],
    "slots_unfilled": ["<slot_name>", ...],
    "reason": "<=30 words: what made this turn (in)sufficient"
  }}
</output_format>

<examples>
  <example>
    <topic>SCREENING FOR EMPLOYMENT HISTORY</topic>
    <interviewer_question>"Tell me about your last job -- when did you
      work there and why did you leave?"</interviewer_question>
    <candidate_turn>"I worked at Target from June 2022 to March 2024.
      I left because my family moved to Pittsburgh."</candidate_turn>
    <result>
      insufficiency_score: 0.0
      slots_targeted: [employer_name, start_date, end_date, reason_for_leaving]
      slots_filled: [employer_name, start_date, end_date, reason_for_leaving]
      reason: "All four targeted slots filled with concrete facts."
    </result>
  </example>

  <example>
    <topic>SCREENING FOR EMPLOYMENT HISTORY</topic>
    <interviewer_question>"And when did you work there?"</interviewer_question>
    <candidate_turn>"Um, it was, like, a while back, I think? Pretty
      recent though, like, you know, before this current thing."</candidate_turn>
    <result>
      insufficiency_score: ~0.85
      d1: 0.95, d2: 0.80, d5: 1.0
      slots_targeted: [start_date, end_date]
      slots_unfilled: [start_date, end_date]
      reason: "No dates given; high filler; both targeted slots unfilled."
    </result>
  </example>

  <example>
    <topic>COMPENSATION</topic>
    <interviewer_question>"What range are you looking for?"</interviewer_question>
    <candidate_turn>"I'm targeting $58-65K base, but I'd consider $55K
      if there's a clear path to $65K within 12 months and the benefits
      cover health and dental."</candidate_turn>
    <result>
      insufficiency_score: ~0.05
      slots_targeted: [acceptable_range_lower, acceptable_range_upper, flexibility_signals]
      slots_filled: [acceptable_range_lower, acceptable_range_upper, flexibility_signals]
      reason: "Concrete range, explicit flexibility conditions; sufficient."
    </result>
  </example>
</examples>"""                  
\end{lstlisting}
\end{tcolorbox}

\subsection{Interview review classification prompt}\label{app:sentimentprompt}
\begin{tcolorbox}[
    breakable,   
    colback=gray!2,
    colframe=gray!40,
    boxrule=0.1pt,
    title={
      \textcolor{black}{\sffamily\textsc{Interview Review Classification Prompt}}
    }
]
\small \ttfamily
\begin{lstlisting}[language={}, basicstyle=\small\ttfamily,
                   columns=flexible]
def make_prompt(text:str) -> str:
    return f"""
You are a sentiment analysis assistant for recruiter feedback.
Classify the following recruiter comment as:
- Positive
- Neutral
- Negative
Do not classify multiple sentiments in a comment
**Output Format**
Positive, Neutral, Negative
**Recruiter Comment:**
\\\{text}\\\
"""
\end{lstlisting}
\end{tcolorbox}

%% file: Sections_appendix/appendix_transparency.tex
\section{Research transparency}
\label{app_sec:research_transparency} 

We pre-registered the experiment at the AEA RCT Registry (trial number \#15385, link: \href{https://www.socialscienceregistry.org/trials/15385}{https://www.socialscienceregistry.org/trials/15385}). The pre-registration includes details on the experimental design, the planned sample size, variables that were expected to be collected, and an outline of the hypotheses and analysis plan. In the following, we describe in more detail the mapping between the paper and the pre-registration. 

\paragraph{Sample size.} As pre-registered, we included in the experiment all applicants who applied between March 7 and June 7, 2025. For this time period, based on pre-experimental data, we calculated an expected sample size for the experiment of around 27,000 applications and stated this number in the pre-registration. However, we also noted that ``The actual sample size depends on several external factors such as demand from the firms commissioning the client companies. Therefore, the sample size is not directly in the researchers' control, meaning the actual sample size may be (substantially) higher or lower than expected.'' Indeed, our realized sample size was substantially higher than the expected one during the pre-registered period in which the experiment took place. 

\paragraph{Treatment conditions.} Our three experimental conditions were implemented as pre-registered. The randomization weights put on each treatment condition were at the discretion of the firm. We stated in the pre-registration that we expect all three conditions to include more than 10\% of applicants in a given month. During the experiment, the firm implemented fixed weights of 60\% \textit{AI interviewer}, 20\% \textit{Human interviewer}, and 20\% \textit{Choice of interviewer}. Realized fractions closely match these numbers. 

\paragraph{Excluding observations.} We pre-registered that we would exclude any applications that were not invited for an interview (and thus not randomized into one of the three conditions). Following this, we excluded 3,828 applications. As pre-registered, for analyses involving the \textit{Choice of interviewer} condition, we focus only on applicants who made a choice.

\paragraph{Key outcome variables.} As key outcome variables, we pre-registered the following variables (in the order in which they appear in the pre-registration): 

\begin{itemize}
\item Whether the interview was successfully completed 
\begin{itemize}
    \item We initially pre-registered a binary variable, but then conducted a more detailed interview-type analysis in Section \ref{app_sec:transcript_topics_types} because we received transcript data. 
\end{itemize}
\item Whether the applicant receives a job offer 
\begin{itemize}
    \item This variable is reported in Section \ref{main_sec:main_results}.
\end{itemize}
\item Interviewer score 
\begin{itemize}
    \item This variable is reported in Section \ref{subsec:interview_scores_comments}.
\end{itemize}
\item Interviewer comment (open-ended text)
\begin{itemize}
    \item This variable is reported in Section \ref{subsec:interview_scores_comments}.
\end{itemize}
\item Whether the applicant accepts an offer conditional on receiving one
\begin{itemize}
    \item This variable is reported in Section \ref{subsec:applicants_reponse_job_offers}.
\end{itemize}
\item Time from initial application until the interview takes place 
\begin{itemize}
    \item This variable was reported in a previous working paper version. The corresponding analysis has been cut for brevity in the current version.  
\end{itemize}
\item Time from initial application until a final decision has been made
\begin{itemize}
    \item This variable was reported in a previous working paper version. The corresponding analysis has been cut for brevity in the current version.  
\end{itemize}
\item Whether the applicant started their job at the respective firm 
\begin{itemize}
    \item This variable was reported in Section \ref{main_sec:main_results}.
\end{itemize}
\item Retention rate: whether the applicant still works at the firm after starting their job.
\begin{itemize}
    \item This variable is reported in Section \ref{main_sec:main_results}.
\end{itemize}
\end{itemize}

\paragraph{Further pre-registered variables.} We further pre-registered to focus on several interview transcript and interview audio variables under the condition that we would receive them, which was not clear at the time of writing the pre-registration. The firm indeed shared the transcripts, but not audio files. Interview transcript variables are analyzed in Section \ref{sec:interview_transcript_analysis}. We pre-registered the analysis of eight transcript variables. We implemented the following six (in brackets we denote the variable names as reported in the paper): vocabulary richness (vocabulary richness score), filler words and hedging (filler word frequency), turn-taking behavior (Number of exchanges interviewer-applicant), response length (Number of questions by applicant), question-answer alignment (linguistic style match index), conversation frictions (discourse marker frequency). Instead of the two pre-registered variables sentiment polarity and redundancy of information shared, we implemented the variables backchannel cue frequency and syntactic complexity score. Our conclusions are similar if we instead use the pre-registered variables. We pre-registered the applicant survey and report its results.

\paragraph{Variables not pre-registered.} Analyses that are based on applicants' standardized test scores, workers' separation reasons, and their performance measures were not part of the pre-registration. The reason is that at the time the pre-registration was written, it was unclear whether the firm would provide data access to these variables. The recruiter survey was not pre-registered. Importantly, the variables and thus the analyses of recruiter behavior in interviews (Sections \ref{subsec:transcript_analysis_recrutiers_standardization} and \ref{subsec:transcript_analysis_recrutiers_adaptiveness}) were also not pre-registered. 